\documentclass[graybox]{svmult}

\usepackage{mathptmx}       
\usepackage{helvet}         
\usepackage{courier}        
\usepackage{type1cm}        
\usepackage{color}
\usepackage{lscape}
\usepackage{wasysym}
\usepackage{supertabular}

\usepackage{makeidx}         
\usepackage{graphicx}        
\usepackage{multicol}        
\usepackage[bottom]{footmisc}

\makeindex             

\begin{document}

\title*{Instrumental Methods for Professional and Amateur Collaborations in Planetary Astronomy}

\titlerunning{PRO-AM collaborations in Planetary Astronomy}

\author{O.~Mousis, R.~Hueso, J.-P.~Beaulieu, S.~Bouley, B.~Carry, F.~Colas, A.~Klotz, C.~Pellier, J.-M.~Petit, P. Rousselot, M.~Ali-Dib, W.~Beisker, M.~Birlan,  C.~Buil,  A.~Delsanti, E.~Frappa, H. B.~Hammel, A.-C.~Levasseur-Regourd, G.~S.~Orton, A.~S{\'a}nchez-Lavega, A.~Santerne, P.~Tanga, J.~Vaubaillon, B.~Zanda, D.~Baratoux, T. B{\"o}hm, V.~Boudon, A. Bouquet, L.~Buzzi, J.-L.~Dauvergne, A. Decock, M.~Delcroix, P.~Drossart, N. Esseiva, G.~Fischer, L. N.~Fletcher, S.~Foglia, J.~M.~G{\'o}mez-Forrellad, J.~Guarro-Fl{\'o}, D.~Herald, E.~Jehin, F.~Kugel, J.-P.~Lebreton, J.~Lecacheux, A.~Leroy, L.~Maquet, G.~Masi, A.~Maury, F.~Meyer, S.~P{\'e}rez-Hoyos, A.~S.~Rajpurohit, C.~Rinner, J.~H.~Rogers, F.~Roques, R.~W.~Schmude,~Jr., B.~Sicardy, B.~Tregon, M.~Vanhuysse, A.~Wesley, and T.~Widemann}

\institute{O. Mousis, J.-M. Petit, P. Rousselot, M. Ali-Dib, F. Meyer, A.~S.~Rajpurohit \at Universit{\'e} de Franche-Comt{\'e}, Institut UTINAM, CNRS/INSU, UMR 6213, Observatoire des Sciences de l'Univers de Besan\c{c}on, France, \email{olivier.mousis@obs-besancon.fr}
\and R. Hueso, A. S{\'a}nchez-Lavega, S. P{\'e}rez-Hoyos \at Dpto. F{\'i}sica Aplicada I, Escuela T{\'e}cnica Superior de Ingenier{\'i}a, Universidad del Pa{\'i}s Vasco (UPV/EHU), Alda. Urquijo s/n, 48013, Bilbao, Spain\\
Unidad Asociada Grupo Ciencias Planetarias UPV/EHU-IAA(CSIC)
\and J.-P. Beaulieu \at Institut d'Astrophysique de Paris, UMR7095, CNRS, Universit{\'e} Paris VI, 98bis Boulevard Arago, 75014 Paris, France
\and S. Bouley \at Universit{\'e} Paris-Sud XI, CNRS, Laboratoire IDES, UMR 8148, 91405 Orsay, France\\
IMCCE, Observatoire de Paris, 77 avenue Denfert-Rochereau, F-75014 Paris, France
\and B. Carry \at European Space Astronomy Centre, ESA, PO Box 78, 28691 Villanueva de la Ca{\~n}ada, Madrid
\and F. Colas, M. Birlan, J. Vaubaillon \at Institut de M{\'e}canique C{\'e}leste et de Calcul des Eph{\'e}m{\'e}rides, UMR8028, 77 Avenue Denfert Rochereau, 75014 Paris, France
\and A. Klotz, D. Baratoux, T. B{\"o}hm, A. Bouquet \at Universit\'e de Toulouse; UPS-OMP; IRAP; Toulouse, France
\and C. Pellier, M. Delcroix \at French Astronomical Society (SAF), Commission of Planetary Observations, 3 rue Beethoven 75016 Paris, France
\and W. Beisker \at International Occultation Timing Association - European Section (IOTA-ES), Germany
\and C. Buil \at Observatoire Castanet, 6 place Cl{\'e}mence Isaure 31320 Castanet-Tolosan, France\\
Association T60, 14 avenue Edouard Belin, 31400 Toulouse, France
\and A. Delsanti \at Aix Marseille Universit{\'e}, CNRS, Laboratoire d'Astrophysique de Marseille, UMR 7326, 13388, Marseille, France\\
Observatoire de Paris-Meudon, LESIA, 5 place Jules Janssen, 92195 Meudon cedex, France 
\and E. Frappa \at Euraster, 1B Cours J. Bouchard, 42000 St-Etienne, France 
\and H. B. Hammel \at AURA, 1212 New York Ave NW, Washington DC 22003, USA
\and A.-C. Levasseur-Regourd \at UPMC (U. P. \& M. Curie, Sorbonne Universit{\'e}s), LATMOS/CNRS, 4 Place Jussieu, 75005 Paris, France
\and G. S. Orton \at Jet Propulsion Laboratory, California Institute of Technology, 4800 Oak Grove Drive, Pasadena, CA 91109, USA
\and A. Santerne \at Aix Marseille Universit{\'e}, CNRS, Laboratoire d'Astrophysique de Marseille, UMR 7326, 13388, Marseille, France\\
Centro de Astrof\'isica, Universidade do Porto, Rua das Estrelas, 4150-762 Porto, Portugal
\and P. Tanga \at Laboratoire Lagrange, UMR 7293, Universit{\'e} de Nice Sophia-Antipolis, CNRS, Observatoire de la C{\^o}te d'Azur, BP 4229, 06304 Nice Cedex 4, France 
\and B. Zanda \at Laboratoire de Min{\'e}ralogie et Cosmochimie du Mus{\'e}um, MNHN, 61 rue Buffon, 75005 Paris, France
\and V. Boudon \at Laboratoire Interdisciplinaire Carnot de Bourgogne, UMR 5209 CNRS-Universit{\'e} de Bourgogne, 9 Avenue Alain Savary, BP 47870, F-21078 Dijon Cedex France
\and L. Buzzi \at Osservatorio Astronomico Schiaparelli, Via Andrea del Sarto, 3, 21110 Varese, Italy 
\and J.-L. Dauvergne \at AFA/Ciel et Espace, 17 rue Emile Deutsh de la Meurthe 75014 Paris, France\\
Association T60, 14 avenue Edouard Belin, 31400 Toulouse, France
\and A. Decock, E. Jehin \at Institut d'Astrophysique, de G{\'e}ophysique et d'Oc{\'e}anographie, Universit{\'e} de Li{\`e}ge, All{\'e}e du 6 ao{\^u}t 17, 4000 Li{\`e}ge, Belgium
\and P. Drossart, J. Lecacheux, L. Maquet, F. Roques, B. Sicardy, T. Widemann \at LESIA, Observatoire de Paris, UMR CNRS 8109, F-92195 Meudon, France
\and N. Esseiva \at Association AstroQueyras, 05350, Saint-V{\'e}ran, France
\and G. Fischer \at Space Research Institute, Austrian Academy of Sciences, Schmiedlstrasse 6, A-8042 Graz, Austria
\and L. N. Fletcher \at Atmospheric, Oceanic \& Planetary Physics, Department of Physics, University of Oxford, Clarendon Laboratory, Parks Road, Oxford OX1 3PU, UK
\and S. Foglia \at Astronomical Research Institute, 7168 NCR 2750E, Ashmore, IL 61912 USA
\and J. M. G{\'o}mez-Forrellad \at Fundaci{\'o} Privada Observatori Esteve Duran, 08553 Seva, Spain
\and J. Guarro-Fl{\'o} \at ARAS, Astronomical Ring for Access to Spectroscopy
\and D. Herald \at International Occultation Timing Association (IOTA), 3 Lupin Pl, Murrumbateman, NSW, Australia
\and F. Kugel \at Observatory Chante-Perdrix, Dauban, 04150 Banon, France
\and J.-P. Lebreton \at LPC2E, CNRS-Universit{\'e} dÕOrl{\'e}ans, 3a Avenue de la Recherche Scientifique, 45071 Orl{\'e}ans Cedex 2, France
\and A. Leroy \at L'Uranoscope de l'Ile de France, Gretz-Armainvilliers, France\\
Association T60, 14 avenue Edouard Belin, 31400 Toulouse, France
\and G. Masi \at Physics Department University of Romeâ ``Tor Vergata'', Viale della Ricerca Scientifica, 1, I-00100 Roma, Italy
\and A. Maury \at San Pedro de Atacama Celestial Explorations, Caracoles 166, San Pedro de Atacama, Chile
\and C. Rinner \at Observatoire Oukaimeden 40273 Oukaimeden, Morocco
\and J. H. Rogers \at JUPOS team and British Astronomical Association, Burlington House, Piccadilly, London W1J ODU, United Kingdom
\and R. W. Schmude, Jr. \at Gordon State College, 419 College Dr.,Barnesville, GA 30204, USA
\and B. Tregon \at Laboratoire Ondes et Mati{\`e}re d'Aquitaine CNRS-Universit{\'e} de Bordeaux1, UMR5798,  351 cours de la lib{\'e}ration 33405 Talence\\
Association T60, 14 avenue Edouard Belin, 31400 Toulouse, France
\and M. Vanhuysse \at OverSky, 47 all{\'e}e des Palanques, BP 12, 33127, Saint-Jean d'Illac, France
\and A. Wesley \at PO Box 409, Campbell, Australian Capital Territory 2612, Australia
}

%
%
\maketitle

\abstract{Amateur contributions to professional publications have increased exponentially over the last decades in the field of planetary astronomy. Here we review the different domains of the field in which collaborations between professional and amateur astronomers are effective and regularly lead to scientific publications. We discuss the instruments, detectors, software and methodologies typically used by amateur astronomers to collect the scientific data in the different domains of interest. Amateur contributions to the monitoring of planets and interplanetary matter, characterization of asteroids and comets, as well as the determination of the physical properties of Kuiper Belt Objects and exoplanets are discussed.}


\section{Introduction}
\label{sec:1}

{Astronomy is one of the rare scientific domains} where amateurs and professionals collaborate significantly. Professional and amateur collaborations (hereafter PRO-AM collaborations) really started in the 19th century, when amateur astronomers could follow their own interests whereas professional astronomers were funded for dedicated tasks (for example, producing tables of stellar positions in order to facilitate navigation). At that time, some rich amateur astronomers even employed professional astronomers to further their astronomical ambitions; these constituted some of the first fruitful PRO-AM collaborations \cite{Boyd11}. However, by the end of the 19th century, a gap started to open between professional and amateur astronomers due to the progressive use of spectroscopy in the field of astrophysics, which required bigger and more expensive telescopes and instrumentation. A strong revival of PRO-AM collaborations occurred since the early 1980s, essentially for two reasons \cite{Boyd11}. First, the growth of Solar System exploration via robotized spacecraft missions motivated the need for round-the-clock monitoring of the planets and the use of historical archives to help understand their structure and evolution. Second, the democratization of digital imaging, the use of more affordable but sophisticated telescopes and the emergence of the internet allowed amateur astronomers to work more closely with professionals. {Nowadays, despite of these advances, the quality of data produced by amateur astronomers remains often a level below those obtained by professionals. However, amateurs can provide professionals with a large quantity of useful and reasonably good quality data taken on the long run. In this context, professional astronomers} realised that it is much easier to collaborate with a network of amateur astronomers spread around the world than persuading the Telescope Allocation Committees of front-line facilities to permit long-term monitoring of objects that may undergo some hypothetical changes \cite{Lorenz03}.
     
\begin{figure}
\begin{center}
\includegraphics[scale=.9]{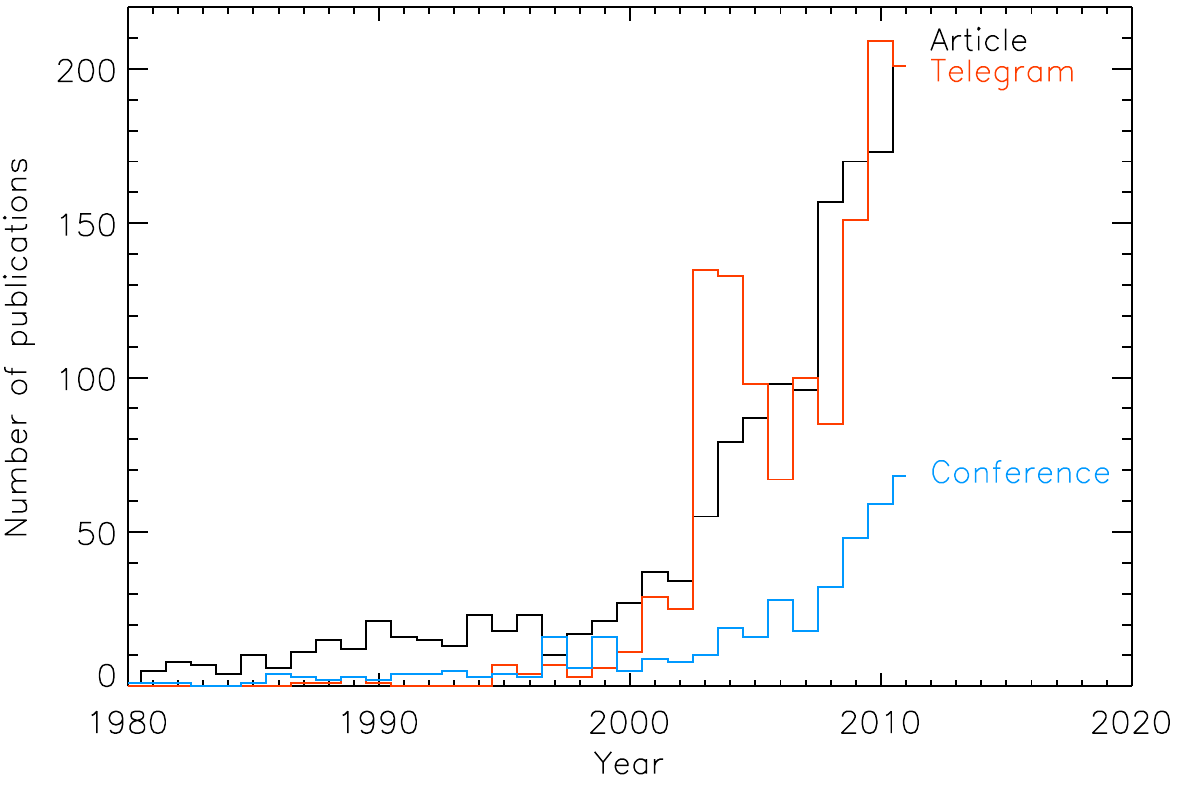}
\caption{Statistics showing a significant increase of the publications involving amateur astronomers over the years. Telegrams correspond to CBET and IAUC. {Conferences are oral or poster presentations at Meetings.}}
\label{rate}      
\end{center}
\end{figure}

Despite the exponential increase in amateur contributions to professional publications over the last decades (see Fig. \ref{rate}), reviews describing the possible fields of PRO-AM collaborations in astronomy are scarce. The aim of the present paper, written both by professional and amateur astronomers, is to address this deficiency in the field of planetary science. It describes the different fields of planetary science in which PRO-AM collaborations are effective {and regularly lead to publications} in order to show to the professional community that amateur collaborations can provide {data useful to their domain}. In addition, this paper can be used by amateurs as a guide to select the field of PRO-AM collaboration to which they could contribute as a function of their motivations and equipments. 

Section \ref{setup} is devoted to the description of the equipment (telescopes, detectors, time synchronization) required by amateurs to perform scientific contributions in planetary science. In Sec. \ref{telluric}, amateur contributions to the observation of Venus and Mars are discussed. PRO-AM collaborations provide useful information to understand their atmospheres, complementing data obtained from orbiters and large telescopes. Section \ref{matter} describes the amateur contributions to the field of interplanetary matter, which spans from the naked eye observation of meteoritic streams to the detection of impactors on other planets via telescopes and video cameras. In Sec. \ref{aste}, we discuss the techniques used by amateurs in the field of asteroid discovery, astrometry and photometry. Techniques used for stellar occultations of asteroids and comets detection in the asteroid population are also addressed. In Sec. \ref{outer}, we address the techniques used in PRO-AM collaborations for monitoring the outer planets and the type of science to which amateurs can contribute. Section \ref{phil} describes the techniques used in PRO-AM collaborations in the field of comet discovery, astrometry, monitoring and photometry. Section \ref{Kuiper} is dedicated to PRO-AM collaborations for the determination of the physical properties of Kuiper Belt Objects and Centaurs via direct observation or stellar occultations. In Sec. \ref{Exo} we discuss the possible contributions of amateurs to the research and characterization of exoplanets through the observation of transits or detection by microlensing. Conclusions are given in Sec. \ref{cls}.

\section{Requirements for observations}
\label{setup}

The choice of digital cameras and the set-up of motorized telescopes play a key role toward the achievement of professional scientific goals. In addition, the development and use of dedicated software is of major importance in standard data processing procedures. In a first step, an appropriate matching of the telescope and the camera is required in order to fit the goals of a given scientific program, since a universal setup does not exist. Together, the coupled telescope and camera constitute the basis of an astronomical setup, but some additional instruments might be added according to the projected scientific goal: a filter wheel (with the appropriate filters), an adaptive optics corrector, or in some cases a spectrometer.  Table \ref{tabsum} summarizes the appropriate equipment for each proposed research topic in this article. This chapter helps in selecting the right instrumentation.

\subsection{Telescope requirements}
\label{size}

In many cases planetary studies require high angular resolution. Different factors control the resolution: diffraction (diameter, obstruction), optical quality (aberrations, glass composition), mechanics (flexures, dilatations, focusing, equilibrium), environmental conditions (turbulences due to the tube, the dome, the building, and the weather). All of these factors must be assessed and the failure to address even one of them directly degrades the final resolution.

The telescope mount is also an important choice. Amateur telescope mounts are generally equatorial and based on a worm drive that has the inconvenient tendency to generate periodic oscillations. The quality of the worm must be measured before buying the mount\footnote{See http://demeautis.christophe.free.fr/ep/pe.htm}. Some motor controllers can correct the periodic error, allowing the use of an imperfect equatorial mount. The next generation is based on direct drive motors and/or absolute encoders on both axes. These technologies avoid periodic errors and should be common in premium amateur telescope mounts in the coming years. Note that the direct drive technology is more sensitive than the worm drive to the positions of the instruments placed on the mount\footnote{http://www.dfmengineering.com/news\_telescope\_gearing.html\#chart}.

All parts of the telescope (tube, mount and pier) must damp vibrations. Even with perfect optics and telescope drive, some factors can produce undesired forces acting on the mechanics (e.g., wind, resonances of proper frequencies). The mechanics must efficiently damp these effects. Another constraint concerns the most frequently encountered classical German equatorial mounts. They are very compact, but the so-called ``meridian flip'' induces a rotation of 180$^\circ$ of the observed field of view. The use of calibration frames (e.g., flat fields) must take into account the tube orientation choice. Many mounts are equipped with a GOTO system that is generally presented as a pad or a computer linked to the mount; it computes the celestial coordinates of the object and send them to the drive motor. It is important to verify that the accuracy of the GOTO system is compatible with the precision needed. 

Recent progress in electronics, mechanics and computer science allows the building of robotic observatories. These remote controlled observatories can provide very high duty cycles by optimizing the ``time on the sky''. When they are autonomous (i.e., no human presence) the requirements on hardware and software are significant. A robotic observatory setup is generally fixed to keep the same calibrations from night to night.

\subsection{Detectors}
\label{ccd}

Many manufacturers promote all kinds of cameras for astronomy. The characteristics described in this section are important to match the science goals. Sections \ref{cmos} and \ref{emccd} describe the camera types that can be used by amateur astronomers. Section \ref{high} is devoted to the specificity of high angular resolution often demanded in planetary science. 

Digital cameras are based on a matrix of pixels that convert photons into electrons. The quantum efficiency of the conversion, the maximum of electrons per pixel, the size and the number of pixels are the main factors of such a matrix. The electronics associated with the chip detector play also an important role with respect to the scientific constraints. The gain gives the conversion from electrons to analog-to-digital units (ADU) and is expressed in electrons/ADU. A high value is usually used for bright objects (planetary surfaces) and a low value is for deep sky. Some cameras allow changing the gain by software, thus giving a high versatility for various topics. The readout noise adds a stochastic component to the signal \cite{Howell06}. Low readout of noise is always preferable, but the value increases as the readout speed diminishes. The thermal noise is very low in recent cameras, but for exposures longer than a few seconds it remains necessary to cool the chip. However, thermoelectric cooling (by Pelletier modules) and air dissipation are sufficient for all cameras used by amateurs.

To obtain accurate photometry of the planets Uranus and Neptune, it can be useful to use a mono pixel detector as a photometer. The science described in Sec. \ref{Uranus} and \ref{Neptune} is obtained with an OPTEC SSP-3 photometer equipped with a S1087-01 photodiode manufactured by Hamamatsu. There is only one readout that generates much less noise than a matrix of pixels. In this case there is no spatial information but the whole-disk brightness of a bright planet is measured with a high signal-to-noise ratio. 

\subsubsection{CCD-- and CMOS--based cameras}
\label{cmos}

Two major technologies are found for digital matrices: CCD (Charge Coupled Device) and CMOS (Complementary Metal-Oxide-Semiconductor). From the astronomer's point of view, CCDs are based on charge displacements, pixel-to-pixel towards a readout amplifier that converts charges into analog voltages. The digital conversion is made by another electronic chip. The pixels of CMOS sensors are able to keep their charge when they are read. This allows an increase of readout speed but a part of the pixel area is used for microelectronics, so the pixel is less sensitive to the photons compared to CCDs. Recent improvements of CMOS, particularly the reduction of the readout noise, led to the concept of sCMOS (the s means Scientific terminology used by some camera manufacturers). There are three main families of charge transfer technologies for CCDs \cite{Martinez98}: the full frame (no frame buffer), the frame transfer (a buffer matrix is used to store the image before reading), and the interline transfer (column buffers store the image before reading). A full frame CCD does {not lose} any area of the matrix to record photons but it must use a mechanical shutter to avoid smearing of charges during the transfer.

The CCD and CMOS technologies continue to improve. The use of microlenses over the pixels now increases their quantum efficiency. Meanwhile CCD chips now often use interline transfer technology which eliminates the need for mechanical shutters. Interline CCDs with on-ship microlenses are currently the basis of analog video cameras used in stellar-occultation observations (see Sec. \ref{occul}) and in other fields (see Sec. \ref{streams}, \ref{showers} and \ref{Moon}), when a fast acquisition with a very accurate timing is required (video astronomy is indeed one of the hardware solutions for this issue; see Sec. \ref{time}). The different models of Watec 902H or Super Circuits PC164C are thereby sensitive and inexpensive cameras. A few other video cameras, such as the Watec 120N or the new Watec 910HX, provide an additional integrating function allowing deeper magnitudes at the cost of time resolution. On the other hand, Digital Single Lens Reflex cameras (DSLR) use mainly CMOS. Manufacturers propose various acquisition functions via a dedicated data processing chip. The price is attractive but the images rate is generally too low for planetary imaging.

\subsubsection{EMCCD technology}
\label{emccd}

One of the main constraints for ground-based, high-angular-resolution planetary imaging is the limitation of angular resolution due to atmospheric turbulence distortion. The main seeing parameters dependence (time, angular isoplanarity patch, and Fried parameter\footnote{Defined as a measure of the quality of optical transmission through the atmosphere due to random inhomogeneities in its refractive index. These inhomogeneities are primarily due to tiny variations in density on smaller spatial scales, resulting from random turbulent mixing of larger temperature variations on larger spatial scales.} dependence) require that, to overtake turbulence limitations without the use of expensive adaptive optics, the solution is to decrease exposure time and, at the same time, increase the sensitivity of the detector. Both CMOS and EMCCD (Electron Multiplying CCD) compensate for these considerations and partially fill the lack of adaptive optics in amateur astronomy. EMCCD technology, described here, increases drastically the sensitivity of CCD imaging systems. 

EMCCD cameras are based on a classical CCD. Between the pixel matrix and the readout gate, a special pixel register is added. The pixels of this register are masked from the incoming light and are polarized with high voltage (typically higher than 50V). Under such a high voltage, the electrons transferred in the register are multiplied by a factor that can reach a few hundreds at the exit of the register. The goal is to multiply each incoming electron to give an output charge constituted of a number of electrons always higher than the readout noise. In this way, the readout noise becomes negligible compared to the initial single electron signal. The gain factor can be tuned by software from 1 (the EMCCD appears to be a classical CCD) to a few hundred. Many physics experiments show that EMCCD technology is one of the best ways to reach ``shot noise'' limitations, instead of read noise in scientific imaging. There are actually two main EMCCD manufacturers currently (E2V and Texas Instruments).

\begin{figure}
\begin{center}
\includegraphics[scale=.65]{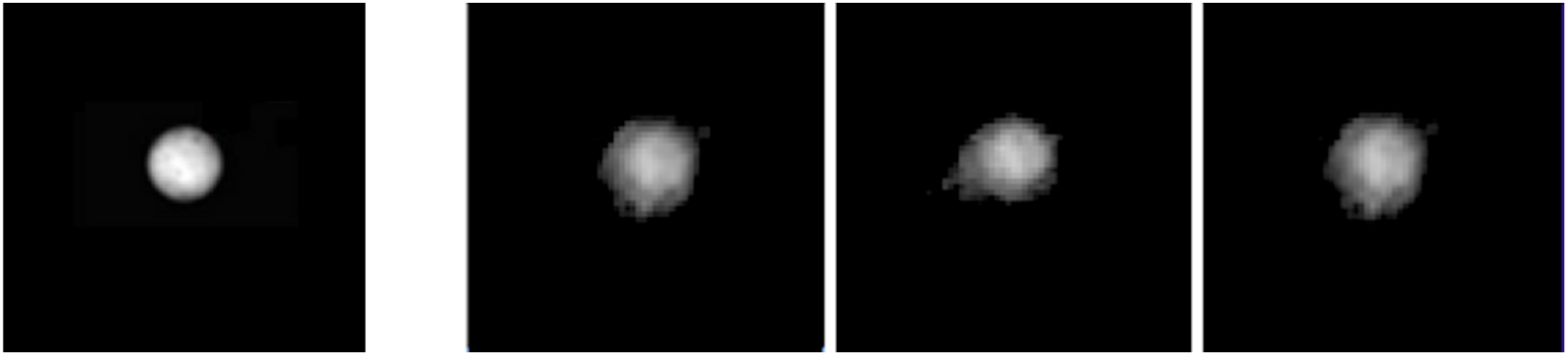}
\caption{Io's ``lucky'' imaging experiment on July 1, 2009 at 5:00 UT (planet size 0.9 arcsec). The 60-cm Newtonian telescope at the Pic du Midi Observatory is equipped with a barlow lens and an EMCCD camera with a 13-nm FWHM H-alpha filter. The exposure time of individual frames is 64 ms. {\bf Left:} image is a simulation of Io's surface at the acquisition date based on probe missions. {\bf Right:} three images selected from the same video run show the predicted features over Io's surface, allowing validation of the acquisition method (credit B. Tregon, Association T60 and Observatoire Midi Pyr{\'e}n{\'e}es).}
\label{io}      
\end{center}
\end{figure}

Already used in the field of amateur speckle interferometry (with a very high magnification involving a low photon number per unit of time and pixel surface), the EMCCD cameras allow reaching an exposure time short enough to take images faster than the atmospheric distortion speed. This regime allows a kind of imaging mode called the ``lucky imaging'' (see details of the technique in Sec. \ref{high}). The number of good quality images obtained suffers then from a very low spatial distortion. A well-known lucky-imaging probability law depends mainly on telescope diameter and Fried parameter knowledge. These ``lucky'' images are often of an outstanding quality. In the field of planetary science, this permits images of highly resolved planetary surfaces in narrow-band filters (a few tenths of nanometers bandwidth). These filters absorb most of the incoming light, and it becomes possible, with the sensitivity improvement, to obtain a quasi-monochromatic image with very high spatial resolution (see Fig. \ref{io}). Another application making use of this sensitivity increase is stellar occultation experiments. Using a 60-cm aperture Newtonian reflector, it is possible to reach magnitude 15.7 at a rate of 25 frames per second, which allows recording stellar occultations by Trans-Neptunian Objects (see Sec. \ref{occultno}).

However, EMCCD cameras still have some limitations in amateur applications. The first is the pixels' well depth, which can rapidly be saturated if the multiplication gain is set too high. This implies that an EMCCD camera for amateur astronomy should be limited to the use of very short exposure times, and so is suitable for only a few types of amateur experiments regardless of standard CCD technology. The second is the loss of linearity at high multiplication gain, which restricts the amplification domain and thus, the photometric accuracy. Another limitation is the speed reachable by the camera (and not only the link speed to the acquisition computer). This makes the EMCCD technology slower than actual sCMOS, which can acquire frames up to 400 fps. EMCCD technology is intrinsically limited to 30 fps in full frames and hardly reaches 100 fps with selection of the region of interest. The last limitation is the price, since these cameras are much more expensive than common CCD cameras.

\subsubsection{High angular resolution}
\label{high}

These last few years, video cameras made major improvements for high angular resolution imaging of planetary surfaces. {To compensate for} the effects of seeing variations, one must increase the frame rate while keeping the noise low. ``Lucky'' imaging is a technique based on these properties. One acquires as many images as possible. After acquisition, one selects the best frames, and finally stacks them. It is even possible to compensate for the residual effect of distortion due to the seeing thanks to the modern software specialized in planetary imaging processing (Registax, AviStak, AutoStakkert!, Iris, Prism, etc).

Fast cameras are the basic hardware for high angular resolution. In the early 2000s, webcams were low-cost imagers providing video mode. The result was impressive, better than what was obtained before with classical CCDs. The reason is that, for planetary imaging, it is better to have many raw images with a high noise level rather than having only few images with low noise because of the ``lucky'' imaging strategy. Moreover, the frame rate is very important because the rotation periods of the giant planets are so short that an acquisition run of ``lucky'' imaging must have a duration that is less than 100 seconds. A classical CCD has a frame rate of about 1 frame/second compared to 5 to 10 for webcams. If only 10\% of the frames reach the desired threshold of quality, it means that 10 frames are good with a classical CCD compared to 100 frames for the webcam. The signal-to-noise ratio (SNR) is proportional to the square mean root of the number of added frames \cite{Howell06}. As a consequence, the SNR of the combined image is {roughly three times} better with webcams. This compensates for the bad readout noise of the webcams. The last virtue of the webcams is that frames are acquired directly with colors. However, webcams are used less now in planetary imaging due to improvements described below.

In 2005, fast black-and-white CCD video sensors, able to run up to 30 to 60 frames per second, became available. The most famous of these are the Imaging Source DMK 21 and DMK 31. The sensors inside are Sony ICX 098 and ICX 204. The Imaging Source software has been adapted to astronomy and the camera is easy to use. But the Sony chips inside have a variable quantum efficiency, typically 50\% at most around 500 nm, and less than 30\% at 700 nm. The readout noise is around 30 electrons. These cameras are affordable, so they are still used by many astronomers who obtain good results.

\begin{figure}
\begin{center}
\includegraphics[scale=.65]{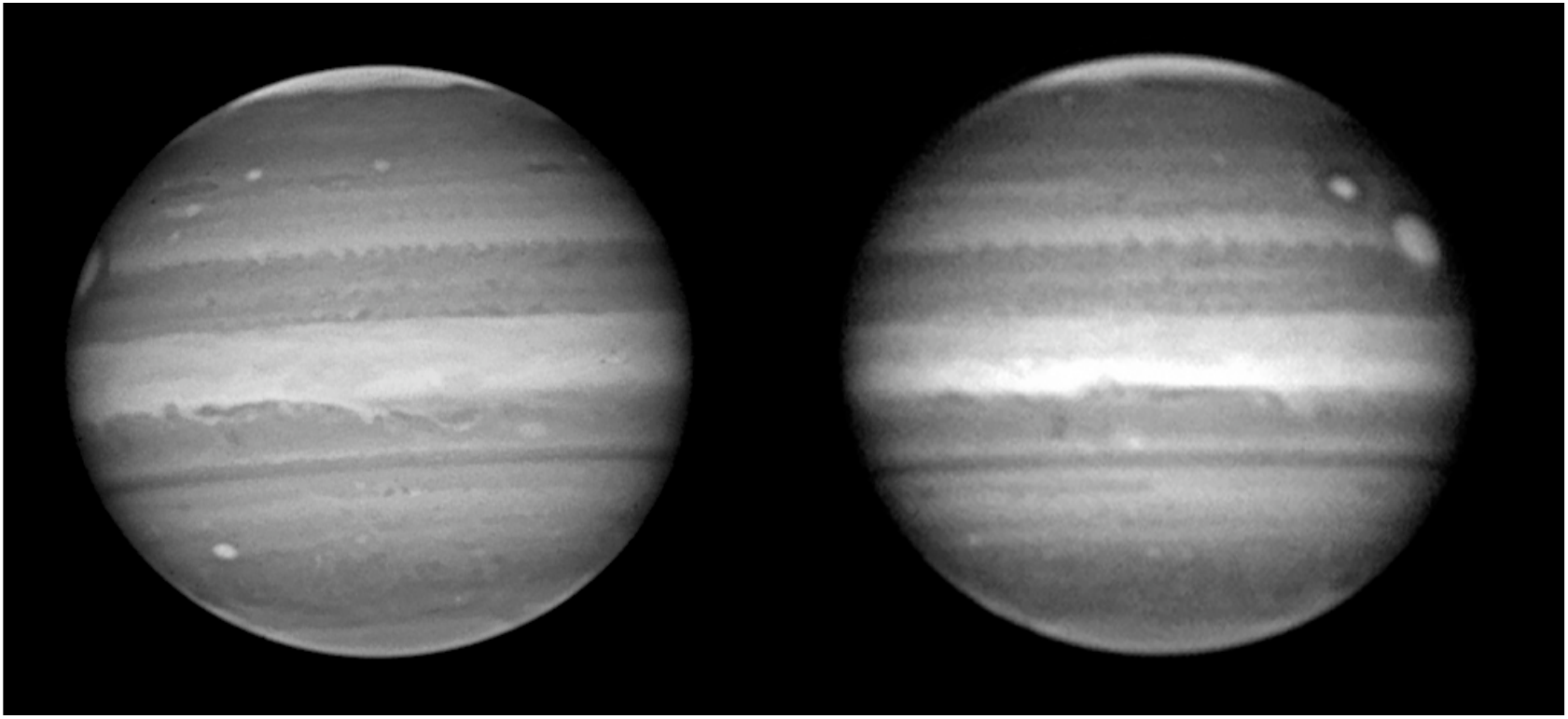}
\caption{Jupiter observed from the Pic du Midi {1.05 m} telescope in the methane band at 890 nm, with a 10-nm FWHM filter. {\bf Left:} Merlin EM247 Raptor Photonics EMCCD camera. {\bf Right:} Sony ICX285 Basler Scout camera (not the same night as the left image). The brightness of Jupiter is low in the methane absorption band and only very powerful cameras must be used to obtain good SNR ratios. This comparison shows the advantage of using an EMCCD camera (left) but the Basler Scout one (right) allows obtaining surprisingly good results at much lower cost (credit: J.-L. Dauvergne, E. Rousset, P. Tosi, S2P, IMCCE and Observatoire Midi Pyr{\'e}n{\'e}es). Details about acquisition techniques are given in Sec. \ref{iot}.}
\label{picdumidi}      
\end{center}
\end{figure}

In 2010, Sony improved the quantum efficiency and the frame rate. One of the most popular camera is now the Basler Ace 640 100gm, with the ICX618 sensor inside. This camera is able to run up to 122 frames per second. The sensitivity is twice that of the previous generation, with moderate quantum efficiency in the infrared part (53\% at 700 nm). With this generation of cameras, the result became so good that the users realised that the resolution of their images was now limited by the refraction of the atmosphere, even when they use selective RGB filters. Consequently, observers started to use some diffraction aberration correctors to compensate for this effect. They are particularly useful on telescopes larger than 200 mm. It is interesting to note that with this kind of sensor, amateurs obtain good results in a difficult domain like observing the methane band of Jupiter at 890 nm (Fig. \ref{picdumidi}). The sensitivity of the previous generation of cameras was far too low at this wavelength to achieve a good spatial resolution.

After having increased the sensitivity, manufacturers are now working to decrease the readout noise. The solution is the sCMOS technology (see Section 2.2.1). Some cameras with a low readout noise of around 1 electron already exist (Hamamatsu and Andor for instance) but are very expensive. Cheaper sCMOS sensors are becoming more and more available. The readout noise is better than that of IXC sensors. For instance the IDS Eye has a readout noise of 10 electrons with the sCMOS chip EV76C661. In the future, we hope to have commercial cameras with very low readout noise at an affordable price. 

\subsection{Spectrometers}
\label{spectro}

Recently, amateur astronomers have starting to use {spectrometers}. This has become possible for three reasons: i) the availability of low readout noise CCD cameras at a reasonable price is a fundamental point, since the light of the spectrum is dispersed; ii) commercial {spectrometers} for astronomy are now available for amateur astronomers; iii) and the number of experiments published with professionals has increased, indicating that the methods used by amateurs was adequate for professional standards. For example, a PRO-AM collaboration in spectroscopy started in stellar physics in the end of the 20th century \cite{Buil11}. In another example, the Be star database of the Observatoire de Meudon\footnote{http://basebe.obspm.fr} is fed by hundreds of spectra per year provided by amateur astronomers.

The resolving power is defined by $R$ = $\lambda$/$\Delta\lambda$ = c/$\Delta v$, where $\lambda$ is the observed wavelength and $\Delta\lambda$ is the resolved element corresponding to the spectral sharpness delivered by the instrument, c is the speed of light, and $\Delta v$ is the resolved element expressed in velocity (km/s). For a given $R$ the size of a spectrometer is proportional to the telescope aperture {(as the image scale at slit is also proportional to the telescope focal length)}, thus allowing amateurs to contribute in spectroscopy. Their telescope apertures are generally smaller than one meter, which allows the use of low-cost compact {spectrometers} that still yield good spectral performance. For instance, an echelle {spectrometer} with $R$ = 10000 equipped with a thorium-argon lamp and linked to the telescope by a glass fiber costs about the same price as a high-quality telescope mount.

The surface composition of a planet produces large spectral features at optical wavelengths. A moderate resolution $R$ $<$ 1000 is enough to study these features. The use of a slit that covers a large field of view allows studying the brightness distribution of the spectral features depending on the distance from the object. This is useful for comets. A recommended {spectrometer} is Shelyak Alpy 600\footnote{http://www.shelyak.com/}, which gives $R$ = 600, a good compromise for most Solar System bodies. This {spectrometer} reaches magnitude 17 over a one-hour exposure at red wavelengths with a SNR of 10 and a 400-mm aperture telescope.

\begin{figure}
\begin{center}
\includegraphics[scale=.65]{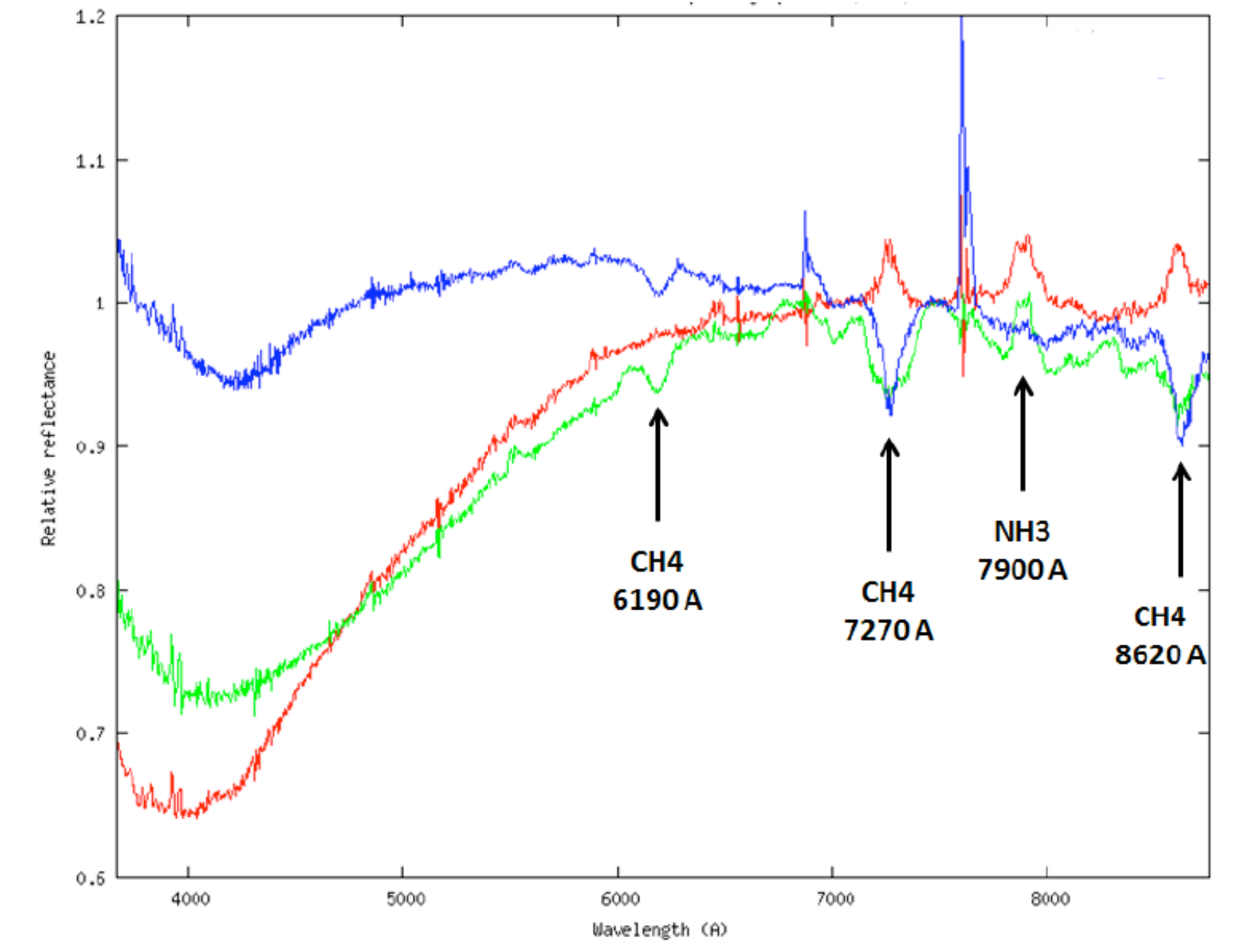}
\caption{Reflectance spectra extracted from selected parts of the Jupiter high-resolution image obtained in 2010 with a Shelyak LISA {spectrometer} and the {1.05 m} telescope of the Pic du Midi Observatory. The resolving power is $R$ = 800. Blue, green and red curves correspond to the South Temperate Belt, North Equatorial Belt and Great Red Spot, respectively. The reference spectrum is taken at the center of the equatorial region (credit C. Buil, F. Colas, J. Lecacheux and Observatoire Midi Pyr{\'e}n{\'e}es).}
\label{lisa}      
\end{center}
\end{figure}

Cameras used by amateur astronomers are based on silicon chips having a band-pass from 370 nm to 1000 nm. Beyond 1000 nm, detectors allowing long exposures are not accessible to amateurs. Despite of these limitations, the scan of a high-resolution planetary image perpendicularly to the slit of a {spectrometer} reveals the spectrum of each pixel of the planetary surface (see Fig. \ref{lisa}). In the case of planetary surface analysis, planets reflect the sunlight and add some spectral features to the solar spectrum. The reflectance spectrum reduction technique consists of dividing the planetary spectrum by the spectrum of a star, which is known to exhibit a spectrum similar to that of the Sun. By this way, it is possible to subtract the Sun's spectral features and to retain only the planetary surface spectral properties. Table \ref{tab:stars} gives a list of stars that are usable for reflectance spectra.

\subsection{Timing}
\label{time}

Accuracy requirements in astronomical observations range from nanoseconds to a few seconds or more depending on the target and the kind of sensor involved (see Table \ref{tabsum}). Commonly available devices (GPS, radio controlled clocks, Internet synchronization) can meet these requirements (even the tighter ones), but represent just the first stage of the process: sensor, acquisition hardware, system clock and software must all be tracked to ensure proper timing control. This section provides a review of commonly used setups and expected performances. 

Most of amateur astronomers manage to synchronize the system clock of their computer using an external time reference. But as soon as software is involved in timing, keeping uncertainty under control becomes a real challenge. A time source commonly used is an NTP server through the Internet. In most situations, dedicated software\footnote{http://www.hristopavlov.net/BeeperSync/} or general purpose timing software\footnote{ see a list in http://www.nist.gov/pml/div688/grp40/softwarelist.cfm} are capable of regularly synchronizing the system clock to UTC with accuracy better than 0.1 sec. This first step is already quite complex on non-real-time operating systems (Windows, generic Unix/Linux) where time management is not a priority process and is dependent on the kernel scheduling that unavoidably impedes the synchronization accuracy. Though Unix/Linux systems are not immune to this problem, they offer a far more comfortable environment to set up a sound implementation of NTP, providing easier and finer control and monitoring of what happens at the system clock level. In any case, people interested in getting the best possible system clock (that means with a negligible contribution to the error budget of the whole software setup) should investigate further the NTP protocol\footnote{http://www.eecis.udel.edu/$\sim$mills/ntp.html} and the best practices to implement it. In the same manner, accessing the updated system clock requires the same precautions. Since a very accurate timing ($<$ 1 sec accuracy) is not necessary in many astronomical observations, it is likely that most CCD imaging software applications are not designed to avoid system interrupt delays when reading the internal clock. Finally, other delays exist when software is involved and whatever the operating system is: delay between the clock read and the acquisition order sent to the CCD camera, delay between the acquisition order and the shutter opening. As a result, and since some of the above difficulties are often not solved, timing driven by software should be used only when the needed accuracy is between one to several seconds. 
 
Circumventing the harshness of software solutions naturally leads to a reliance on hardware to do the timing. This is required for stellar occultation observations, astrometry of very close near-Earth objects, and to a lesser extent for observations of meteoroid streams, fireballs, or impact flashes on the Moon. With hardware solutions, an absolute timing accuracy at the 0.01 sec level can be reached. Such performances are easily obtained with video camera recording by timing odd and even field exposures. Times are directly inserted into each video field composing a frame using a video time inserter (VTI). The timing is based on the vertical sync pulses (V-sync) which occur within 1 millisecond around the times of the exposures. In the case of integrating video camera models, measurable delays of the time-stamping need to be taken into account\footnote{http://www.dangl.at/ausruest/vid\_tim/vid\_tim1.htm}. Temporal reference can be provided by the accurate GPS 1-pulse-per-second (PPS) signal which is extracted from some GPS receivers (e.g., the Garmin GPS 18x). 

When standard CCD imaging or digital cameras are involved, the best solution is the direct timing of the shutter opening/closing. This is generally obtained using a GPS board capable of reading and timing a trigger coming from the shutter. This solution requires a calibration of the delay between this trigger and the real opening/closing of the shutter.

\section{Monitoring of terrestrial planets }
\label{telluric}

Amateur observations of the terrestrial planets Mercury, Venus and Mars are performed on a regular basis. Observations of Mercury are difficult due to the small maximum elongation from the Sun: it reaches only 28$^\circ$ in the most favorable cases, thus the planet is always at low elevations relative to the horizon. Amateur observations of Venus and Mars provide useful information for understanding their respective atmospheres, complementing data obtained from orbiters or large telescopes. Active collaborations exist in the three cases between professionals and amateurs but we restrict this section to a description of Mars and Venus observations which provide more extensive scientific cases.

\subsection{Venus }
\label{Venus}

Venus has a dense and warm atmosphere that is completely covered by clouds. The clouds display high-contrast features in UV light, which are marginally observable in violet wavelengths. Convective-like features with a horizontal scale of a few hundred kilometers are observable in tropical latitude. A large-scale horizontal ``Y''-shaped cloud feature is generally visible extending from the equator to mid-latitudes. The cloud patterns can be observed repeatedly by amateurs. Two intriguing characteristics are immediately evident: the global superrotation of the atmosphere which is much faster than the surface, and the nature of the ultraviolet colorant that makes the upper clouds well contrasted at UV wavelengths. Additionally, the atmosphere is highly variable, both dynamically (requiring extended periods of observations) and chemically (requiring spectroscopic observations). Ground- and space-based observations in UV (reflected light on Venus day-side) or in IR wavelengths (thermal radiation from the lower atmosphere escaping from the night-side) have produced significant results in studying the dynamics of Venus' atmosphere at different vertical layers \cite{Sanchez-Lavega08a,Moissl2009,Hueso2012}. Composition measurements are provided from spectroscopy at different wavelengths (UV, IR or millimeter ranges). The variation of some constituents like CO, OCS, SO$_2$ \cite{Collard1993,Encrenaz2012} is related to dynamical processes and their study is realised from observations via spectroscopic systems.

Large amateur or semi-professional facilities such as 2-m class telescopes, with near-IR imaging or spectroscopic cameras and their possibility of reading 20 days or more of continuous observing appear as a crucial step to complement observations obtained from spacecraft such as Venus Express. ESA has created the Venus Ground-Based Image Active Archive\footnote{http://www.rssd.esa.int/index.php?project=VENUS} \cite{Barentsen08}, which is an online archive of ground-based amateur observations of Venus motivated by the Venus Express mission. Many amateur observations were also acquired at the time of the June 2012 transit of Venus, thus increasing the interest in this planet by both the general public and amateur astronomers.

Understanding atmospheric processes requires long-term monitoring of the planet. Although there is now a wealth of longitudinally averaged data on the zonal cloud-level winds, the rapid variability of these winds and even their organization in local time and latitude still require observations at several timescales (from one hour to several days). The same rapidly variable distributions are observed in near IR and thermal IR, illustrating the chemistry evolution for  trace species. 

\subsubsection{Observing Venus from the ground}

Venus is never observed at solar elongations superior to 47$^\circ$. As a result, it often lies at a close angular distance from the Sun and is seldom seen during full night time. However, due to its extremely high brightness, Venus observations can be obtained during daytime. Venus' orbit has one or two greatest elongations per terrestrial year, the mean synodic period being 584 days. The western (morning) elongation occurs 0.4 terrestrial year after the eastern (evening) one; the following eastern elongation takes place 1.2 years after the last western occurrence. Elongations are governed by a long-term cycle of 8 years, and a given configuration will be repeated almost exactly after a few years. 

\subsubsection{How to observe?}

Images of Venus can be secured with any of the usual instruments used in the amateur world. However, it should be noted that with exceptions, the best UV images have been acquired from open-tube designs or with non-refractive correcting plates (newton, cassegrain, dall-kirkhamÉ) and by high-end apochromatic refractors of at least 15-cm diameter (provided that their glasses allow efficient UV transmission), because high optical quality in very short wavelengths is easier to achieve with such instruments.

The basic technique adopted for high-angular-resolution planetary images is to take short movies of the planet with webcam or camcorders, then to choose the best frames (i.e., those least degraded by poor seeing) and add the sharper frames to compose the final picture. For useful results, black and white cameras have an advantage for Venus (see Sec. \ref{ccd}). Cameras equipped with a color CCD have poor sensitivity for UV imaging; and the very low level of contrast of details that can be detected at longer wavelengths than near UV (from 400 nm to 1000 nm) also requires high-contrast cameras. Useful Venus images are obtained almost exclusively via relatively narrow band filters rather than through integrated (visible) light. 

A near-UV filter is recommended. A few filters peaking around 350--360 nm (with FWHM $<$ 100 nm) are available at moderate to high prices. An interesting alternative to the UV filter is the use of the very affordable Wratten 47 (W47) ``violet'' filter. This filter peaks at 380 nm with a FWHM of around 100 nm, and still transmits light between 400 and 450 nm, at wavelengths where the CCD is much more sensitive than below 400 nm. As a result, it can produce images of the UV markings with better sharpness and resolution than a strictly UV-pass filter, although they are of slightly less contrast. The W47 filter however requires the parallel use of an IR-blocking filter because the glass strongly leaks infrared light above 700 nm. 

Another recommended filter is a generic near-infrared long-pass filter for day-side imaging. A large number of models are available in the market. Experiments done in the amateur community over the last decade prove that filters with a transmission cut-on at $\sim$800 nm give images of slight, but noticeable, higher contrast on Venus than filters transmitting from 700 nm. 

\begin{figure}
\begin{center}
\includegraphics[scale=.5]{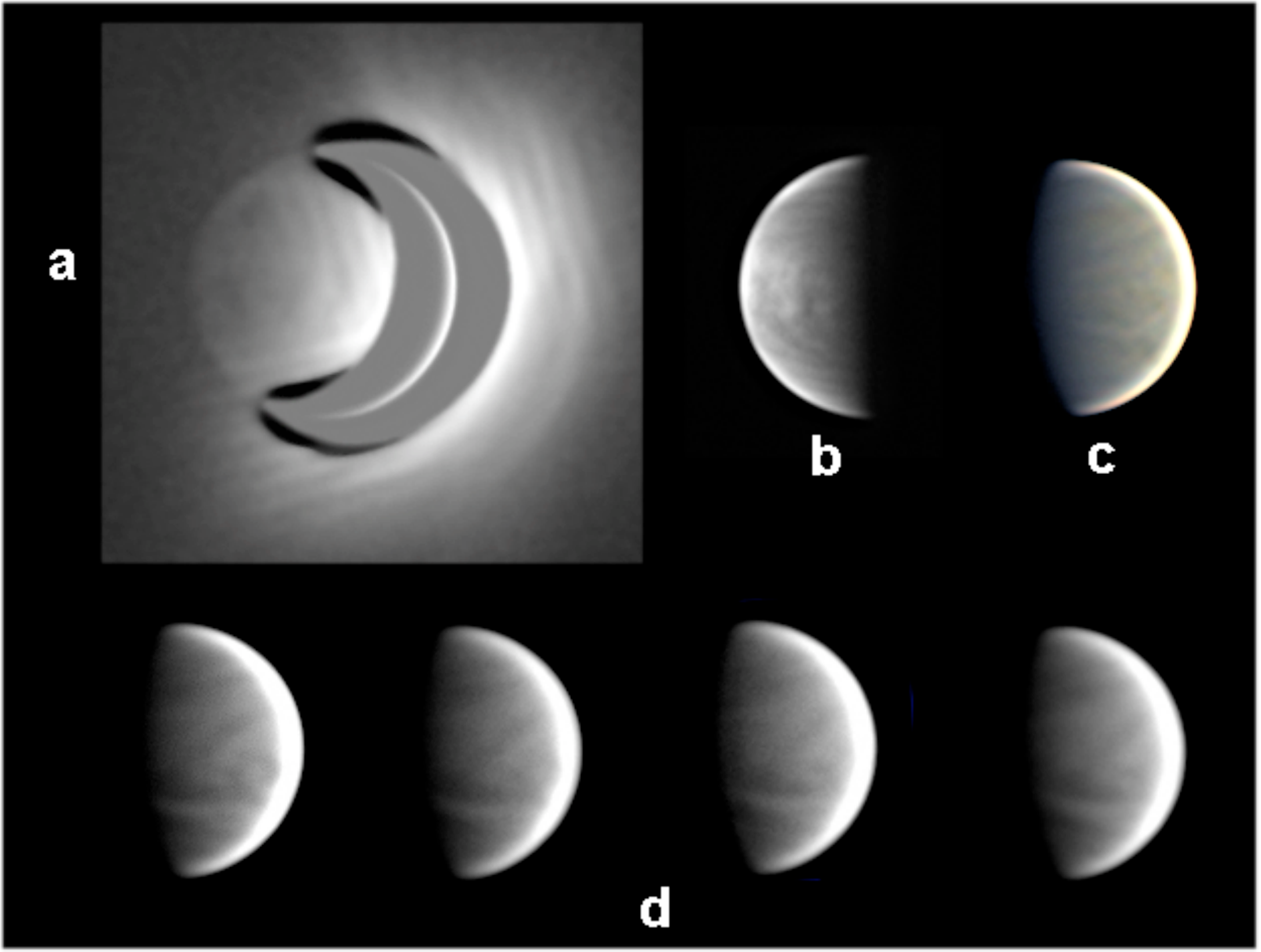}
\caption{Examples of Venus images taken by amateurs showing different features. {\bf a)} 1-micron image of the thermal signal from the surface, showing Phoebe Regio and Beta Regio as dark patches acquired with a 28-cm SCT on May 24, 2012 (credit J. Boudreau). {\bf b)} UV images acquired with a 15-cm apo refractor on March 28, 2012 (credit C. Viladrich). {\bf c)} Enhanced RGB image taken with a 25-cm Gregorian telescope on September 16, 2012 (credit C. Pellier). {\bf d)} A series of near-IR images (850 nm+) taken with the 60-cm Cassegrain of the St V{\'e}ran observatory on September 28, 2012 over more than 3 hours, showing the rotation of the atmosphere (credit G. Monachino and the Astroqueyras association).}
\label{Venus_im}      
\end{center}
\end{figure}

Another filter to get is an infrared filter with transmission centered around 1000 nm (1 micron) to image the thermal emission from the surface. Such filters (like the Sch{\"o}tt RG 1000) can be a bit difficult to find but they are inexpensive. Examples of observations in these wavelength ranges are shown in Fig. \ref{Venus_im}.

\subsubsection{What to observe?}
\label{Venuswhat}

$\bullet$ \textbf{Observation in UV (dayside):} UV light records the so-called ``UV markings'', that are induced by the absorption of a still unknown chemical component at the upper layers of the Venusian atmosphere ($\sim$65--70 km). UV surveys provided the first detection of the 4-day rotation of its atmosphere \cite{Boyer1961}. Long-term studies of UV features can be useful to detect unusual events such as the brightening events observed in 2010 and trace the overall dynamics of the upper cloud layer. In 1793 Schr{\"o}ter found the so-called ``Venus phase anomaly'': the Venus phase, i.e., the fraction of the illuminated disk visible from Earth, is 6 days, far from the theoretical value. Further investigations and interpretations could be achieved by amateurs using different filters to calculate the relative gap between observation and theorical value, as it seems to vary especially between red/green and blue/violet light \cite{Sohl1993,Heath2005}.\\

$\bullet$ \textbf{Observation in near-IR (dayside):} Near-infrared wavelengths ($>$ 700 nm) record absorption features at a lower atmospheric level (60--65 km). Although lower contrast, they are still easy to record with amateur equipment because IR wavelengths are less influenced by our atmosphere (better seeing, less scattering), and because cameras usually have high IR sensitivities. A long-term survey of features observed in near-IR is interesting as they trace the atmospheric dynamics at a slightly lower altitude and less is known about these features in comparison to UV details. Measurements of the rotation period of the planet at those wavelengths have been already carried out but merit further study \cite{McKim07}.\\

$\bullet$ \textbf{Observation in near-IR (thermal emission on the night side):} At 1000 nm, the thermal signal emitted by the surface of the planet can be recorded from Earth thanks to the low absorption of Venus CO$_2$ at this wavelength \cite{Lecacheux1993}. Correlations of dark areas recorded on images with the Magellan altimetry map of the Venusian ground could allow one to identify possible ``contaminations'' of the thermal signal by transient, low clouds.\\

$\bullet$ \textbf{Possible observations at visual wavelengths:} Visual wavelengths (400--700 nm) recorded with RGB filters are also showing some details of extremely low contrast on the dayside. If albedo markings observed in blue light (400 to 500 nm) are identical to those imaged in UV (with much reduced contrast), a long-term survey of details in green (500 to 600 nm) and red (600 to 700 nm) could also be interesting as they do not correlate exactly with features observed in adjacent bands. On the night-side of Venus when the planet is observed as a crescent, the controversial Ashen light \cite{McKim07} would also be an interesting subject of study.

\subsection{Mars }
\label{Venus}

The Mars observation season spans a period of about 10 months centered around the opposition date, which is encountered every 26 months. At opposition, Mars' size lies in the 15--25'' range, allowing this planet to reach a visual magnitude of -2.9. {Mars is the only planet whose solid surface can be seen and charted in detail in visible light from Earth}, making it a popular target for high-resolution imaging by amateurs. However scientific interest of Mars observations from the ground resides in the atmospheric phenomena, which determine the presence of clouds, changes in the surface albedo patches, which track the seasonal and inter-annual redistribution of dust; and the evolution of polar cap cycles. 

Amateur observations of Mars continue to contribute to Mars research by complementing spacecraft data and offering global (both spatial and temporal) coverage from the ground. Areas of particular interest are those where global coverage is required and high-resolution is not needed. These include (i.e., \cite{Parker1999}): a) Mars weather and clouds; b) regional or global dust storms; c) Unusual high-clouds observed at the limb of the planet; and d) long-term evolution of polar caps. Also, long-term local albedo variations are of great interest since they trace the modulations of dominating winds (activating the dust storm sites) over several years/decades \cite{Geissler2005, Fenton2007}. We stress the importance of ensuring the continuity of the observational record from the ground (essentially covering the last ~140 years), which constitutes the base of long-term studies of the planet \cite{Fenton2007}.

Similar techniques and equipment as those detailed for Venus observations are used for the observation of Mars. The most common filters for Mars observations are R V B filters for albedo examination and I images. Images in the R or I filter are useful for surface features and images in the B filter are best suited for mapping the clouds and fogs. Also, the dust clouds during storms are best mapped in R. The observation of Mars is best done with images taken at regular intervals during several hours; and because the planet's rotation differs only by 40 minutes from ours, an Earth-based observer must wait one month to observe the entire longitude range. Thus, global coverage requires the cooperation of a worldwide network of observers.

The International Society of Mars Observers (ISMO) publishes monthly reports about the Martian weather and other areas of research achievable by amateurs, and the Mars section of the British Astronomical Association\footnote{http://britastro.org/baa/} (BAA) publishes complete reports for a whole apparition. The spacecraft exploration of Mars strengthened the collaboration between professional and amateur astronomers, resulting in collaborations such as the International Mars Watch program, which grew strong in support of Mars Pathfinder and Mars Global Surveyor in the late 90s and is still active now\footnote{http://elvis.rowan.edu/marswatch/news.php} with the goal of supporting the Mars Science Laboratory.

\section{Interplanetary matter}
\label{matter}

Interplanetary solid matter consists of a large amount of tiny dust particles (micrometeoroids) and of a few larger extraterrestrial fragments (meteoroids). Small interplanetary dust particles are well known to produce meteor showers, whenever swarms of such particles enter the Earth's atmosphere with high velocities. General information about interplanetary dust and meteors showers can be found, respectively, in \cite{Grun01,Jenniskens08}. A large part of interplanetary dust originates from dust ejected from cometary nuclei, with also a significant contribution from dust released by asteroidal collisions \cite{Lasue07,Nesvorny10}. Since dust particles are slowly spiraling towards the Sun (under Poynting--Robertson effect), they build up a lenticular cloud, with increasing density towards the Sun and the near-ecliptic invariant plane of the Solar System, so-called the zodiacal cloud. Solar light scattered on interplanetary dust particles forms the zodiacal light, which appears, to the naked eye, as a faint cone of light above the western horizon in the evening or above the eastern horizon before sunrise (at least whenever the ecliptic is high above the horizon and in complete absence of any light contamination). Its study is of importance not only for Solar System science, but also for the detection of exoplanets (which may be surrounded by exo-zodiacal clouds) and of faint extended astronomical sources (such as distant galaxies).

Larger extraterrestrial fragments present in the interplanetary medium may be revealed through bolides and fireballs, induced by the entry of meteoroids in the Earth's atmosphere, as well as by impacts of meteoroids on other Solar System bodies, e.g., giant planets or our Moon or even Earth. These events are rare, hard to predict, and often chaotic - setting a limit on the amount of data that professionals can acquire. The general public may play a role in the video recording of terrestrial bolides and fireballs (later leading to fair orbital determinations), as for instance illustrated with the Peekskill event (October 1993, USA) or the Chelyabinsk event (February 2013, Russia). Amateurs have an important role in this field by helping the professional community in the scientific characterization of such phenomena, thus providing links between the impactors and the properties of their parent bodies. 

The help of amateurs is also extremely valuable for finding meteorites on Earth. Meteorites are the surviving parts from a meteoroid after ablation and fragmentation in the atmosphere and impact on Earth (or on another planet). They are time capsules from the beginning of the Solar System, yielding a chronology of the first $\sim$100 Myr and appear to come mostly from asteroids, although some younger meteorites, originating from Mars and from the Moon, have also been identified. Asteroidal meteorites show an amazing diversity in their texture and mineralogy, and illustrate the geologic diversity of the small bodies in our Solar System. These samples are invaluable in providing a detailed, albeit biased, history of Solar System evolution. In the following sections we explain how a PRO-AM collaboration helps the advancement of our knowledge in this area of astronomy.

\subsection{Meteors and meteoritic streams}
\label{streams}

While the name ``meteor'' has been used to describe any atmospheric phenomenon, it is mostly used at present to represent the effect produced by an extraterrestrial fragment entering Earth's atmosphere, becoming incandescent by friction, and inducing a fast-moving fireball or streak of light. Extraterrestrial fragments mostly come from comets or asteroids, thus providing testimonies of the Solar System formation about 4.6 billion years ago, and on small bodies structure and fragmentation processes. Indeed, comets release dust and may fragment when getting closer to the Sun on their elliptical orbits, while asteroids may suffer collisions. Such events were much more frequent in the early era of our planetary system. As a consequence, learning about the formation of meteoroids today can teach us about what happened in the distant past. The impact of an asteroid or of a fragment with size larger than tens of meters (fortunately) seldom occurs. However, everyday, hundreds of micrometeoroids enter the Earth's atmosphere. It is estimated that a total of 13,000 metric tons of material falls on our planet each year \cite{Duprat01}. 

Meanwhile, meteoroids are far too small to be detected by classical astronomical observations. In particular, they are too small to be detected by an optical telescope directly and too large to scatter sunlight efficiently. {To date, the only means to detect meteoritic streams directly is to observe them from space observatories at infrared wavelengths (around 24 $\mu$m) \cite{Vau10}.} As a consequence, because most of our observations are restricted to the Earth's atmosphere, our knowledge of the meteoroid environments of the Solar System is very poor. Attempts to detect meteors in solar system atmospheres have been extremely difficult to date, largely because instruments were not designed with such detections in mind \cite{Domokos07}. {In this context, meteors are the only indirect way to detect the presence of meteoroids and derive invaluable information on the formation and destruction mechanisms of their parent comet or asteroid. Moreover, individual meteors happen at unpredictable times, so the only way to identify a meteor shower (especially low-level ones) is to monitor the sky continuously. In turn, identified showers lead to the discovery of the parent comet or asteroid. Once the parent body is known, prediction of future meteor showers can also be performed.} There is still much to be done in this field, both from professional and amateur observers.

\subsubsection{What can amateurs bring to the topic?}

Amateurs have an extremely long history in this field since meteors can be witnessed by the naked eye. As a consequence, this field is probably one of the most ancient in astronomy.  Today amateurs are very well organized thanks to the International Meteor Organization (IMO)\footnote{http://www.imo.net/}. Born in the 1980s, this organization gathers hundreds of observers around the planet and organizes an annual conference where both amateurs and professionals can share their knowledge and experience. Why is such global organization important? Even professional cannot provide continuous monitoring, such as aircraft campaigns to observe meteor showers \cite{Jenniskens02,Vaub12}. Enthusiastic amateurs provide global coverage by continuously observing, achieving higher numbers to improve statistical studies. Since most of the data are publicly available on the Internet, anybody is free to use them and analyze them in order to find new meteor showers, or look for parent bodies. Results can easily be published in WGN journal of the international meteor organization \cite{Greaves12}. 

\subsubsection{When to observe?}
On average, there are between 4 and 10 meteors visible by naked eye per hour at any time of the night. As a consequence, anybody can observe at any time, provided that the sky is dark and clear. Chances are that such observations will catch what we call sporadic meteors, i.e., meteors not belonging to any particular shower. Meteor showers, as designed by the IAU (International Astronomical Union), correspond to streams of meteoroids following parallel orbits; these are often found to originate from a comet. The major showers are quite well known: the Perseids in August, the Geminids in December, the Leonids in November and so on. Exceptional showers also happen from time to time. The last one prior to this publication was the 2011 Draconids shower. During such occasion, amateur and professional astronomers often travel across the world in order not to miss such a unique opportunity \cite{Vaub12}. 

\subsubsection{How to observe?}
As mentioned previously, the easiest way to observe a meteor shower is simply using the naked eye. For decades, this was the only way to observe. Such observations require concentration as well as an efficient way to record the data. The knowledge of the sky and the meteor showers helps to distinguish meteors belonging to showers from sporadics. The location of the radiant (region in the sky where the meteors of a given meteor shower seem to come from) greatly helps. Photography is the next natural technique, one which has been used for decades. More recently, film photography has been replaced by digital photography. Repeated exposures of a few seconds along with a fast lens allows anyone to catch meteors, especially during showers. The latest technique is the video, allowing an absolute time-resolved observation. A high-sensitivity or intensified camera, coupled with detection software (such as MetRec \cite{Molau99}, UFOCapture\footnote{http://www.sonotaco.com/}, MeteorScan \cite{Gural97} or ASGARD \cite{Weryk08}) allows one to set up an experiment able to observe every night. Optimal set-up is a double-station observation, observing the same portion of the atmosphere using cameras located between 60 and 130 km away from each other. In this way, the 3-D trajectory of the meteor can be reconstructed and the orbit computed. Subsequently, a full analysis of the lightcurve as a function of the altitude and the atmospheric pressure can be performed, providing us with an estimate of the meteoroid strength and structure.

Radio observations are also possible, using a simple dish and receiver. The principle is to observe a distant transmitter, usually invisible from the receiver station. When a meteor appears between the two stations, the signal is reflected by the plasma, and becomes detectable from the receiver. This technique is known as a ``forward scatter'' observation and allows 24/7 monitoring of meteors, whatever the weather. However, only a poor determination of the direction and velocity are achieved.

\subsubsection{How can amateurs contribute to the data?}
Visual observations have to be sent to the International Meteor Organization in order to be of use for the scientific community. An online form is available in several languages\footnote{http://www.imo.net/visual/report/electronic}. An automated preliminary data analysis allows one to directly follow the evolution of a meteor shower with time. In the case of a major event, a full analysis follows such preliminary reduction. Video data can be automatically shared thanks to online databases such as the IMO VMDB (Visual Meteor Datebase)  or the French BOAM (Base des Observateurs Amateurs de M{\'e}t{\'e}ore)\footnote{http://www.boam.fr/}. Care must be taken when setting up such databases: in order to be useful for the scientific community, recommendations regarding the data to be saved were described in \cite{Koschny09}. The goal is to have enough elements in order to judge the quality of an observation, and to draw conclusions based on complete confidence in the data set used. For example, amateur software might not provide the uncertainties of the measurements so at a minimum the observer has to mention that they are not calculated.

\subsubsection{Future plans}
In the field of meteors, the collaboration between amateur and professional astronomers will continue. Thanks to amateurs, more and more cameras around the world provide not only global coverage, but also continuous meteor survey. By combining the data from several years, amateur and professional astronomers are able to identify otherwise unrecognizable meteor showers. The most important point is to always refer to the IMO since it is the world center of both amateur and professional astronomers working in the meteor field.

\subsection{Fireballs and meteorites recoveries}
\label{showers}

This chapter has an obvious connection with the previous one, as bolides and fireballs are simply bright meteors. We focus here only on extremely bright events leading to meteorites. For professionals as for amateurs, watching the sky in search of fireballs and watching the ground in search of meteorites are two distinct domains with few connections, and they are therefore done by distinct teams. We know that meteorites originate from the Solar System mainly from asteroids, from comets, and occasionally from impacts on the Moon or on Mars. Most meteorites are found without a proper observation of their fall, making it impossible to compute an accurate orbit and hence to determine their source region. On the other hand, astronomers have accurate orbits for about one million asteroids, and they can determine dynamical families of objects coming from the same catastrophic event. Connecting the worlds of fireballs and meteorites would be important since we know little about asteroidal matter as well as about meteorite orbits. Reliable orbits, i.e., orbits with an accuracy better than 1 AU for the semi-major axis, are known for only a dozen of meteorites.

\subsubsection{Connecting asteroids and meteors, an open scientific domain}

In the past years the main goal of space missions Hayabusa (JAXA) and Stardust (NASA) was return samples from a Solar System object to Earth. The goal is the same for the future missions OSIRIS-REx (NASA), Hayabusa 2 (JAXA) and possibly MarcoPolo-R (ESA), aimed at pristine Near Earth Asteroids \cite{Barucci11}. These sample-return missions make it possible to study extraterrestrial materials with the most complex analytical tools available on Earth, so that their nature and perhaps their origin may be investigated. On the other hand, collecting meteorites is an inexpensive way to reach this goal if their orbits can be determined. Their origin might not be as precisely known, but the study of numerous meteorites compared to the small number of samples that can be collected by expensive space missions will allow us to address statistical questions. Among the great major issues of meteoritic and asteroidal sciences are the assignation of a meteorite class to an asteroidal family, the number of parent bodies represented by the samples in our collections, the source of iron meteorites, etc. Another mystery is the dynamical mechanism that delivers meteorites to Earth. We know that Near Earth Asteroids (see Sec. \ref{NEOs}) come mainly from the Flora family (inner asteroid belt) and that this region could not be a main contributor to meteoritic material \cite{ver}. This dilemna can be answered through the determination of many meteorite orbits so that the possible meteoroid streams may be studied.

\subsubsection{Fireball observation network}

Ondrejov astronomers (Czech republic) were the pioneers in fireballs networks. Their network was based on photographic plates, allowing them to find the Pribram meteorite in 1959, and the Moravka meteorite in 2000. US astronomers developed the Prairie network in 1960 and found the Lost City meteorite 10 years later. Unfortunately, the network stopped its activity after this recovery. The Innisfree and Peekskill meteorite orbits were determined by luck because amateur witnesses used camcorders. It is important to note that the efficiency of these two pioneering networks was very low, mainly due to the photographic technique that did not allow real-time observation. Video techniques have become more popular, allowing amateurs to perform accurate measurements. Indeed, since 2000, video observations have become predominant. This technique is used both by professionals (Canadian Fireball Network) and amateurs (European fireball network) and is based on the idea of using ``fish eye'' lenses that cover all the sky. The typical network spacing is about 100 km, allowing a highly accurate measurement by triangulation for meteorite recovery. Such a network density is hard to achieve by professionals only: the main problem lies in the logistics, as each observing location must be managed by humans for efficiency. This is less of a problem for amateurs because each participant has to manage only one camera. The difficulty for amateurs lies in the networking, a difficulty that can now be solved with the Internet. In summary, amateur observations can play an important role, but only if they are included in a network. We thus encourage observers to contact an association such as the IMO.

\subsubsection{Observation configuration}

\begin{figure}
\begin{center}
\includegraphics[scale=.165]{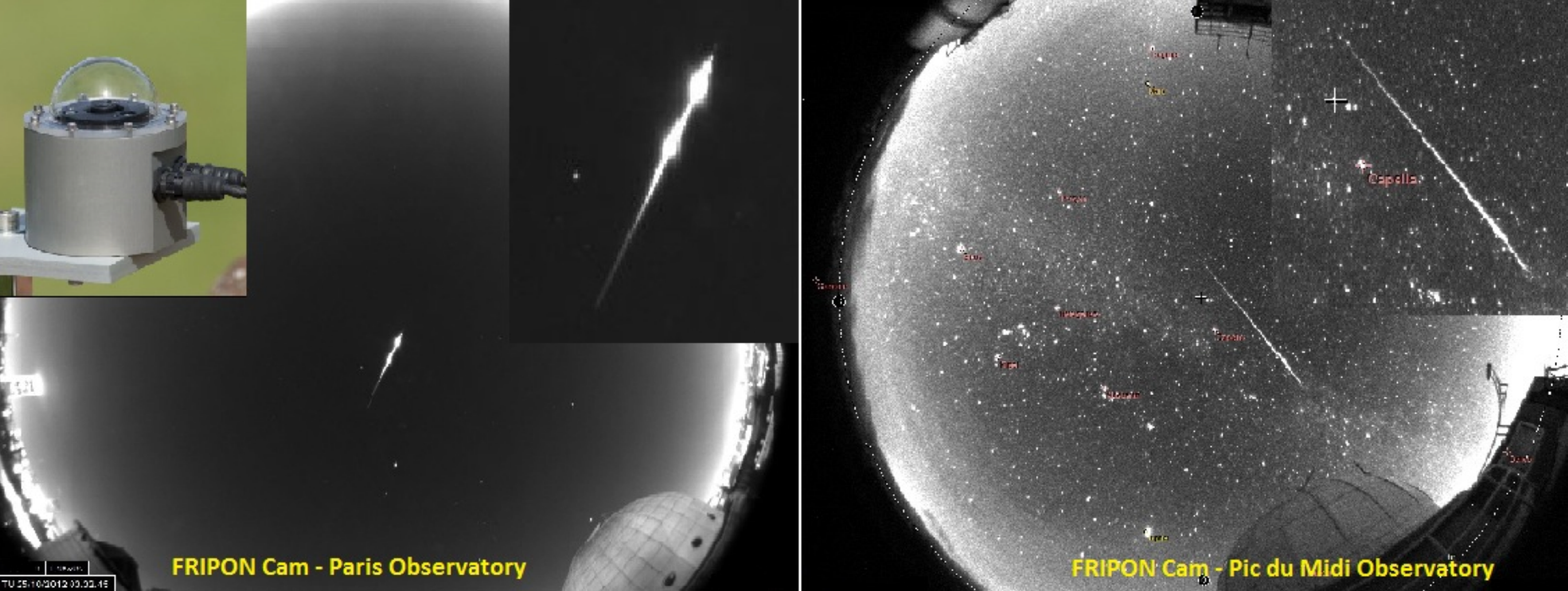}
\caption{Two fireball detections: on the left over the polluted sky of Paris and on the right over the Pic du Midi observatory. {The comparison shows that fireball detections can be made even in light-polluted areas.} The fish-eye camera with its housing can be seen on the left.}
\label{fig:fire}  
\end{center}
\end{figure}

To be effective, a monitoring network must be dense {(about one station every 100 km, since the meteors occur at 100 km of altitude)}. Every new observing station is welcome, the cost for each being about 1000 euros at current prices. Our experience showed that the efficiency of a station depends mainly on the availability of the observer, this criterion being even more important than the weather or light pollution. Fireballs as faint as magnitude 10 are easily detected from light polluted cities (see Fig. \ref{fig:fire}). This point is important for amateurs who live mainly under bright skies. The video technique has been predominant since 2000, but cameras based on CMOS chips seem to be the future for several reasons :

\begin{enumerate}
\item Resolution often better than 1 million pixels
\item Frame rate until 50 fps
\item Anti blooming features
\item Low noise
\end{enumerate}

\noindent Given the data volume, it is impossible to store full-night observations. The basic algorithm of fireball detection is based on the subtraction of the previous image from the current one. Because the motion of the sky is slow, the only remaining objects are transient events such as meteors. This approach may appear simple in theory, but it is important to avoid false detections such as airplanes, satellites, storms, birds, etc... Several software packages (UFO capture\footnote{http://sonotaco.com/soft/e\_index.html}, Asgard\footnote{http://meteor.uwo.ca/$\sim$weryk/asgard/}, Metrec\footnote{http://www.metrec.org/}) can perform this task.

\subsubsection{Meteorite orbitography}

Orbit determination must be done before the body is slowed down by the dense atmosphere, namely before it descends to an altitude of 80 km above the Earth's surface. Orbit determination requires a position and a velocity at $t_0$. The position is quite easy to determine with an accuracy of a few hundreds of meters. The velocity is more difficult to measure, as it must be done with a few frames, the problem being that the determination of the orbit semi-major axis is mainly dependent on that velocity. In the end, velocity uncertainty compromises the proper determination of the origin of the meteorite, making it the Achilles' heel of the method. A solution can be to use radar observations combined with optical, the idea being that the geometry is determined by the optical network and the velocity by radar observations.

\subsubsection{Meteorite recovery}

The ability to recover meteorites from a field is based on the quality of the computed orbit needed to determine the strewn field. The Canadian Fireball network \cite{Brown2011} succeeded in the determination of an ellipse of 1 km by 5 km for the Grimsby meteorite, a surface sufficiently well defined to organize a recovery campaign. Fireballs usually become dark, i.e., not visible, between 20 km and the ground, this step being called the ``dark flight''. During this time the meteorite's trajectory is sensitive to the wind. We thus need a model of the atmosphere to determine the shift of the strewn field compared to the case of a static atmosphere. One other problem is to estimate whether the fireball will end as a meteorite or disintegrate to dust. The analysis of the lightcurve will play an important role: there is a great chance of getting a meteorite if the fireball is still visible at an altitude of 20 km. In the end, it is essential to collect fresh material, making it necessary to organize the recovery campaign within 24 hours. This campaign must comprise several dozens of searchers. This is an area where help from amateurs might thus play an invaluable role.

\subsubsection{Conclusions}

PRO-AM connections are important if one wants to develop a dense observation network for the discovery of fireballs and accurate measurement of their physical properties. Professionals must use the data to compute accurate orbits, so that the origin of the meteorites and the location of their strewn field may be determined. They also must collect data from a variety of sources in order to decide whether or not to organize a recovery campaign. Amateurs can play an essential role in helping collect the material quickly before it is deteriorated by terrestrial alteration due to atmospheric conditions. To conclude this section, meteorite science is a good field for amateurs and professionals to working together to answer important questions about the origin of the Solar System.
     
\subsection{Giant-planet impacts}
\label{impacts}

Impacts had a profound influence on the evolution of the Solar System. Their remnants in the forms of craters are found on nearly all solid bodies in our Solar System.  Because of their great gravitational attraction, the giant planets are the most likely place to witness impacts, despite their lack of solid surfaces.  For Jupiter, recent years have shown dramatic evidence for impacts into its atmosphere of both previously identified (Comet Shoemaker-Levy 9 - hereafter SL9) in 1994, see \cite{Harrington04}) and unexpected bodies.  In the latter category four events have been recorded between 2009 and 2012, all of them discovered by amateur astronomers. The large number of amateurs observing Jupiter results in a nearly continuous monitoring of the planet during its apparition that greatly exceeds the number of observations obtained from professional telescopes.

\begin{figure}
\begin{center}
\includegraphics[scale=.170]{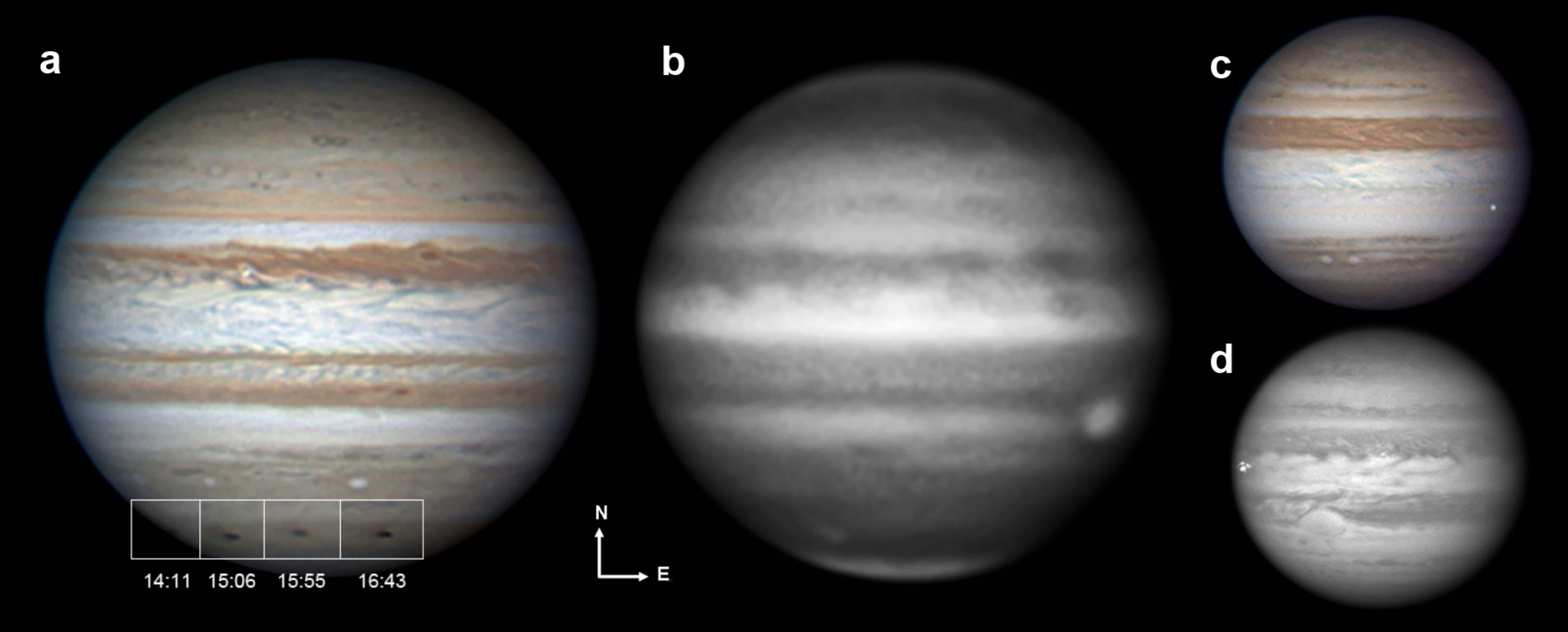}
\caption{{\bf a)} Series of images taken on July 19, 2009 revealing the 2009 Jupiter impact (credit A. Wesley). The dark debris cloud is the size of the European continent. {\bf b)} Images acquired on July 21, 2009 in a methane absorption filter where the impact debris stands out as a bright feature high in the Jovian atmosphere (credit D. Parker). {\bf c)} Flash bolide on images acquired on June 3, 2010 (credit A. Wesley). {\bf d)} Flash bolide on images acquired on September 10, 2012 (credit G. Hall).}
\label{fig:Impacts_Figure1}    
\end{center}   
\end{figure}

On July 19, 2009, an unknown body collided with Jupiter on its night side near 55$^\circ$S planetocentric latitude and 305$^\circ$W System III longitude \cite{Sanchez-Lavega10}. The object left a large-scale dark debris cloud observed on the planet for months. The first observations of the impact were obtained by A. Wesley from Australia. On June 3, 2010, a bright bolide flash was detected also by A. Wesley above Jupiter's clouds that left no detectable influence on the atmosphere \cite{Hueso10a}, followed by a similar event on August 3, 2010 and a third event on Sept. 10, 2012 (see Fig. \ref{fig:Impacts_Figure1}) . All of them were  confirmed by observations acquired by at least two observers. These impacts were unexpected because a large impact such as the SL9 in 1994 was assessed as a very rare event \cite{Harrington04} and smaller impacts such as those producing short bolide flashes were not considered as detectable from ground-based observations. For the 2009 impact, unlike SL9, none of the actual impact phases was observed. Nevertheless, significant information on the impact aftermath was obtained from several spectroscopic and imaging studies of the resulting thermal energy, composition and particulate debris. The impactor size has been estimated to be $\sim$ 0.5--1 km \cite{Sanchez-Lavega10}, based on similarities of its visible debris with respect to ``intermediate'' SL9 fragments. The possibility has been proposed that the impacting object had a significant stony component, quite different from the icy composition of SL9 \cite{Hammel10}.  

Infrared observations \cite{Fletcher11a, Orton11} confirmed this interpretation and suggested that the body was less icy than SL9 and compositionally more like an asteroid. Differentiating between such bodies is important, because Jupiter should have cleared out all asteroids from its orbit long ago. Cometary impacts are estimated to be 1,000--10,000 times more likely than asteroidal impacts \cite{Schenck04}. If this is true, then either (i) the 2009 impact was a statistical fluke, (ii) Jupiter-family comets are heterogeneous in composition, with deep interiors than cannot be detected from spectroscopy, or (iii) there is a distinct population of asteroids among bodies classified as comets, as suggested by the suspected existence of a continuum between some asteroids and comet nuclei.

Identifying the sizes of the impacting objects serves as a primary proxy for the size distribution of the large population of bodies in the outer Solar System that are too small to be detected directly.  Thus, not only do measurements of impacts provide quantitative insights into the range of Jupiter's gravitational influence, but they have the potential to determine properties of the groups from which the impactor might have originated: main-belt asteroids, quasi-Hilda comets or Jupiter-family comets, Jovian Trojans or Centaurs.

\subsubsection{How can amateur astronomers contribute?}

The observing time allocated to professional astronomers by large observatories is given competitively, thus that they are only able to observe Jupiter at most for a few days per year or the equivalent number of hours. In contrast, the large number of amateur astronomers obtaining regular observations of Jupiter and Saturn allows a nearly continuous monitoring of these atmospheres, which increases the probability of detecting random rare events. It comes as no surprise that the large impact in 2009 was discovered by an amateur and that only 7 individual amateurs (A. Wesley, C. Go, M. Tachikawa, K. Aoki, M. Ichimaru, D. Petersen and G. Hall) were successful in detecting the three flash bolides. In fact, the key to detecting impact events, particularly the short-lived bolide flashes, is monitoring the planets as continuously in time as possible. 

There are two basic observation sets that detect impacts. First, if the bolide flash only lasts for 1--2 seconds, it requires continuous filming of the planet at a high frame rate (see Sec. \ref{ccd}). Small telescopes equipped with webcams or video recorders are able to perform such detections. One of the bolides was detected with a modest telescope of only 15-cm but larger apertures (30--35 cm) are preferred to better characterize the light-curve of the flash.  Second, detection of dark debris fields within the atmosphere produced by a larger impact can be made by any standard telescope plus CCD imaging. Modest equipments can also contribute to the study of the aftermath of such events \cite{Sanchez-Lavega10}.  Impacts leave traces of particulates in the upper atmosphere \cite{Perez-Hoyos12, dePater10}; therefore images using filters that systematically block out light reflected from deeper clouds, e.g., a narrow-band 890-nm filter centered in a gaseous absorption feature of methane, could be considered a ``smoking gun'' that differentiates a dark feature that is intrinsic to Jupiter from an impact related ``scar.''

\subsubsection{Software support}

Impacts are rare and important events that mobilize both professionals and amateurs. The experience obtained in the previous four impacts indicated that a quick message to Jupiter researchers and amateur networks (see detailed information in Sec. \ref{outer}) ignites a large number of observations that can probe the nature of an impact or its atmospheric response to a large impact. 

Video monitoring is essential to detect the short flashes extending only 1--2 seconds. The lightcurves of the flashes allow the energy released by the impact and the size of the impacting object to be determined. Free software developed by amateurs can be downloaded from the Planetary Virtual Observatory and Laboratory website (PVOL)\footnote{http://www.pvol.ehu.es}. The software is capable of doing an automated search for impact flashes during any video segment. This is valuable for amateurs who do not have the time to examine what are often hours of observations at the frame-by-frame resolution for anomalous bright spots. Information about other software projects related with bolide searches on Jupiter are also available on that webpage. For intermediate size objects, i.e. between those producing short flashes and those producing large-scale debris fields such as the 2009 impact, fast follow ups in methane band absorption filters may confirm the presence or absence of particulate materials. For large impacts, the dark debris fields are advected over weeks or months by the Jovian circulation at levels close to the tropopause, allowing the study of the dynamics of this altitude.

\subsection{Impact flashes on the Moon}
\label{Moon}

\begin{figure}
\includegraphics[scale=.65]{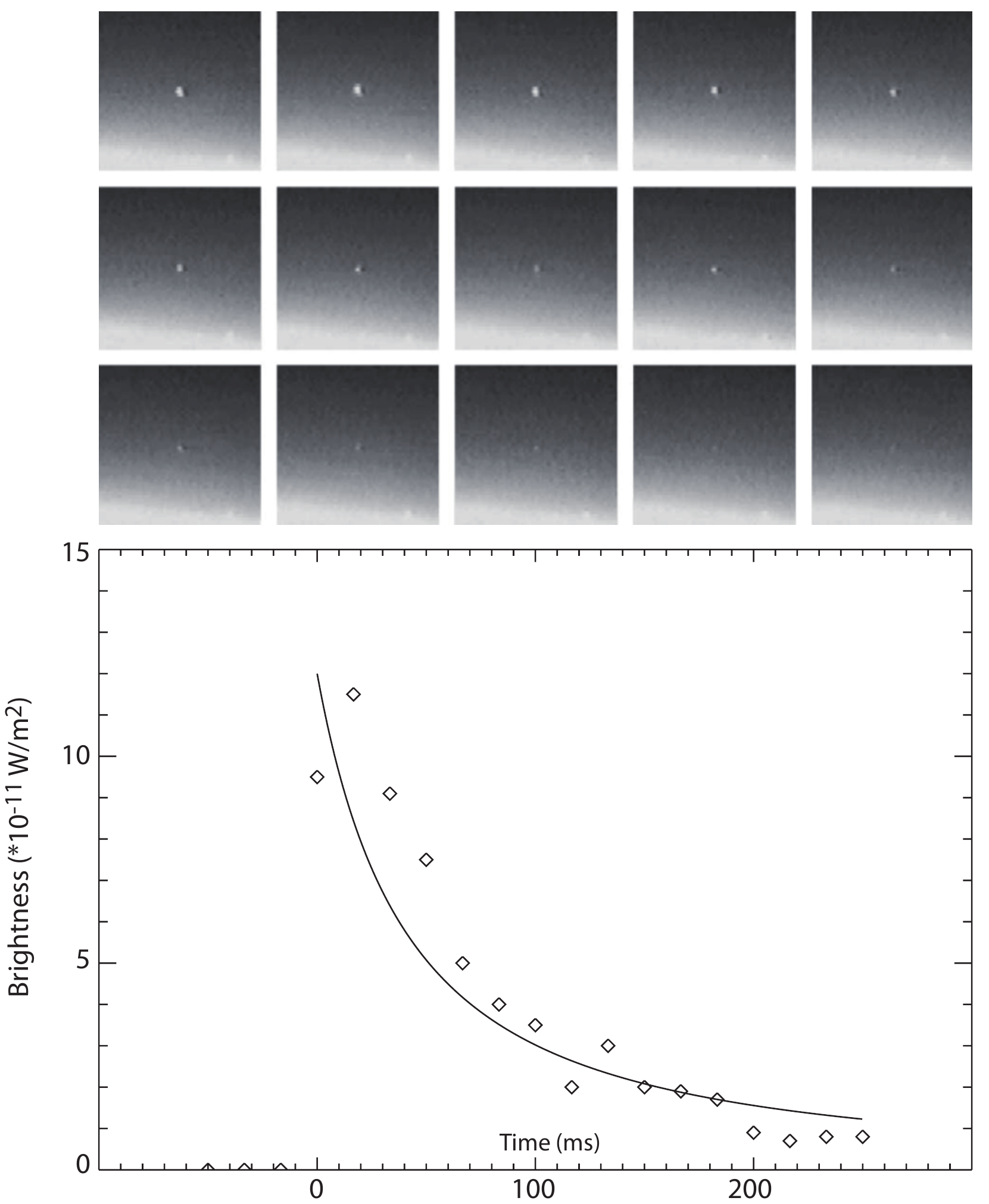}
\caption{Example of a time sequence at 60 images per second of a lunar impact flash, with corresponding brightness curve (open diamonds) (data from \cite{Yana02}). The black line represents a model of impact-generated emission associated with a cloud of optically thin cooling droplets (a droplet radius of 60 micrometers has been used to fit the data, and the thermal evolution has been calculated following equations presented in \cite{Bou12}). {The light curve of the flash was obtained using a black-and-white CCD video camera (Ikegami ICD- 42DC, CCD: TI-TC277-40) attached directly to a 20-cm Newtonian telescope with no filter.}}
\label{fig:flash}  
\end{figure}

Transient changes at the Moon's surface have been reported for several centuries, in most cases using relatively modest instruments run by professional or amateur astronomers \cite{Cam72}. These events are generally referred as Lunar Transient Phenomena (LTP). In the last two decades, LTP were recorded using commercial video cameras, and may be now accurately defined as transient luminous events occurring on the non-illuminated fraction of the lunar disk with a magnitude ranging from 3 to 10 that typically vanish in a fraction of second (see Fig. \ref{fig:flash}). The cause of these phenomena has been now clarified. They are seen as the result of hypervelocity impacts (11--72 km/s) of small fragments of asteroids or comets at the surface of Moon \cite{Ortiz00}. The term ``lunar flashes'' or ``impact flashes'' is thus now commonly used. Details on the origin of the increase in brightness remain however under debate. The release of kinetic impact energy is known to induce melting, vaporization and even ionization of the target rocks, all these phases being involved at some stages in the origin of the radiation \cite{Nem98,Artem01,Yana02}. Recently, the photometric curve describing the radiation peak and its subsequent decay, and the correlation between duration and magnitude of these events, have been explained to first order as the thermal emission of an optically thin expanding ejecta cloud of micrometer-sized liquid droplets \cite{Bou12}.

Regular monitoring of the lunar surface from ground-based observatories distributed around the world is essential to constrain the amount and size distribution of interplanetary matter entering the Earth-Moon system. Such data are critical to quantify the present impact hazard at the surface of the Moon. Considering the technical simplicity and inexpensive cost of the equipment required to produce data worthy of scientific analysis (see below), amateur astronomers can play a major role in this field of research by joining professional observational networks . 

\subsubsection{PRO-AM collaborations in lunar flashes detection}

In the late 1990s, and during the Leonids 1999 and 2001, two international teams including amateurs and professionals performed the first recordings of lunar flashes. The first team, located in the United States, contributed to the development of networks dedicated to the observation of these phenomena (ALPO -- Association of Lunar \& Planetary Observers; IOTA -- International Occultation and Timing Association) and reported the first observations \cite{Dun00,Cud02,Cud03}. The expertise acquired by several amateur astronomers involved in the early detections allowed them to participate in the creation of a group of professional observers based at the NASA Marshall Space Flight Center, which is still active \cite{Coo06,Coo07,Su08}. The second group located in Spain was composed of astronomers from different Spanish laboratories and amateurs including observers from the observatory of Mallorca \cite{Ortiz00,Ortiz02, Ortiz06}. In the 2000s, the International Meteor Organization has also shown activity in this field. PRO-AM collaborations have also allowed people in Japan to detect several lunar flashes during the 2004 Perseids \cite{Yana06} and the 2007 Geminids \cite{Yana08}. Since then, several detections were also performed  by groups of French and Italian amateur astronomers \cite{Ba12,Spo12}.

\subsubsection{When to observe lunar flashes?}

Detection of lunar flashes is only possible on the non-illuminated fraction of the Moon since the illuminated side is too bright compared with lunar flash magnitudes. During gibbous phases of the Moon, the lit side prevents lunar flashes from being observed on the non-illuminated side. Just after this phase and before the new Moon, Earthshine (the indirect illumination of the lunar surface by reflected Sun-light from the Earth) may also limit detection. The period immediately around the new Moon is also not optimal as the Moon has low elevation above the horizon and may not be observed anymore after astronomical twilight or before astronomical dawn. Optimal periods of observations therefore extend from a week before new Moon (last quarter) to a week after new Moon (first quarter) excluding 2--4 days centered around the new Moon.  Optimal conditions of observations depend on the location of the observation point at the surface of Earth, and also vary with season, and should be computed from an astronomical ephemeris\footnote{http://www.imcce.fr/en/ephemerides/}. At mid-latitude regions, it is possible to search for impact flashes for up to 20--30 hours per month (best conditions are naturally in winter). 

\subsubsection{Performing the observation: technique for video detections}

The following list includes the required equipment for lunar flashes observations, which meets criteria for scientific analysis. These criteria include the capability to determine the location and time of an impact flash on the lunar surface, and a calibrated photometric observation of the luminous event. 

\begin{enumerate}
\item The camera: a lunar flash is a very short event (typically a few tens of ms). The frame rate of the video camera is therefore a critical parameter and should be faster than 25 frames per second. The inexpensive black-and-white Watec 902H and 120N (1/2'' sensor) cameras have been successfully tested for this kind of observation. Such cameras have a wide range of other applications (such as observations of meteors or stellar occultations by asteroids).

\item The telescope: in order to perform a global monitoring of lunar flashes, the field of view should be comparable to that of the Moon (30 arcminutes).  With a 1/2'' sensor, a 30' field of view implies a short focal length telescope (between 70 cm and 1 m). Newtonian telescopes from 15 to 25 cm in diameter with F/D 4 can be very efficient. A 20-cm Schmidt Cassegrain telescope with a F/6.3 focal reducer can be used but the field of view is too small for global monitoring. A good and well-tested solution combines a 35-cm Celestron (C14) with the Hyperstar optical system (F/1.9), thus reducing the focal length of the instrument to a value of 68 cm.

\item Time recording: impact flashes may be recorded simultaneously by several observers, which provide an essential confirmation of the nature of the event against other potential phenomena (cosmic rays, reflections from space debris, ...). Recording the time of the event is therefore a critical aspect of the observation. Solutions for this problem are easily implemented using the computer clock updated at a NTP sever or a GPS signal inserted into the video signal of the camera (see Sec. \ref{time}).

\item Detection: with this minimum equipment, a few detections per lunar month may be achieved under 100\% clear weather conditions during the appropriate periods of observations. This number may be increased by focusing on meteor showers which display generally higher rates than sporadic impacts. Continuous observations and post-processing of the data is the best solution. Software such as Lunarscan or UFOcapture are generally used to search for changes between individual images. The characteristic of the detected changes are then analyzed to confirm the detection of an impact flash (intensity, duration, detections from several telescopes ideally placed at different locations).
\end{enumerate}

\subsubsection{Future plans for PRO-AM joint observations of lunar flashes}

Today, the most efficient observation program is run in the U.S. with more than 260 detections in 7 years. However, these detections are only performed in one region of the world and necessarily represent only a fraction of the total number of fragments of asteroids and comets hitting the Moon every year. Amateurs are welcome to join professional observing programs in order to increase the number of detections substantially. An international network (ILIAD -- International Lunar Impact Astronomical Detection) is currently being created by a group of French scientists \cite{Bou12}. Observers in Morocco and Mongolia have already joined it. This network seeks to expand in the coming years, and volunteers and initiatives from various amateur observatories are welcome. By participating in such a project, amateur astronomers can also cooperate with professionals by writing publications or participating in international conferences. Today, camera technologies rapidly change, and camera systems will emerge that are increasingly suitable for the observation of lunar flashes. Cameras will be more sensitive, faster, and will cover wavelengths outside the visible spectrum. All these improvements should allow both professionals and amateurs to increase the number of detections.

\section{Observations of asteroids}
\label{aste}

\begin{figure}
\begin{center}
\includegraphics[scale=1.00]{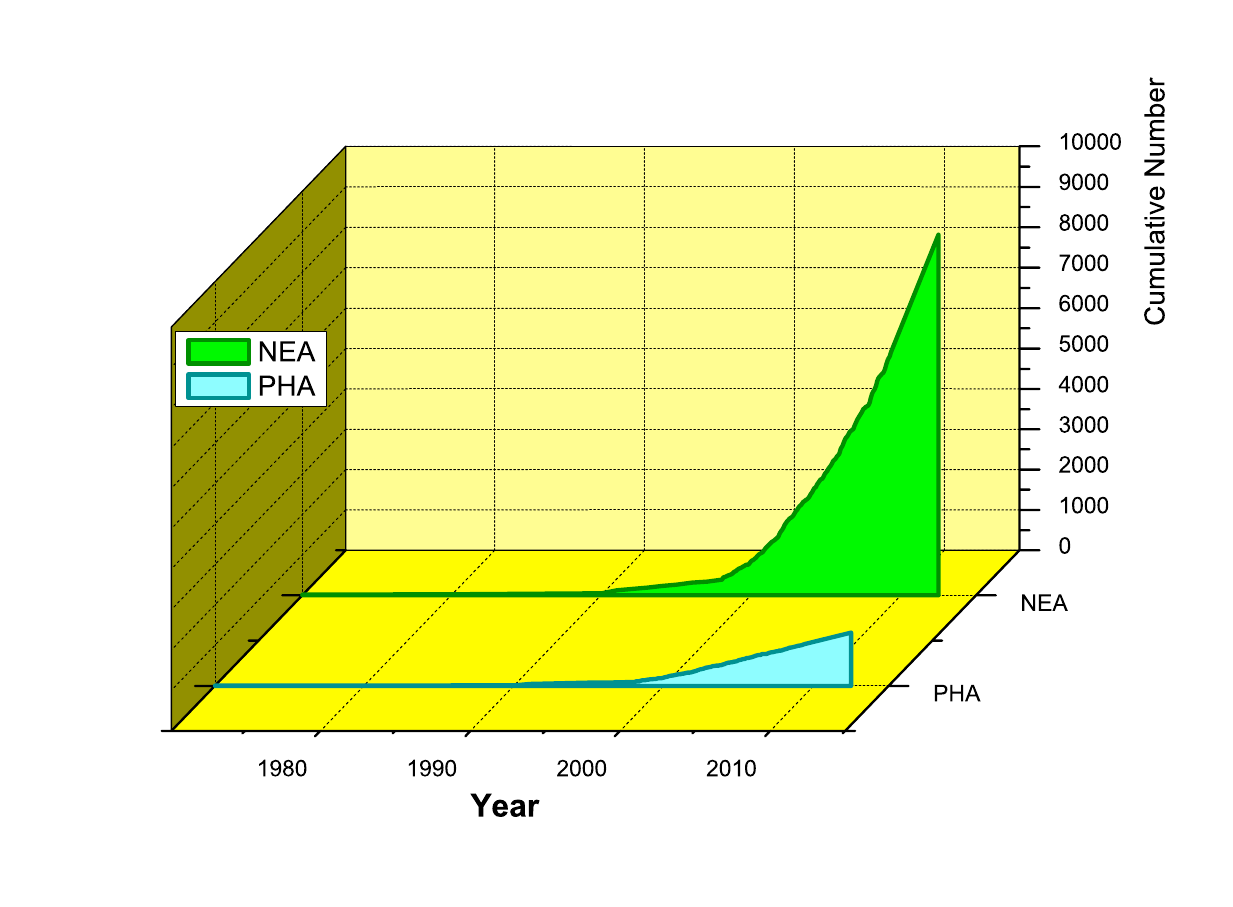}
\caption{Statistics of discoveries of NEAs and PHAs as presented in 2012. The last two decades show the large interest of professional and amateurs for discovery and precovery objects of our Solar System.}
\label{stat}      
\end{center}
\end{figure}

As of this writing, the asteroidal population contains more than 600,000 discovered objects\footnote{An up to date list is available at http://cdsarc.u-strasbg.fr/cgi-bin/nph-Cat/txt/max=588132?B/astorb/astorb.dat}. Most of them are located between Mars and Jupiter, in the so-called the Main Belt (MB) (Trans-Neptunian Objects and the Centaurs are discussed in Sec \ref{Kuiper}). Approximately 10,000 objects are intersecting the orbits of terrestrial planets (Fig.\ref{stat}). These are the so-called Near-Earth Asteroids (NEAs). More than one thousand of these NEAs have Minimum Orbit Intersection Distance (MOID) below 0.05 AU with respect to Earth: these objects are called Potentially Hazardous Asteroids (PHAs). Two distinct groups of asteroids are also orbiting on trajectories similar to that of Jupiter $60^\circ$ ahead and behind the planet, i.e., the so-called Greeks and Trojans groups. 

The first asteroid discoveries during the 19th century initally generated a high involvement of the research community but astronomers progressively lost interest in their study during the following decades. In the 1970s, lunar exploration showed a huge discrepancy between the number of fresh impact craters and the known number of NEAs. This led to new surveys of asteroids, mainly aiming at detecting these bodies, and that exponentially increased the size of their population \cite{mb00}. For a few decades, stellar occultations and radar have been used to access information about asteroid shapes. In the 1990s, CCD technology replaced progressively the photographic searches \cite{mb0}, and photometric methods have been developed to derive the physical properties of the asteroids. In the following sections we introduce these techniques and propose how they can be used by amateurs in order to make real contributions in the field. 

\subsection{Discovery and astrometry of Near Earth Asteroids}
\label{NEOs}

Most of the current discoveries of NEAs are made by large asteroid surveys that are associated with the NASA Spaceguard Survey Program. The number of discovered asteroids grows continuously: fainter objects are discovered in the Main Belt, as well as NEAs observed in more favorable geometries (when they come close to the Earth). In the case of discoveries, measuring the positions of objects (astrometry) is fundamental for establishing their orbital elements. Gravitational fields of the Sun and major planets, mutual encounter between asteroids, non-gravitational Yarkovsky/YORP effects \cite{mb1,mb2} will perturb the orbit of these objects. As a result, the orbits of these bodies becoming increasingly uncertain with time, the accuracy of their ephemeris decreases. Hence astrometry needs to be done continuously in order to maintain and improve the accuracy of ephemeris. In the last two decades {we have witnessed} a democratization of instruments (telescopes), detectors (CCDs) and techniques of observations. Here we discuss how amateur and professional astronomers can work together in the field of asteroid discovery, recovery, and precovery\footnote{Precovery or pre-discovery is the process of finding the image of an object in archived images of the sky obtained prior its discovery.}. 

\subsubsection{Detection of asteroids}
\label{astdet}
\begin{figure}
\begin{center}
\includegraphics[scale=.50]{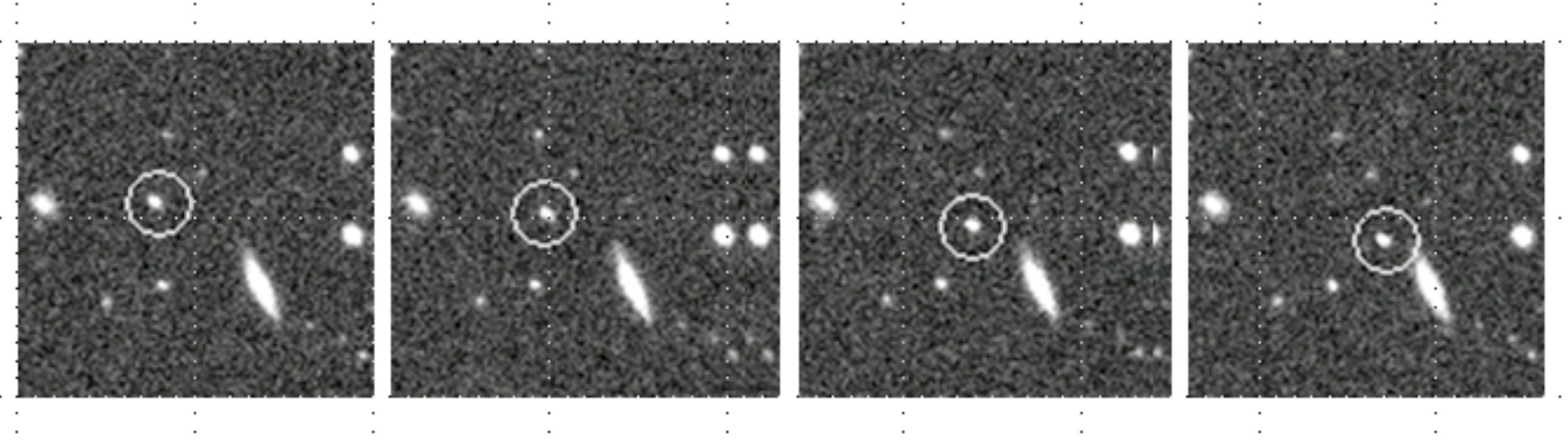}
\caption{Differential movement of asteroid (165660) 2001 LE18, in the images taken at Oukaimeden Observatory, Morocco. Four images of 180 seconds each, spanning 15 minutes in total are enough to observe the differential movement of this Solar System object in reference to the fixed stars and galaxies. The object was observed using the 0.5m F/3 Newtonian telescope equipped with an SBIG STL11k camera (credit C. Rinner).}
\label{blink}      
\end{center}
\end{figure}

An asteroid can easily be detected in a star field. Today, several software packages allow the automatic detection of moving objects in a set of CCD frames (large surveys and surveys with huge amounts of data use dedicated automated pipelines for detection of Solar System objects). Then the software provides the opportunity to confirm manually the reality of the detected object through either individual sub-frames around the moving object or through an animation of the successive frames (blinking). The blinking technique (Fig.~\ref{blink}) is applied to register a series of images of the same field which contains the object. The purpose of blinking is to identify an object which presents a differential movement compared to the stars in the field.

If an object appears with a differential movement and is not found in the catalogue of asteroids\footnote{the catalog of asteroids can be accessed from http://cdsarc.u-strasbg.fr/cgi-bin/nph-Cat/txt/max=588132?B/astorb/astorb.dat while the ephemeris of objects can be obtained from http://vo.imcce.fr/webservices/skybot/}, it might be a newly discovered object. Measuring its positions (astrometry) and reporting these measurements to the IAU Minor Planet Center (MPC)\footnote{http://www.minorplanetcenter.net/} then becomes a critical task. In the case of discoveries of NEAs, because the objects have large differential movements and a very tight observational window for small telescopes (the apparent magnitude could change by several units in a few days), having a fast automated pipeline for data reduction and astrometry measurements is essential in reporting the results as quickly as possible.

Depending on the differential movement of the asteroid, its apparent magnitude, and the aperture of the telescope, the exposure time of an image could be between 20 and 240 seconds (above this range of values, the CCD chip can saturate or the asteroid's trajectory segment might be too extended to be detected by the software). Usually, the image of the asteroid will be a small segment compared to the point-like images of stars. If the asteroid is very faint, an alternative strategy could be to track the object following its differential movement ({\it pencil-beam} search). Thus, the stars will be represented by trails, while the object will be a faint point-like source. The major failure of tracking on differential movement for fast objects is the lack of adapted procedures (Point Spread Function, pinpoint, centroids,...) to compute the coefficients of astrometric calibration automatically.

\subsubsection{Data-mining of asteroids}

There are several international initatives for data-mining of asteroids in archives that were initially devoted to scientific programs oriented to cosmology, star structure and evolution. Two of these initiatives are cited here as representative of collaborations between professional and  amateurs astronomers: Euronear\footnote{http://euronear.imcce.fr/tiki-index.php?page=MegaPrecovery} and the Spanish Virtual Observatory initiative for NEAs\footnote{http://www.laeff.cab.inta-csic.es/projects/near/main/?\&newlang=eng}.

Astronomical databases produced by professional observatories can be accessed via Internet for Solar System objects searches. Precoveries and recoveries of MB asteroids and NEAs are activities adapted to data-mining. Serendipitous encounters of asteroids in the archives can be retrieved by comparing their ephemeris with the epoch when the images were obtained \cite{mb5}. The presence of objects in an archive is a function of their limiting magnitude and for this reason systematic inspection of candidate images must be done \cite{mb6}. The images containing asteroids are then used for astrometry.

\subsubsection{Pipeline for astrometric measurements of asteroids}
\label{pipe}
Several software programs hav been developed to perform astrometry. They generate an output file in the format of a MPC report. Once the images are recorded, the specific steps for astrometric reduction are:\\

\begin{enumerate}
\item Preprocessing of the images (cleaning images using calibration images)
\item Running the detection software
\item Confirm the reality of detected objects
\item Check for fast movers\footnote{http://www.minorplanetcenter.net/iau/NEO/PossNEO.html}
\item Send the list of detected/confirmed objects to the MPC, flagging possibly interesting objects.
\end{enumerate}

Amateur and professional astronomers involved in the EURONEAR network \cite{mb3,mb4} use the Astrometrica\footnote{http://www.astrometrica.at/} software developed by Herbert Raab for astrometric data reduction. The software allows both quasi-automatic and manual manipulations of images, astrometric measurements of asteroids, as well as the email sending of MPC reports. Additionally, stacking procedures for increasing the S/N ratio and the use of several astrometric catalogues (UCAC2, UCAC4,  USNO, NOMAD,...) are available. The choice of this software was motivated by the user-friendly interface and the possibility of quick training of persons involved into data-reduction processes for each session of observations. Several other programs devoted to astrometry (MaximDL, astrometry.net, Tangra, Prism, C2A,..) can also be considered for astrometric data-reduction.

\subsubsection{Amateur contributions}

Technically, we estimate that the equipment level needed for performing good astrometry of asteroids is fairly accessible. A 30-cm telescope equipped with a CCD camera with a field of view larger than 60 $\times$ 60 arcmin is a good start. However, this equipment will be constrained in terms of discoveries due to the limiting magnitude of objects. The observers can start training (observing and data reduction procedures) with objects from the Main Belt with well-known orbits. Once the good feedback and methods of reporting astrometry data are acquired, the observers will be able to start hunting for new objects.

\subsubsection{Valorizing the observations}

\begin{figure}
\includegraphics[scale=.99]{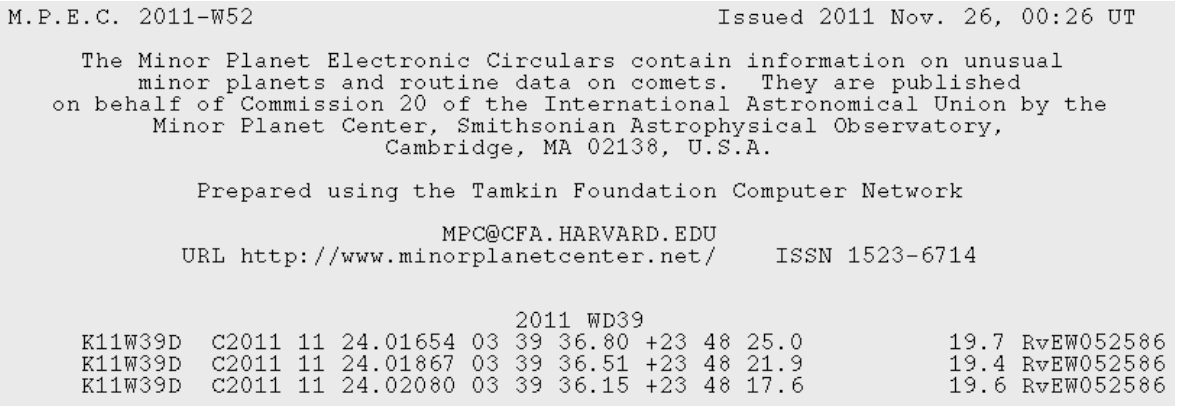}
\caption{Header of the MPEC 2011-W52 announcing the new object 2011 WD39. This electronic telegram was edited 30 min after the report sent by the group of observers from the {1.05 m} telescope at Pic du Midi Observatory, France, on November 26, 2011 (credit: M. Birlan, F. Colas, M. Popescu and A. Nedelcu).}
\label{mpec}
\end{figure}

Astrometry of asteroids is centralized by the MPC. An automatic update of the NEA confirmation page\footnote{http://www.minorplanetcenter.net/iau/NEO/ToConfirm.html} is made each time a new discovery is reported. A new designation is assigned after the reception of observations by one or several observers/telescopes from at least two nights. If the discovery is confirmed, a new Minor Planet Electronic Circular (MPEC) containing the provisional denomination of the new asteroid is also edited by the MPC (see Fig.~\ref{mpec}).

\subsection{Lightcurves of asteroids}
\label{light}

\begin{figure}
\includegraphics[width=\textwidth]{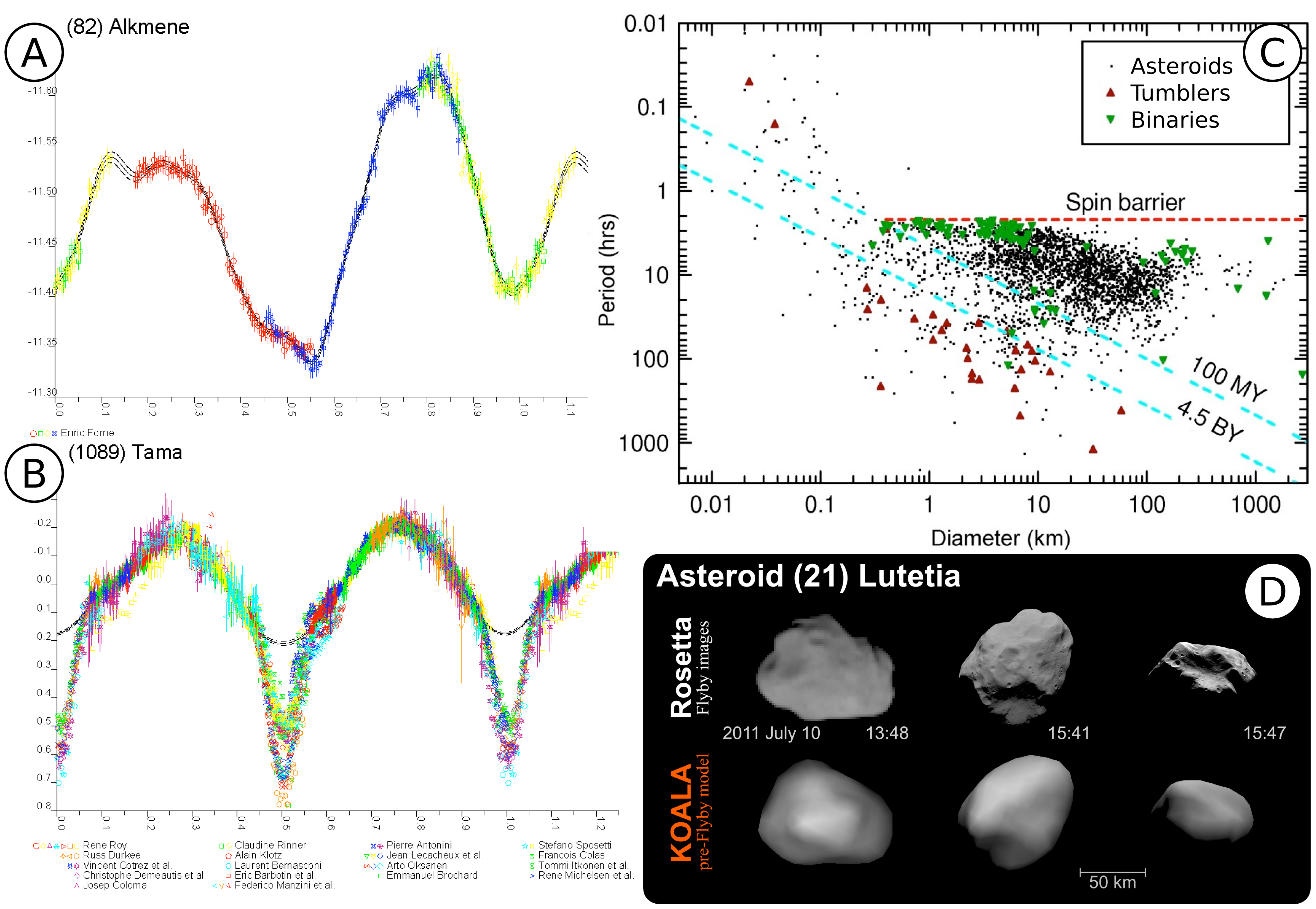}
\caption[Example of asteroid lightcurves and studies]{%
A few examples of lightcurve observations and derived quantities. \textbf{A)}~Composite lightcurve of (82) Alkmene obtained over 4 nights of observation (from CdR database). The flux variations are directly linked with the changing illumination of the asteroid.
\textbf{B)}~Composite lightcurve of the binary asteroid (1089) Tama. The strong dips result from the mutual eclipses between the two components of the system, superimposed over the rotation-induced lightcurve (adapted from \cite{Be06}).
\textbf{C)}~Rotation period vs diameter of about 3\,000 asteroids (adapted from \cite{Wa09}). The so-called ``spin barrier'', given by the balance between self-cohesion and centrifuge acceleration, is clearly visible. Note how most of the known binaries rotate with a period close to the limit.
\textbf{D)}~The shape model of (21) Lutetia, obtained with the KOALA multi-data inversion algorithm by using many lightcurves from amateurs \cite{ca10} compared with the images returned during the flyby of the asteroids by the ESA Rosetta mission \cite{ca12b}.
\label{fig: light}
}
\end{figure}

Time-series of photometric observations (lightcurves) of asteroids are the most efficient way to derive their global physical properties such as rotation period, orientation of the spin axis, 3-D shape, and multiplicity (Fig.~\ref{fig: light}). These basic properties are the key to understand the whole asteroid population, its evolution, and its links with meteorites. For instance, the spin (period and orientation) and shape are among the main parameters of the non-gravitational forces (YORP effects) that slowly change the spin and orbit of the asteroids with time and are responsible for meteorites delivery to Earth \cite{ver}. Alternatively, the study of multiple asteroids is the most precise way to determine the asteroids' density, which may be one of the most fundamental parameters to constrain their interior and bulk composition \cite{ca12}.
However, we have access to these quantities for only a tiny fraction of the half a million asteroids known to date. Indeed, the current method to derive period, spin, and 3-D shape (limited to convex hulls) requires numerous lightcurves, taken over several apparitions, to cover many Sun-asteroid-Earth geometries \cite{Kaa01,Kab01}, leading to the publication of models for 300 asteroids \textsl{only} (see DAMIT\footnote{http://astro.troja.mff.cuni.cz/projects/asteroids3D} \cite{Du10}). This can be partly solved by the use of sparse photometry, characterized by a delay between two photometric measurements larger than the rotation period \cite{Du05,Ha11}. However, due to the difficulty of determining the rotation period using sparse data only, ``traditional'' lightcurves are still required.

\subsubsection{The rise of amateurs in asteroid photometry}

Asteroids are often used to teach astrometry and photometry, because of their short-term variability in both position and flux. These very characteristics have also made amateurs interested in their observation. This increasing interest in observing asteroid lightcurves by amateurs occurred in the late 1990s, with the advent of technology such as less expensive telescopes and cameras. Ironically, a significant fraction of the professional community was slowly turning from the asteroids to concentrate on Trans-Neptunian objects at that time. Amateurs have therefore been the main observers of asteroid lightcurves for about a decade.
  
Several organizational initiatives flourished, the most notable being CdR \cite{Be06} and LightCurve Data Base (LCDB) \cite{Wa09} with thousands of lightcurves of asteroids each. Collaborative efforts, including joint campaigns of observations, have been organized by amateurs with great results (see for instance \cite{Ste12}; the best opportunities of observation are published tri-monthly \cite{Wa12}). An increasing number of amateur-led studies, including observations, period analysis, and shape modeling are published in the \textsl{Minor Planet Bulletin}\footnote{http://www.minorplanet.info/mpbdownloads.html}. In the meanwhile, a few professionals have been involved with the amateur community, proposing targets for observations and organizing co-publications. Numerous small main-belt binaries have thus been discovered and characterized \cite{Pra12} and the period, spin, and shape of few tens of asteroids were determined \cite{Du07,Ha11}.

\subsubsection{What and how to observe?}
With several hundreds of asteroids brighter than V\,=\,16 at any time, targets are available for all equipment, from modest aperture (20-cm) to large telescopes, with CCD cameras (preferentially without anti-blooming to ensure a linear response in photometry). If most asteroids are suitable for the purposes of observation and data reduction, some overarching structure for target selection is highly desirable. Indeed, additional lightcurves of (4) Vesta will for instance bring no further knowledge on the asteroid, since it has already been observed from the ground and visited by a spacecraft. A few recommendations on the target selection and cadence of observation are listed below.

\begin{enumerate}
\item \textbf{Target:} 
Possibly the best option to choose a target is to be registered in an active mailing list of observers, such as CdR\footnote{http://obswww.unige.ch/$\sim$behrend/page$\_$cou.html} or CALL\footnote{http://www.minorplanet.info/call.html}. Otherwise, any target listed in the latest issue of the \textsl{Minor Planet Bulletin} under the \textsl{Lightcurve Photometry Opportunities} section can be selected. This list contains tens of targets brighter than V$\sim$15.
\item \textbf{Sampling:} If the period is already known, a cadence of observations below 2--3\% of the period is highly desirable. If the period is yet to be determined, then observations should be taken every few minutes.        
\item \textbf{Coverage:} As a general rule, the longer the asteroid is observed the better. A long session during one night, covering as much as possible of the rotation period, or of the eclipsing events in the case of a binary asteroid, yields more information than many short slots of observation. For period determination, one or few consecutive nights are generally enough. For shape or orbital modeling, frequent monitoring is required. The same asteroid should therefore be observed every few weeks, during its whole apparition. For any binary or shape modeling targets, multi-apparition data are also required.
\item \textbf{Photometric accuracy:} Because relative (as opposed to absolute) photometry is sufficient for the analysis of most asteroid properties, including the complex 3-D shape modeling and orbit determination, each asteroid lightcurve can be very valuable. The relative precision should however not be cruder than 0.05 or 0.1 mag (typically achievable with 1\,min exposures on a V = 12 target with a 20--30 cm aperture). Note that the filter (``color'') is not relevant to study the shape nor the binarity.  One should therefore pick a filter at will, for example Johnson V/R or Gunn g/r, and stick to it consistently.
\item \textbf{Archiving:} Because old data are crucial for analysis, we encourage any observer (amateur, professional, teacher) to feed  their observations to archiving portals such as LCDB\footnote{http://minorplanetcenter.net/light\_curve}, where their contributions will be archived and properly referenced for future use.
\end{enumerate}

\subsection{Stellar occultations}
\label{occul}

Diameters and shapes are physical parameters crucial to understand the mechanism of formation, collisional disruption and evolution of asteroids. Currently known diameters have been measured mainly indirectly, by the application of thermophysical models to ground-based and space-based infrared observations. This is the case, for example, of the sample of asteroid observations by the WISE telescope (Wide Infrared Survey Experiment). However, due to several uncertainty sources, thermal infrared size can be affected by relevant dispersion and/or systematic errors \cite{ca12}. The best calibrations of thermal infrared sizes are probably obtained from well observed stellar occultations by asteroids, as shown by \cite{Shev06}. 

Beside the few objects visited by space missions, asteroid sizes can be derived by speckle imaging \cite{Drummond89}, stellar occultations, disk-resolved imaging (from the ground or HST \cite{Thomas05,Marchis10}), radar Doppler-echoes (NEAs) \cite{Ostro02}, interferometry in the visible with the Fine Guidance Sensors (FGS) mounted on the Hubble Space Telescope \cite{Tan01,Hestro02,Tan03} or in the mid-infrared from the ground \cite{Delbo09}. Most of these techniques are very time consuming and critically applicable to restricted categories of objects. Until now they have provided precise results on a very small number of bodies, with the notable exception of stellar occultations by asteroids, resulting in $\sim$60 diameters observed over the last 15 years.

\begin{figure}
\begin{center}
\includegraphics[scale=.48]{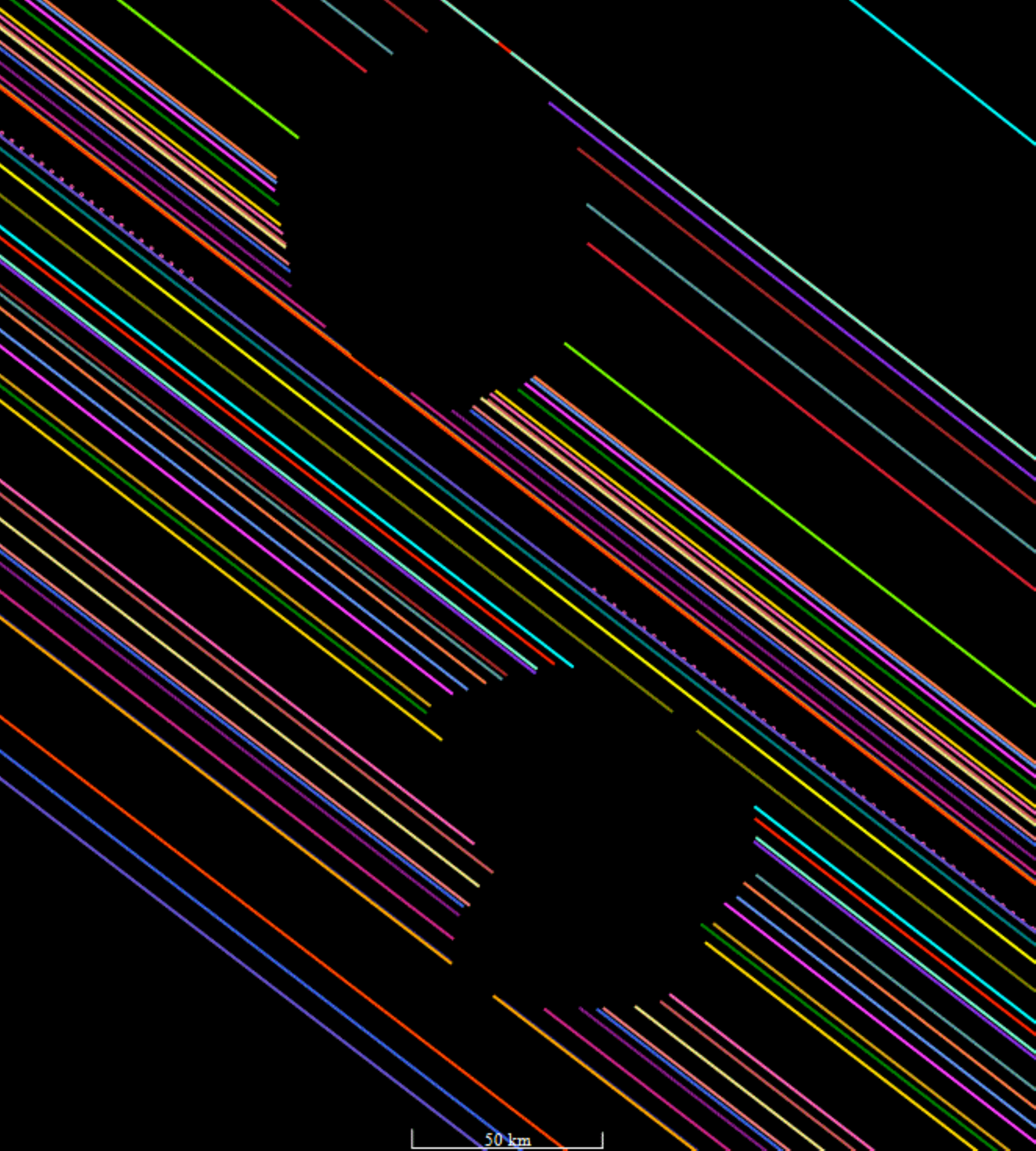}
\caption{Profile of the binary asteroid 90 Antiope obtained by observing the occultation of the star LQ Aqr on July 19, 2011. This figure shows the milliarcsecond resolution achievable on asteroid silhouettes with the occultation technique. Colored lines represent the different observers distributed within or around the predicted path of the asteroid's shadow. Each line corresponds to a single observation of the target star over time and is interrupted when the star is occulted by the asteroid. The observations are reduced using the Besselian fundamental plane as the reference plane. For each observed event by each observer, the observer's location is projected onto a moving reference frame corresponding to the ephemeris motion of the asteroid's shadow on the plane. The result above is finally displayed in the sky plane.}
\label{fig:antiope}
\end{center}
\end{figure}

Stellar occultations rely upon the detection of the extinction of light due to the asteroid passing in front of a star (see Fig. \ref{fig:antiope}). The uncertainty on the derived size is thus linked to the timing of the occultation, to the position of the observers relative to the shadow center and to the amount of flux drop during the event. Provided that the orbit of the occulter and the position of the target star are known with sufficient accuracy, the observability of an event depends mainly upon the brightness of the star, and not that of the occulting asteroid. For this reason, events involving Trans-Neptunian Objects are also accessible to telescopes of modest diameter \cite{Ortiz12}.

It is worth mentioning that the only techniques for directly detecting concavities are stellar occultations (which can be applied to any asteroid), photometry of mutually eclipsing binary asteroids (see e.g., \cite{Bart12}), and radar ranging (which is mainly limited to NEAs and the largest Main Belt asteroids). Several publications deal with results obtained for asteroids by occultations (see for example \cite{Shev06}, \cite{Des08}, \cite{Dur11}).

\subsubsection{Organization and planning}
\label{organ}

Amateur astronomers have always played --and still play-- a major role in asteroid occultation prediction, observation and data reduction. This is essentially due to the dense geographical coverage needed to get useful results (with the average observer spacing smaller than the target size) and to the fact that occultations by asteroids have been considered an inefficient technique for several years, due to the prediction uncertainty arising from uncertainties in the positions of both the asteroid and the star. Passionate amateurs, willing to accept a large fraction of negative results, have thus pioneered the field. Over the years, the exploration of new techniques, the development of hardware and software tools, and the collection and archiving of data have been mainly driven by amateur astronomers, in some cases supported by active collaborations with professionals. 

The Hipparcos catalogue resulted in a major improvement in prediction accuracy, as it removed systematic zonal errors in star catalogues that affected both the positions of stars, and asteroids. The availability of the Tycho and Hipparcos catalogues, and subsequent catalogues based on the Hipparcos reference frame, resulted in a 10-fold increase in the annual number of observed occultations over the period 1997 to 2003. 

Today, predictions for asteroids are based on Tycho, Hipparcos and UCAC catalogues, for a total of about 4 $\times$ 10$^6$ stars at V $<$ 12, usually computed by the specialized program {\sl Occult} by D. Herald, available from the website of the International Occultation Timing Association (IOTA)\footnote{http://www.occultations.org/}. The same program can perform sophisticated operations of event selection, data reduction and access to past observations. Events selected on the bases of tight observability criteria are made public by IOTA to a diverse community of amateur and professional astronomers under the form of tabulated ephemerides, star finding charts and maps of shadow paths. 

An important part of planning is to coordinate the placement of observers across the predicted path. This is frequently achieved using the OccultWatcher program\footnote{by H. Pavlov: http://www.hristopavlov.net/OccultWatcher/OccultWatcher.html}, which coordinates observers wherever they are located on the Earth without the need for any direct interaction among them, and/or dedicated mailing lists (PLANOCCULT and IOTAoccultations mailing lists for European and American observers, respectively). However, despite the vast amount of information and tools available on the web, there is an evident need of more observers, as active ones cover just a small portion of the Earth's surface. This research field thus represent an interesting and promising opportunity for amateurs. 

A typical site on Earth using the most commonly available predictions for asteroids has $\sim$50 opportunities of observations per year\footnote{This rough estimate results directly from the prediction of occultation using the currently usual approach (asteroids $>$ 40 km, Hipparcos stars...). This estimate can also be confirmed by running through the list of the predicted events.}, about half of which occur in good geometric conditions (night-time, star high above the horizon, no moon).

Only a fraction of them will typically produce positive events, but also negative events have their own importance, as they can put upper limits on the object size when some positive chords are detected elsewhere, at adjacent sites. Also, observers far from the predicted centerline can still have chances of positive results when the uncertainty is large, or when an unknown satellite is present.

Currently, the accuracy of predictions for Main-Belt asteroids with excellent orbits is about 100\ km on the Earth's surface. As a result, observations of occultations of asteroids smaller than $\sim$30--40 km  have a low probability of success, as the asteroid diameter is much smaller than the uncertainty in the location of the path. When targets of special importance are a candidate for an occultation, ``last-minute'' astrometry is sometimes performed with professional telescopes in the hours/days preceding the event, with the occulted star and the asteroid being imaged together to eliminate any local errors in the stellar catalog. For a small number of asteroids with satellites (including binary asteroids), separate predictions for the satellites are possible in the rare cases when their orbits are known. In such situations, occultations by the satellites can be used to improve the orbits with spectacular results.

The future availability of stellar and asteroid astrometry by the Gaia mission is expected to reduce the prediction uncertainty in the path location to only a few km. This will make it possible to set up many observers to record an occultation with high confidence, allowing a detailed profile to be measured. Current plans, including the measurement by Gaia of bright stars, will mitigate their degrading position accuracies, a result of the uncertainties in old catalogues such as Tycho2, UCAC4 and PPMX, used to determine proper motions. Today, the observational results are collected by four regional coordinators: Australia/New Zealand: J. Talbot (Royal Astronomical Society of New Zealand)\footnote{http://occsec.wellington.net.nz/aboutus.htm}; Europe: E. Frappa (Euraster)\footnote{http://www.euraster.net/}; Japan: T. Hayamizu (JOIN, Japan Occultation Information Network); USA: B. Timerson (IOTA). The observations are periodically uploaded to the Planetary Data System for diffusion to the scientific 
community\footnote{http://sbn.psi.edu/pds/resource/occ.html}. 

\subsubsection{Observing strategy} 

Occultation observing is both a matter of general strategy and of specific techniques applied to the single observing station. Concerning the strategy, we can distinguish (a) the regular survey mode from (b) the focused campaign. In (a) the observer chooses the events to be observed from a given site (often a fixed telescope) while in (b) portable equipment is used to cover events of special interest by putting several observers across the predicted shadow paths. Case (a) is often suitable for occultations with a path uncertainty much larger than the asteroid size, since a displacement of the sites would only improve the probability of positive detections. (b) requires the development of several stations with a more intensive effort, but it can be highly rewarding especially if the target has a specific interest (e.g., binary asteroids).

The observation technique relies upon fast photometry and accurate absolute timing of the observations. For the occultation by a typical Main-Belt asteroid moving at 15 km/s, observed using video at a frame rate of 10 frames/sec, the uncertainty on the occultation of each chord extreme will be around 1.5 km, representing 5\% of the size of a 30-km body. An absolute timing accuracy at the 0.01-sec level should be the target. Such performances are usually obtained by sensitive and inexpensive analog video cameras (see Sec. \ref{ccd}) either connected to a PC through a frame grabber or to a video recorder. For timing, event recording at the hardware level is the only accurate option to avoid biases introduced by unpredictable delays between the software/operating system and the shutter opening/closing (see Sec. \ref{time}). Data reduction usually proceeds with an automated relative photometry of the video by comparing the target brightness to other sources in the field\footnote{Standard programs for this task include ``Limovie'' (http://astro-limovie.info/index.html) and ``Tangra'' (http://www.hristopavlov.net/Tangra/Tangra.html)}.

Alternative acquisition techniques can be adopted by using digital cameras, in fast imaging mode or in ``Track Delay Integration" mode (i.e., by shifting the charge on-chip toward the read-out register, at an appropriate constant rate). Alternatively, telescope tracking can be stopped or run at modified speed, with the image being recorded using a standard CCD imaging camera with the shutter opened and closed at known times\footnote{http://www.asteroidoccultation.com/observations/DriftScan/Index.htm}. One of the most notable, systematic surveys adopting non-tracked images is run by the automated TAROT telescopes North and South (A. Klotz, E. Frappa -- results on the Euraster website). 

Typical analog video cameras as those mentioned above are sensitive enough to observe stars at V$\sim$12 with 0.04 sec integration and a 20-cm telescope at f/3.3 -- a configuration easily obtainable with commercial Schmidt-Cassegrain telescopes with a focal reducer. Compact camera lenses with wide fields and ``fast'' focal ratios (for example 85\ mm f/1.4) can reach V$\sim$11 with 0.32 sec integration, and are often found in portable equipments. Camera lenses can also be used for deploying pre-pointed acquisition stations. Sometimes a single observer will set up well over 10 stations spread over many tens of km across the predicted path. Recent experiments, performed in particular in the USA, have shown that this approach can be very efficient when the predictions are sufficiently precise.

\subsection{Search for comets hidden in the asteroid population}
\label{main}

{The orbits of asteroids and comets are dynamically discriminated using the Tisserand parameter ($T_J$) with respect to the gravitational influence of Jupiter. This parameter is defined as:

\begin{equation}
T_J = \frac{a_J}{a} + 2 \sqrt{\frac{a}{a_J} (1 - e^2)} \cos  i,
\end{equation}

\noindent with $a_J$ Jupiter's semi-major axis, $a$, $i$ and $e$ the minor body's orbital semi-major axis, inclination and eccentricity, respectively.} It is a constant of motion during a close approach between Jupiter and an interplanetary body, and it provides a way to connect the post-encounter dynamical properties with the pre-encounter ones. {Minor bodies with $T_J < 3$ are considered as comets, whereas those characterized by $T_J > 3$ are identified as asteroids \cite{Levison96}. On the other hand, this is not an absolute rule since some comets have $T_J > 3$ and some asteroids display $T_J < 3$.}

Comets are also observationally defined as objects displaying a bound, detectable coma, which is due to the temperature driven sublimation of volatile gases, lifting up dust grains from the nucleus. When the dust/gas production is important enough, the comet displays a huge tail that can be several millions of kilometers long. Recently, cometary tails were detected around some Main Belt asteroids (e.g., with T$_J$ $>$ 3), blurring the secular definition of a comet (see \cite{Hsieh04} for the first example). Yet, some objects discovered with a T$_J$ $<$ 3 have an asteroidal appearance (this is the topic of this section), and are therefore listed among asteroids, although they belong --dynamically speaking-- to the comet world. {Hence discovering faint cometary activity is the only way to secure the physical status of the observed small body.} These recent discoveries tell us that definitions have to evolve with the progress of science, and that a new vision of the comets/asteroids populations will soon emerge.

Hunting {cometary activity} in the asteroid population in a systematic way is important to cast some light on the different sub-populations of comets and their possible dynamical reservoirs, to understand in what conditions cometary activity can occur, to identify the corresponding physical and chemical mechanisms at work, and ultimately to constrain models of Solar System formation and evolution. This valuable systematic search for cometary activity can rely on a wide network of amateur observers. They can significantly contribute to the comet discovery effort and provide particularly interesting targets for subsequent in-depth studies by professional astronomers.

\subsubsection{The T3 project, a worldwide PRO-AM collaboration}

\begin{figure}
\begin{center}
\includegraphics[scale=.40]{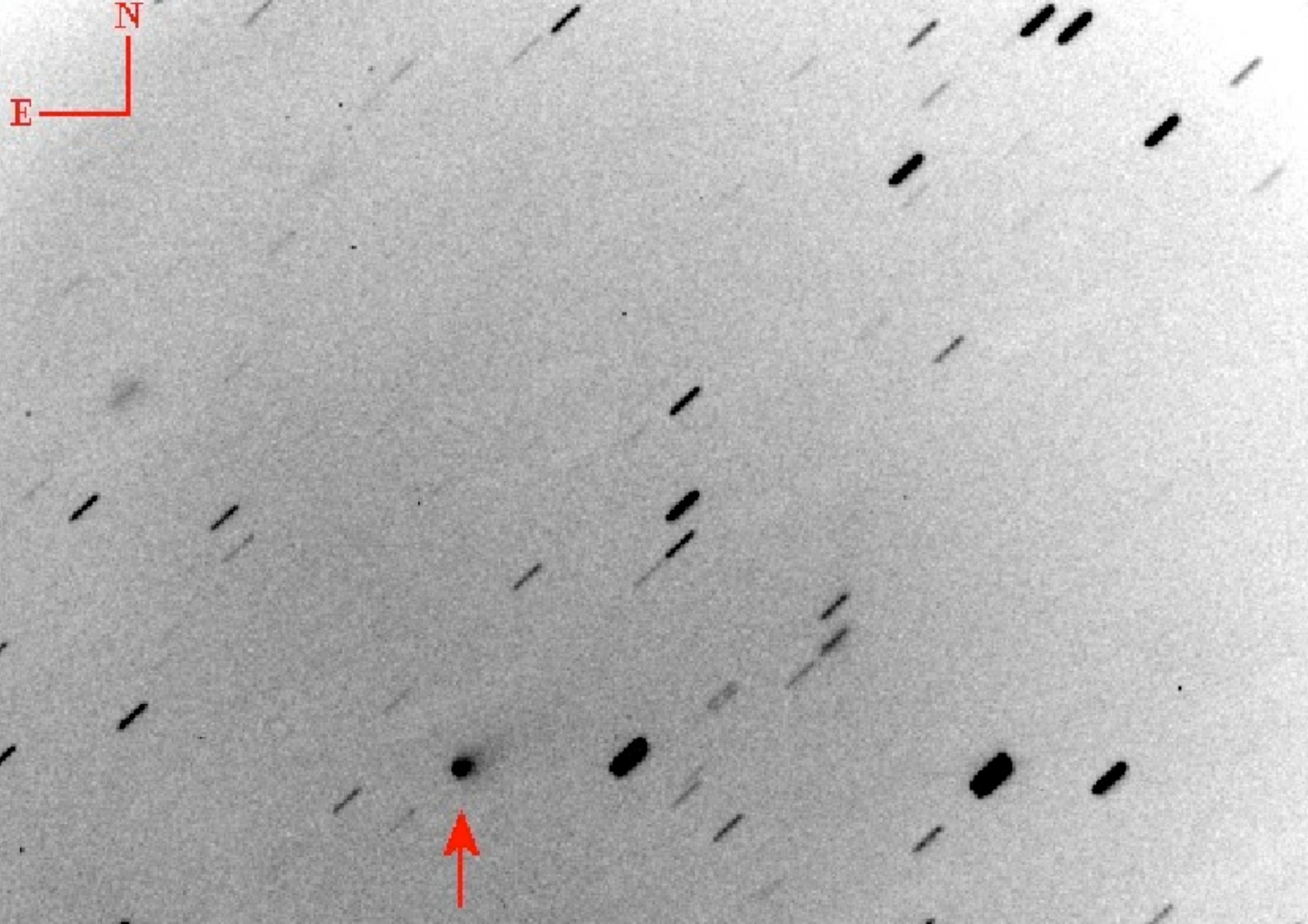}
\caption{Comet C/2005 YW (LINEAR) in the R-band taken on October 11, 2006 from the Schiaparelli Observatory, Varese, Italy (MPC 204) (credit L. Buzzi). The cometary appearance is obvious (at 2.11 AU heliocentric distance) on this R-band magnitude-15 object. This stack is the sum of 50 exposures of 15 sec each, taking into account the 1.54''/min comet apparent motion. Images were taken with a 0.6 m f/4.64 telescope + CCD SBIG ST10-XME, 1.5''/pixel, the resulting field of view is 14 $\times$ 9.8 arcmin.}
\label{main:fig1}      
\end{center}
\end{figure}

The T3 project (named after the $T_J=3$ boundary between asteroids and comets) was born at the end of 2005 thanks to a collaboration between the Physics Department of the University of Rome and several amateur astronomers in Italy. It started with the first coma detection on asteroid 2005 SB$_{216}$ \cite{BuzziMPEC,FogliaIAUC} on amateur images (the technique is described below), soon confirmed by astronomers at the Institute for Astronomy at University of Hawaii, USA \cite{BuzziMACE}. Professional confirmation is crucial in the process of cometary activity detection, in order to discard false positives. After a presentation at the Meeting on Asteroids and Comets Europe (MACE) 2006, many observers joined the program, and the project became worldwide with a network of both professional and amateur observatories. In Italy, the observations are conducted on two telescopes from the Schiaparelli Observatory, MPC 204, (0.4 m and 0.6 m in diameter; see Fig. \ref{main:fig1}). The 2 m Faulkes Telescopes on Mauna Kea (Hawaii, USA) and Siding Springs (Australia) are involved in the project. USA teams also contribute from the 0.5 m I-NET telescopes, MPC H06 (New Mexico), the Astronomical Research Institute, (Illinois), with 0.6 and 0.8 m telescopes and the Kitt Peak National Observatory 1.3 m telescope (Arizona). From ESO/Chile, some observations are conducted in La Silla, with the TRAPPIST 0.6 m and the Swiss 1.2 m Euler telescopes.

\subsubsection{The observing planner and technique}

The ideal goal is to observe all asteroids with a $T_J<3$ and a constraint on the magnitude limit, solar elongation and Jupiter MOID. Candidates fulfilling the right criteria are automatically extracted from two lists: the Minor Planet Center Orbit (MPCOrb) database\footnote{http://www.minorplanetcenter.net/iau/MPCORB.html}, which contains orbital elements of minor planets that have been published in the MPC circulars, and the MPC Near Earth Object Confirmation Page (NEOCP). The latter is checked on a daily basis and the candidate list is immediately sent to the observers, as time is a critical factor for observation. Indeed, if a coma is detected, an IAU circular can be directly published (electronic telegram, CBET\footnote{http://www.cbat.eps.harvard.edu/cbet/RecentCBETs.html}), stating the comet discovery. From this screening step, an \textit{``Observing Planner''} is issued to the team twice a month, indicating the asteroid designation, perihelion date, $T_J$, number of observed oppositions, orbital semi-major axis, eccentricity and inclination, current sky position and magnitude, geocentric and heliocentric distances, solar elongation and Jupiter MOID. The probabilities of the source regions (Outer Main Belt or Jupiter Family) of NEAs are also indicated \cite{Bottke02}.

The observations should be performed under good seeing conditions (which depends on the observer's location). A first set of typically 30 images should be obtained. Integration time should be set to limit the trailing effect on the asteroid for a given exposure, and typically ranges from 30 to 120 sec (sometimes up to 5 minutes) depending on the apparent brightness of the target, so as to reach a minimum signal-to-noise ratio (SNR) of 10. No particular filter is required, in order to reach the maximum SNR. All the satisfactory images should be bias, dark and flat-field corrected and stacked according to the asteroid's apparent motion using Astrometrica or an equivalent software. A second set of images should be obtained within the same night to reduce the number of false positives in case of faint background source contamination on the first series, in particular for average seeing sites.

\subsubsection{The detection method}

\begin{figure}
\begin{center}
\includegraphics[scale=0.70]{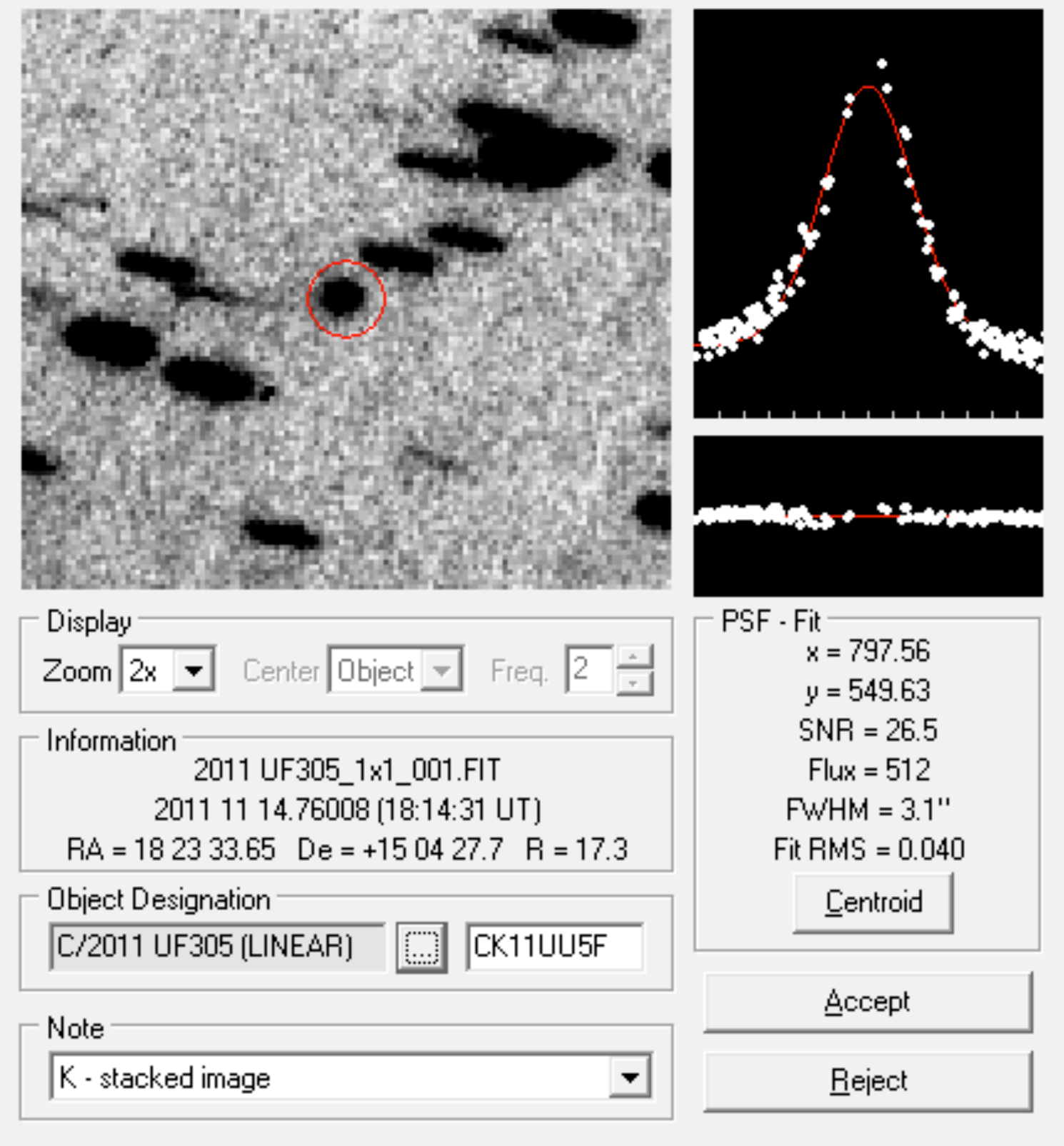}
\caption{Example of a radial photometric profile FWHM estimation from a stack of images for comet C/2011~UF$_{305}$ (LINEAR), with a R-band magnitude of 17.3.}
\label{main:fig2}       
\end{center}
\end{figure}

If a cometary feature is obvious by visual inspection of the stacked image, the observer sends a message to the MPC CBAT (Central Bureau for Astronomical Telegrams) and to the team for a rapid and independent confirmation. If the cometary appearance is not obvious, the Full Width at Half Maximum (FWHM) comparison method is applied \cite{Masi}. The radial photometric profile's FWHM of the asteroid is measured as well as the one from nearby stars (on a stacked image centered along the stars, e.g., with zero motion; see Fig. \ref{main:fig2}). If the FWHM of the asteroid is significantly larger (at least 25$\%$ greater) than the one from the stars, a coma can be suspected, in particular if the results from the different asteroid stacks are similar. The corresponding image should be circulated within the team, along with the SNR and FWHM measurements for further observations. The coordinator will eventually request a professional confirmation for the amateur confirmed targets, in order to send the definitive report to MPC. If no coma is detected from the first visual inspection and FWHM study, confirmation of negative detections via the amateur network are similarly important.

\subsubsection{Main results and perspectives of the T3 project}

Since 2005, eight comets have been identified in the asteroid population thanks to the T3 project: P/2005~SB$_{216}$, P/2005~YW, P/2002~VP$_{94}$, P/2010~WK, P/2010~UH$_{55}$, P/2011~UF$_{305}$, P/2011~FR$_{143}$, and C/2011~KP$_{36}$. The asteroids were initially discovered by automatic surveys: LONEOS, LINEAR, SpaceWatch and Mt Lemmon. A number of other comets were also identified from the screening of the Near Earth Object confirmation page at MPC: in 2012, 12 comets were detected, and this number is still increasing, demonstrating the efficiency of this PRO-AM network.

To make the discovery process even more reliable, the team is collaborating with R. Miles (Golden Hill Observatory, UK) to set up a second photometric method to provide a confirmation of the cometary objects with a slightly different approach \cite{Miles09}. The object's integrated luminous flux is measured with increasing circular apertures (curve of growth) and compared to the same measurements performed on nearby stars. This method, also referred as ``aperture photometry'', permits a normalization of the photometry to constant seeing conditions. This strongly limits the false alarms due to the contamination of the FWHM measurements by the degradation of the seeing during a series of observations.

Observers interested in participating in the T3 Project will find additional information and instructions to join the program at http://asteroidi.uai.it/t3.html.

\section{Imaging, spectroscopic and photometric measurements of outer planets}
\label{outer}

The giant planets Jupiter and Saturn are among the favorite targets of amateur astronomers, offering outstanding science subjects on which amateurs and professionals regularly collaborate. In fact, amateur contributions are now regarded as an essential tool to study the atmospheres of Jupiter and Saturn for the following reasons:

\begin{enumerate}
\item They provide a long-term global view able to support high-resolution regional observations from a spacecraft. This is clearly illustrated by the demand of amateur support for the Juno  mission science, particularly when the spacecraft arrives at Jupiter in the summer of 2016;
\item They allow prediction of the locations of features of interest, helping in planning the use of professional telescopes;
\item Visible observations provide the visible context for remote sensing at other wavelengths;
\item Amateur observations often allow identification of transient phenomena that could not be caught by pre-planned spacecraft observations;
\item They allow long-term tracking of seasonal changes, or large-scale weather phenomena.
\end{enumerate}

We are living in a golden age of observations of the giant planets that has arisen from advances in imaging techniques and low-cost cameras.

\subsection{Image observing techniques}
\label{iot}

\begin{figure}
\begin{center}
\includegraphics[scale=.30]{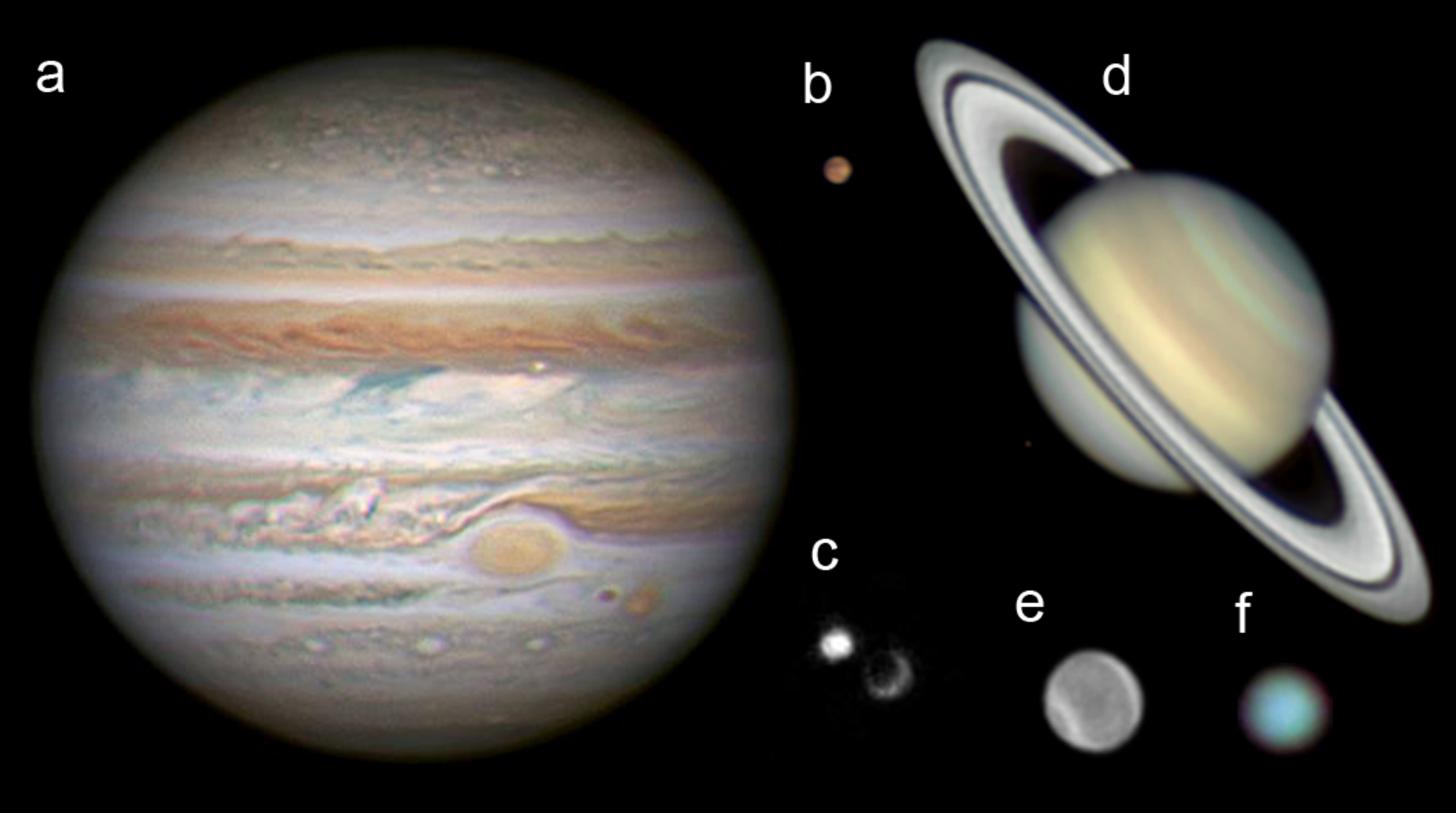}
\caption{Examples of images of the Giant Planets and their satellites. {\bf a)} Jupiter imaged with a 35-cm telescope on November 23, 2012 close to opposition (credit A. Bianconi). {\bf b)} Details on Ganymede observed with a 28-cm telescope on December 2, 2011 (credit M. Kardasis). {\bf c)} Single frame of a movie of a rare Io-Ganymede eclipse observed with a 28-cm telescope on August 16, 2009 (credit C. Go). {\bf d)} Saturn image captured with a 30-cm telescope on May 2, 2012 (credit E. Morales). {\bf e)} Uranus image obtained with a 35-cm telescope on September 8, 2012 with an IR filter integrating light for 45 minutes (credit D. Peach). {\bf f)} Neptune image acquired with a 25-cm telescope on August 11th, 2012 (credit C. Pellier). Differences in images size correspond to the diffraction limit of a 35-cm telescope and show the relative degree of detail available in each objects. Panels {\bf c)}, {\bf e)} and {\bf f)} are displayed with a 2x zoom to show better the details available in the original images.}
\label{fig:Giant_Planets_Figure1}     
\end{center}  
\end{figure}

Traditionally, visual observations resulted in astronomical drawings of the changing clouds in these atmospheres. The transition to amateur photography of the planets in the 1960--1985 was followed by digital observations with CCD cameras (80--90s) and continued at the beginning of the 21st century with high speed CCD cameras that resulted in a high-resolution image revolution. Amateur astronomers were the first to film the planets using the ``lucky'' imaging method \cite{Law06} to produce nearly diffraction-limited images. This technique consists in obtaining a video recording with short-exposure frames (typically 1/10 to 1/60 for broad-band filters and depending on the luminosity of the object) in order to freeze the effect of atmospheric turbulence (see Sec. \ref{ccd}). Freely available software written by the amateur community such as Registax\footnote{Written by C. Berrevoets. Available on: http://www.astronomie.be/registax/} or Autostakkert\footnote{Written by E. Kraaimkap. Available on http://www.autostakkert.com/} can be used to select the best-quality frames and stack them into a high-resolution image that can be processed to bring out atmospheric details on the order of the diffraction limit of the telescope. An observer equipped with a 35-cm aperture telescope can produce images with a spatial resolution of 0.4 arcsec in the visible range which translates into images of Jupiter, Saturn, Uranus and Neptune with an effective resolution of 115, 50, 9 and 7 resolution elements, respectively. Most observers will produce images that oversample the diffraction limit by a factor of 3--5 resulting in visually appealing images. Figure~\ref{fig:Giant_Planets_Figure1} shows relevant examples of images obtained by amateur astronomers of the Giant Planets and the Jovian satellites.

There is no ideal telescope for planetary imaging but most observers use Schmidt-Cassegrains. Cameras should have relatively small pixels of the order of 5--8 $\mu$m and small read-out noise. Additionally, Barlow lenses are generally used to increase the effective focal length of the telescope and produce higher resolution images. For systems where the final focal length is too short for the camera pixel size (typically when the FWHM of the Airy disk at the focal plane is smaller than 2 pixels), the final size of the image can be increased at the processing step with the drizzle algorithm \cite{Fruchter2002} available on Registax and Autostakkert when the video recording is long enough. The drizzle algorithm shifts and recenters the final image considering a pixel grid with a smaller pitch and higher-resolution than the original. 

Particular care needs to be taken to have the telescope perfectly collimated and well thermalized with its environment. Larger-diameter telescopes are more difficult to thermalize and may require more cooling. Observations at low elevation angles may benefit from the use of Atmospheric Dispersion Correctors (ADC) but these are generally not used by most amateurs due to their relatively high cost. Cameras need to be able to film at rates of 15 frames per second (fps) or higher (60--100 fps ideally) and motorized filter wheels are needed if the observer wants to compose color composite images or change the filters during the same observation run without the risk of adding dust to the optical system. 

High-resolution images of Jupiter and Saturn are now obtained by a large number of amateur observers. Images in broad-band visible and near-IR filters trace the dynamics of these atmospheres and even resolve details on Jupiter's satellites. Because of the planet's rotation the video observations acquired to stack a single channel stacked observation are limited to a certain duration before the rotation smears the details. Typical acquisition times are limited to less than 3 minutes for Jupiter and 4 minutes for Saturn. However the freely available software WinJupos\footnote{Written by  G. Hahn. Available on http://jupos.privat.t-online.de/} allows compensation for planetary rotation on Jupiter images and allows stacking of images obtained over as much as 10--15 minutes. Images must be processed carefully to bring out the fine-scale details and a combination of deconvolution techniques, high-pass filters and wavelet filters allows one to process the initially blurred stacked images. Each observer generally perfects his/her own processing techniques, rendering images with a personal touch in the degree of processing. Image processing strongly modifies the reflectance of the cloud features and does not allow one to calibrate these images in absolute intensity or reflectivity. This, together with the common use of broad-band filters, makes very difficult to use these observations for analysis of the vertical cloud structure based on radiative-transfer models. Co-registered stacked images without processing can be used for that purpose but generally require a calibration source. A technical description of photometric calibration of amateur images of the giant planets is presented in \cite{Mendikoa12}.

Images acquired in wide-band filters can be used to construct RGB or Luminosity-RGB color composite images. Narrow-band filters in the near UV and in the strong 890-nm methane absorption band trace higher levels of the atmospheres of Jupiter and Saturn where the contrast is dominated by the presence of upper hazes. Observations in narrow-band filters require longer acquisition times for each frame resulting in darker images and less capability to reach the diffraction limit of the telescope. Although only a limited number of amateurs own sets of filters in these wavelengths, their observations are very valuable since they sample different altitudes compared with the more usual broad-band visible filters.  Finally, although Uranus and Neptune are still difficult targets, images of their disks can be obtained with 30--50-cm telescopes. Infrared cut-off filters around 680 nm are able to resolve bands of Uranus but each frame needs to be significantly longer and the total acquisition time can be as high as 35--45 minutes. Additionally, photometric and spectroscopy measurements of brightness variations in Uranus and Neptune may be  used to study their atmospheres and the onset of convective events with smaller size telescopes. 

\subsection{Spectroscopic and photometric observations}
\begin{figure}
\begin{center}
\includegraphics[scale=0.21]{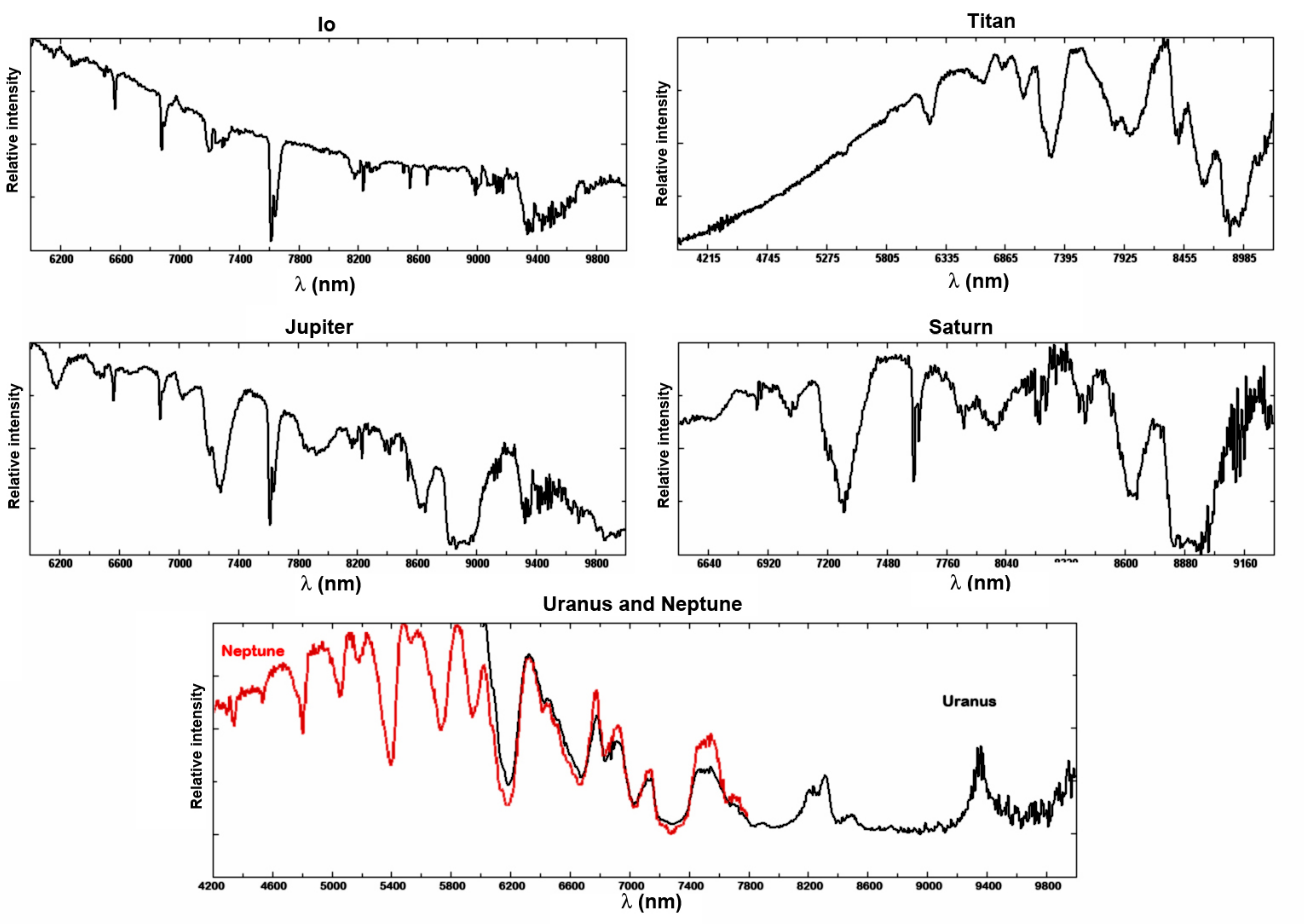}
\caption{Examples of spectra of giant planets and satellites (credit J. Guarro-Fl{\'o}). {Two 40.6-cm Schmidt-Cassegrain telescopes were used to acquire these spectra. Both telescopes were equipped with spectrographs using a grating of 600 lines per millimeter and a slit of 30 microns. Visual spectra were acquired with a camera AUDINE KAF-1603 ME with pixel size of 9 microns and the infrarred spectra with an ATIK 314L camera with pixel size of 6.45 microns.}}
\label{fig:Planets_spcetroscopy_figure}     
\end{center}  
\end{figure}

The use of sensitive CCD detectors and the recent availability of low-cost versatile spectrometers aimed at the amateur community have also resulted in advances in spectroscopic observations of the giant planets. Although only a few amateurs regularly obtain such observations, they can obtain spectra more regularly than the scarce observations performed using professional telescopes. Uranus, Neptune and Titan are particularly interesting targets because low-resolution spectroscopy or broad-band photometry at methane absorption bands can be used to inspect changes in the atmospheres of these objects caused by convective eruptions or changes in the bands of the planets \cite{Lorenz03}. Although Neptune is a challenging target and Titan presents the additional difficulty of scattered light from Saturn, significant results are achievable by amateurs monitoring long- and short-term changes in these atmospheres. Jupiter and Saturn offer easier targets with easily identifiable ammonia and absorption bands but with lower scientific interest when compared with data obtained from imaging or high-spectral resolution spectra from professional telescopes (see Fig. \ref{fig:Planets_spcetroscopy_figure}). We refer the reader to the previous section \ref{spectro} for details on spectroscopy techniques. In principle, large volcanic eruptions on Io could be detected from spectroscopic observations with amateur equipment.

\subsection{How to contribute}

\begin{figure}
\begin{center}
\includegraphics[scale=0.24]{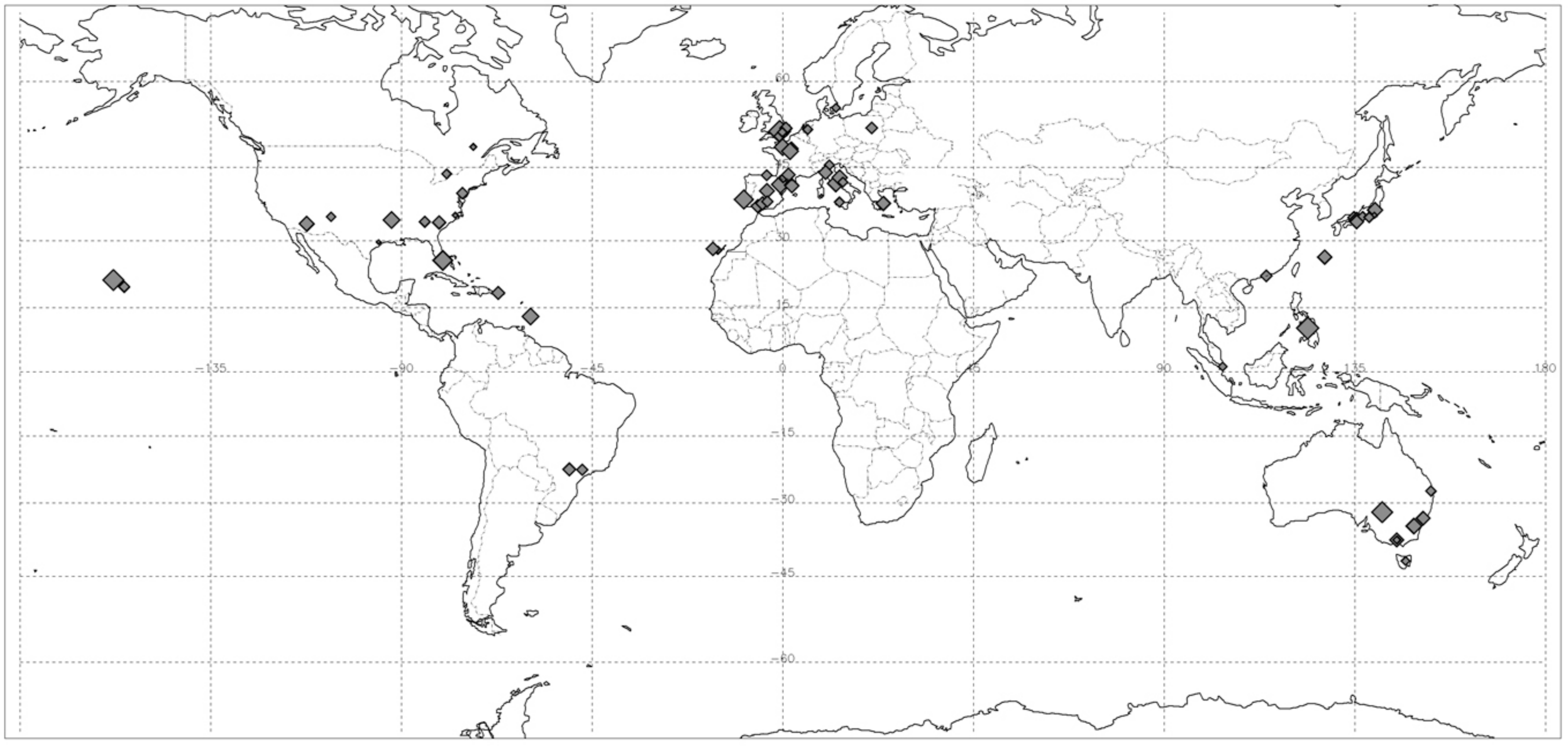}
\caption{Geographical location of about 80 prominent observers contributing to the IOPW-PVOL database. The size of each point is a measure of the number of Jupiter image contributions. Images supplied from Far East and Australia, Europe and North and South-America can monitor Jupiter  continuously.}
\label{fig:IOPW_Observers_location}       
\end{center}
\end{figure}

Broad PRO-AM collaborations have been underway for the last 25 years under the International Outer Planets Watch, which currently hosts a large database of giant planets observations performed by amateurs. The database, called the Planetary Virtual Observatory and Laboratory (PVOL)\footnote{http://www.pvol.ehu.es/pvol/}, is documented in \cite{Hueso10b}. Additional databases mainly in the amateur community store many individual observations and are commonly consulted by professionals (Association of Lunar and Planetary Observers in Japan (ALPO-Japan)\footnote{http://alpo-j.asahikawa-med.ac.jp/indexE.htm}, Soci{\'e}t{\'e} Astronomique de France (SAF)\footnote{http://www.astrosurf.com/saf/SAF} and Association of Lunar and Planetary Observers (ALPO)\footnote{http://alpo-astronomy.org/ALPO}).  News about current topics of interest are posted regularly on that site and detailed reports on the Jovian atmosphere are posted at the BAA website regularly. The distributed geographical location of observers allows for global monitoring of Jupiter and Saturn close to their opposition (see Fig. \ref{fig:IOPW_Observers_location}). {Continuous observations represent several observations a day, which are currently achievable by a large network of amateur observers. The Jupiter and Saturn planetary periods of 10 hours are well suited to observations from America, Europe, the Middle East and the Far East (Japan, Phillipines, Australia). Strategic points such as Hawaii or the Middle East are covered by a very small number of observers. Typically, such networks have performed more than 15 observations per day close to Jupiter's opposition in the last few years.} The freely available WinJupos software can be used to navigate ground-based images of the giant planets, project them into different geometries and obtain measurements of atmospheric details.

\subsection{Jupiter} 
\label{Jup}

Because of its large size in the sky, ranging from $\sim$35 to 50 arcsec, the planet Jupiter has been one of the favorite targets of amateur astronomers. The study of the morphology of the Jovian clouds and their movements have been practically in the hands of amateurs for more than a century. The best accounts of these observations are summarized in the books by Peek \cite{Peek58} and Rogers \cite{Rogers95}. Amateurs currently use the techniques previously described, allowing dynamical studies of the atmosphere. Traditionally amateur associations have conducted qualitative descriptions of Jovian cloud morphology variability as well as quantitative measurements of the dominant zonal motions of the features, with continuous descriptive records by the BAA (UK), ALPO (USA), ALPO-Japan, SAF (France). These historical works can be found in their publications (Journals, Memoirs and Bulletins) and updated reports on the current state of the Jovian atmosphere in their webpages. Additionally, for two decades the amateur JUPOS project\footnote{http://jupos.org} has been measuring Jupiter images coming from worldwide historical and current observations, collecting them in a complete database of positional data allowing more detailed dynamical studies of the atmosphere.

In what follows we describe the target studies of the amateur community in Jupiter, leaving apart the contribution to impacts that has been treated previously.

\subsubsection{Studies of atmospheric features}

A major contribution of the amateur community to Jovian studies has been the classification of the rich variety of Jovian cloud morphologies and the identification of their pattern evolutions and life cycles in the visual range (mostly covering the 400--800 nm wavelength range). Typically this has been done at a maximum resolution of $\sim$1,000 km on the Jovian disk, enough to resolve most of the planet's major atmospheric features. The continuous long-term coverage is important because the atmosphere undergoes a variety of large-scale climatic cycles lasting 1--2 years which repeat, regularly or irregularly, at intervals of years or decades.  Moreover, there is presently very little ground-based professional imaging capability available for Jupiter in the visible waveband, so amateur images are most often the only sources of a continuous record. 
 \newline

$\bullet$ \textbf{Major planetary scale disturbances:} Jupiter experiences episodic planetary-scale disturbances that produce albedo changes in the dominant bands of the planet from â``zones'' (high albedo at visible wavelengths) to ``belts'' (low albedo). The two best known examples are the South Equatorial Belt Disturbances (SEBD) at latitude $16^{\circ}$S and the North Temperature Belt Disturbances (NTBD) at $23^{\circ}$N. In each case, after gradual conversion of the dark belt to a quiescent and zone-like state, the active phase starts with one or more convective events that transform these latitude bands from a zone to a belt-like aspect in a matter of months when a turbulent pattern of features propagates eastward and/or westward from the sources as driven by the wind shears. Quantitative descriptions from data obtained by amateur observations can be found for the South Equatorial Belt (SEB) quiescent phase in \cite{Fletcher11b,Perez-Hoyos12}, for the SEBD in \cite{Sanchez-Lavega96a,Sanchez-Lavega96b}, and for the NTBD in \cite{Sanchez-Lavega91,Sanchez-Lavega08b,Garcia00,Barrado09}. 

Other characteristic belts that experience major changes are the South Temperate Belt (STB) at $31^{\circ}$S with fades and bright cloud eruptions, and the North Equatorial Belt (NEB) at $10^{\circ}$N with abundant bright storm activity (``rifts''), rare fades and northward albedo extensions \cite{Sanchez-Lavega90}. The asymmetry between the life cycles of the SEB and NEB is one of the major areas where amateurs can make important contributions.\\

$\bullet$ \textbf{Vortices:} Most oval shaped features we see in Jupiter are vortices that show different sizes and colours (from ``white'' to ``brown'' and ``red''). Anticyclones dominate in number and are located in latitudes where the speed of the zonal wind is close to zero. The most famous and best studied is the Great Red Spot (GRS) with its large size and well contrasted red colour.  The amateur contribution to the study of this vortex has been extensive, including its long-term history and length variations (roughly from 40,000 km at the end of the 19th century to 20,000 km at present), its 90-day zonal oscillation \cite{Trigo-Rodriguez00}, and its rare interactions with smaller ovals \cite{Sanchez-Lavega98}, examples of which led to targeting of specific observations with the HST \cite{Hueso09} and New Horizons spacecraft \cite{Cheng08}. Other anticyclones well studied by amateurs were the three long-lived white ovals at latitude $33^{\circ}$ S whose merger formed a single vortex called BA \cite{Sanchez-Lavega99,Sanchez-Lavega01}, which itself turned red several years later. Amateur contributions have been important in studying the changes in the long-term motions of BA \cite{Garcia-Melendo09} and in identifying its colour changes \cite{Perez-Hoyos09,dePater10b,Wong11}. Other traditional targets of amateur observations are small white and red anticyclones \cite{dePater10b} and the classical ``barges'' (persistent cyclones over large periods) at $16^{\circ}$ N.\\

$\bullet$ \textbf{Waves and other disturbances:} Some of the conspicuous features long studied by amateurs are now thought to be large-scale wave-phenomena in Jupiter's atmosphere. This is the case of the northern plumes and dark projections at $7^{\circ}$ N, whose long-term evolution can be studied in detail from the amateur data base \cite{Arregi06}. Amateurs have also contributed to the knowledge of the South Equatorial Disturbance (SED) at $7^{\circ}$ S \cite{Rogers05,Simon-Miller12} and South Tropical Disturbance (STrD) at $22^{\circ}$ S, that are perhaps examples of modes 1 and 2 equatorial and tropical waves.  Outside the visible range, amateur methane-band images have also been combined with professional infrared data to analyze the dynamics of upper-level waves on the NEB, producing conclusions that would not have been possible with either data set alone \cite{Rogers04}. 

\subsubsection{Zonal wind measurements}

\begin{figure}
\begin{center}
\includegraphics[scale=.25]{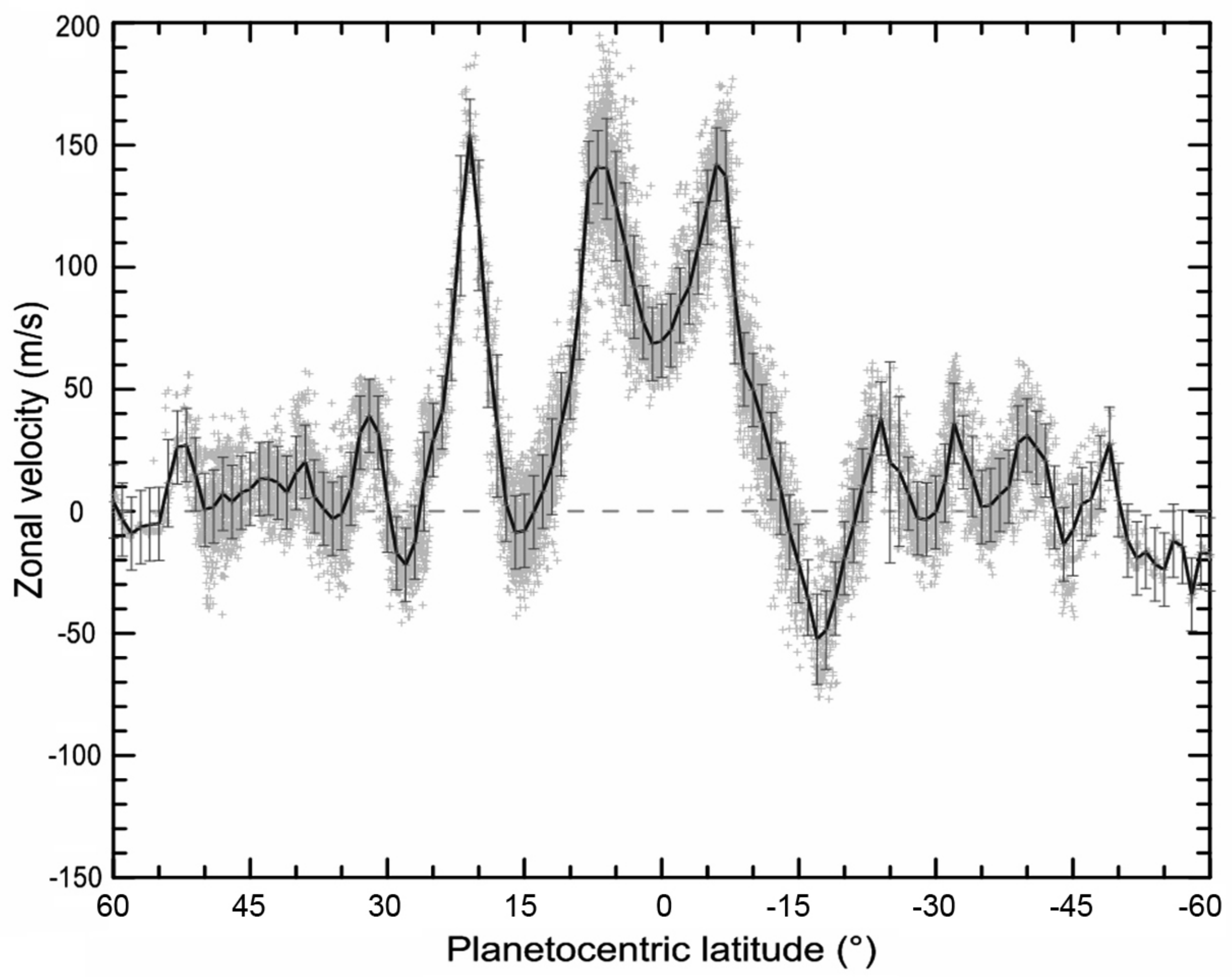}
\caption{Jupiter zonal wind velocity profile derived from amateur images (IOPW-PVOL database) and based in images from September 2011 to December 2011 (prepared by N. Barrado-Izagirre). Each dot represents a correlation wind measurements. The black line is the mean value and error bars represent the standard deviation of measurement over a latitudinal bin of 0.3$^{\circ}$.}
\label{fig:Jupiter_Figure1}   
\end{center}    
\end{figure}

East-west drift rates of visible features have been routinely retrieved by amateurs since the 19th century. Tracking of specific long-lived atmospheric features over dozens or hundreds of days were possible. This method determines velocities with a small error of $<$1 m/s and with a latitudinal resolution of $1^{\circ}$. However, because only large features could be tracked, the speeds did not necessarily refer to the local zonal winds but to specific large features, and the peaks of many jets could only be detected intermittently if at all. True zonal wind profiles could only be established by spacecraft imaging, until recent years. However, the high resolution of amateur images now makes it possible, using image pairs separated by 10--20 hours, to correlate the brightness profiles along latitude circles, allowing one to retrieve zonal wind profiles with a resolution of 0.3$^{\circ}$ in latitude and $\sim$5 m/s in velocity (see Fig.~\ref{fig:Jupiter_Figure1}). Ideally, this requires full mapping of the planet made by compositing images as Jupiter completes a rotation (which requires multiple observers distributed in longitude on Earth) and careful correction of limb darkening effects. Current zonal wind retrievals are very promising for future studies on wind profile changes in relation to morphology changes, and for establishing the amplitudes and temporal scales of the variability of wind velocities. 

\subsubsection{Quantitative photometry and spectroscopy}

The characterization of global albedos and colour changes of belts and zones and other major features can be obtained from amateur photometric images (prepared from the raw and unprocessed frames). Unfortunately, commonly used broadband Red, Green, Blue filters (RGB) that approximately match the Johnson B, V and R bands, are not very well suited for retrieving physical information of the vertical cloud structure. Because of the prevalence in the giant planets of Rayleigh scattering at short wavelengths (380--450 nm), a well suited filter for photometry is an UV one (Johnson U). However, because of the reflectivity decrease of the planet at these wavelengths and lower quantum efficiency of most camera detectors, useful images can only be taken with telescopes with diameters of 30 cm or larger. The same occurs with the widely used narrow filter centered at the 890-nm methane absorption band in which the images give information on the optical depth and vertical distribution of clouds and hazes. Future studies by amateurs equipped with telescopes with diameters above 35 cm may also benefit from including narrow filters centred at the weaker 725-nm methane band and in the adjacent continuum at 750 nm. Good images with these filters and careful calibration using standard stars of solar type or calibrated by reference to professional observations could be used for absolute photometry and radiative transfer modelling of Jovian clouds \cite{Mendikoa12}.

\subsection{Saturn}
\label{Sat}

Saturn subtends near 20 seconds of arc when close to opposition, atmospheric details have an intrinsically lower contrast and fainter surface brightness than for Jupiter. Nevertheless the same techniques used for imaging Jupiter, Venus and Mars work for Saturn although a larger-aperture telescope is needed to resolve the faint details of its atmosphere. Except for the latitudinal banding, Saturn usually has a characteristic dull appearance with few meteorological structures observable from the ground (the exceptional Great White Spots are discussed below). A 15-cm refractor may begin to resolve details such as the Cassini division in the rings and the differences between the bright equatorial zone and the rest of the atmosphere. Larger telescopes (20--28 cm) are able to resolve small scale storms in the disk, monitoring the global convective activity of the planet. The current generation of fast cameras allows observers to track even some of the cloud features not directly associated with storm activity. The demonstration by several amateurs that they could regularly detect the atmospheric features occasionally observed at high-resolution by the Cassini spacecraft triggered a renewed interest in observations of the planet that peaked again with the onset of the December 2010 Great White Spot (GWS) \cite{Fischer11a,Fletcher11c,Sanchez-Lavega11}.

\subsubsection{Saturn's storm activity}
Saturn shows less frequent convective storms than Jupiter, typically with smaller size and lower frequency and intensity. Mid-latitude storms have developed yearly in the so-called ``storm-alley'' at $35^{\circ}$S planetocentric ($41^{\circ}$ planetographic) latitude from 2002 to 2009 during southern hemisphere summer and early autumn (see Fig. 20). Cassini observations have produced high-resolution views of these 3,000 km size storms. They produce intense electric activity from electrostatic discharges \cite{Fischer08} and visual lightning \cite{Dyudina2010}. The same kind of features had been observed at high resolution by the Voyager spacecraft flybys in 1980--1981 \cite{Smith1981,Smith1982} at $35^{\circ}$N planetocentric latitude \cite{Hunt1982} over the northern hemisphere summer hinting to a seasonal cycle of convective activity. Storms on Saturn may endure several months and, while the Cassini spacecraft has studied some of these storms at high resolution on particular dates, the characterization of their long life cycles requires the long-term monitoring provided by ground-based observers. 

Since Cassini orbit insertion in 2004, there has been very active and efficient cooperation between researchers associated with Cassini's Radio and Plasma Wave Science (RPWS) instrument and amateurs. Alerts are issued when Cassini's RPWS detects Saturn Electrostatic Discharges (SEDs) allowing amateurs to observe the storm in visible wavelengths, usually within less than 2 to 3 days, hence providing accurate positions in latitude and longitude, and measurements of drift rates \cite{Fischer11b}.

\subsubsection{Saturn's Great White Spots}
Monitoring of Saturn by amateurs has resulted in discoveries of the onset of the Great White Spots of 1990 \cite{Hale90} and 2010, the latter at the same time as the Cassini RPWS instrument \cite{Fischer11a,Sanchez-Lavega11}. The 2010/2011 GWS was the first storm to be detected in the northern hemisphere at the beginning of northern springtime, and it developed 10 years earlier than expected from previous GWSs which appeared in late Saturn summer \cite{Sanchez-Lavega1994}. Images provided by amateurs spotted the storm on the first day of its activity (5 December 2010, observations by T. Ikemura) and tracked its evolution nearly continuously over 8 months, allowing a high-temporal resolution and long-term monitoring of its activity at cloud level (see Fig. 21), while Cassini instruments were able to study it a very high spatial and spectral resolution at less frequent intervals. Amateur images also provided a direct comparison between the visible albedo at the main cloud level, the hazes structure close to the tropopause with observations at the 890 nm methane band \cite{Sanz-Requena2012}, and the thermal field at the tropopause and above as observed by Cassini and large professional telescopes \cite{Fletcher11c,Fletcher12}. This multi-wavelength, multi-layer long-term sort of comparison is impossible with spacecraft or typical ground-based observatories alone. The storm ceased its activity in July/August 2011 and the abundant turbulent features observed at cloud level largely dissipated over 2012, leaving only small traces of the past activity (see Fig. 21c). However, the equatorial GWS experienced a revival in 1994 \cite{Sanchez-Lavega1996c} and amateurs are well equipped to monitor possible convective activity over the planet as the seasons proceed on Saturn.

\subsubsection{Other topics of research}
The quality of ground-based observations such as those presented in Fig. 21 warrant that other scientific subjects can be treated. Amateur images have already been able to monitor the activity of 'spokes' in the co-rotation zone of the rings in 2010 and 2012 after Saturn's 2009 spring equinox. Hypotheses for spoke creation include small meteors impacting the rings and electron beams from atmospheric lightning propagating to the rings \cite{Jones2006}. {As Saturn's North hemisphere receives more and more sunlight in the next few years, amateurs have been able to observe Saturn's north polar hexagon regularly since early 2013, constraining its overall rotation rate \cite{Sanchez14}.}

\subsection{Uranus and Neptune}
\label{UraNep}

Observations of the ice giants Uranus and Neptune are particularly challenging for amateurs. Their large heliocentric distances cause the planets' apparent disks to be too small to be well resolved under typical seeing conditions: Uranus and Neptune subtend on average only 3.8 and 2.4 arcseconds, respectively.  The ice giants are also relatively faint (visual magnitudes +5.3 and +7.8 respectively at opposition), and interpretation of photometry and spectroscopy is challenging due to a paucity of context data.  Nevertheless, both Uranus and Neptune exhibit significant atmospheric variability when observed from large telescopes even at visible wavelengths, and thus PRO-AM collaborations have ensued for these distant planets. These studies fall into several categories: visual reports and imaging, photoelectric photometry, spectroscopy, and satellite occultation observations. We first discuss Uranus, which has been of significant interest in recent years, and follow with Neptune.  We conclude with a few tips for amateurs interested in ice-giant observations.

\subsubsection{Uranus}
\label{Uranus}

$\bullet$ \textbf{Uranus visual and imaging studies.} Although Uranus was generally bland in images taken by the Voyager spacecraft in 1986, historical records of past equinoctial times suggested that discrete features were sometimes bright enough to see with small telescopes at visible wavelengths \cite{Alexander1965}. Thus, Uranus has long been a tantalizing target for amateurs.  S. O'Meara noted a bright spot on the seventh planet in September of 1981, from which he determined a rotational period of 16.4 hours \cite{Omeara1985}.  This period is consistent with features tracked in subsequent Voyager, Hubble, and Keck imaging \cite{Hammel2005}.  F. Colas and J. L. Dauvergne recorded images at the Pic du Midi observatory \cite{Sromovsky2012}, and others have drawn and imaged Uranus as well (Fig.~\ref{fig:uranus_neptune_IMAGES}; see also \cite{Arditti2009}). In 2007, Uranus reached its first equinox since the advent of modern astronomical imaging (the last equinox was 1965). Professional telescopes revealed a striking upsurge in activity in the years surrounding equinox, though most of it required the exquisite spatial resolution and sensitivity of the Hubble Space Telescope and the Keck 10-m \cite{Hammel2009,dePater2011,Sromovsky2012}. In October 2011, near-infrared images acquired with the Gemini telescope revealed an extraordinarily bright feature \cite{Sromovsky2012}.  An alert went out to the amateur community, because an amateur detection could trigger a ``Target of Opportunity'' proposal with the Hubble Space Telescope \cite{HammelHST2009}. In spite of attempts by some of the best amateur observers, the detections were marginal with smaller telescopes. Successful observations of the feature were obtained with the {1.05-m} telescope at the Pic du Midi observatory and the Very Large Telescope, and these served to predict the feature's position. Subsequent observations from Keck and Hubble revealed that the feature had diminished in brightness \cite{Sromovsky2012}. The episode does demonstrate that some visible-wavelength features are occasionally within the reach of amateurs with large telescopes, consistent with the historical visual observations. This was confirmed in 2012, when amateur images taken with instruments from 25 to 40 cm detected details on the planet in the near infrared (Fig.~\ref{fig:uranus_neptune_IMAGES}). We thus encourage amateurs equipped with telescopes of ~25 cm or larger to monitor Uranus for features.  Alerts for confirmed features should be sent to the professional astronomical community.\\

$\bullet$ \textbf{Uranus photometry.} Several amateurs have carried out whole-disk brightness measurements of Uranus \cite{Schmude2012}.  In all cases, they used an SSP-3 solid-state photometer along with filters transformed to the Johnson B, V, R and I system (individual transformation corrections were made for each telescope-photometer-filter system). As of October 2012, the Remote Planets Coordinator of the Association of Lunar and Planetary Observers (ALPO) had received 1054 brightness measurements of Uranus, mostly at Johnson V (58 \%) with a smattering of other wavelengths: B (24\%), R (9\%) and I (9\%).  The upper panel of Fig.~\ref{fig:uranus_neptune_PHOTOMETRY} compares average normalized magnitudes in the V filter from amateurs compared with the long-term results from Lowell Observatory; the amateur results are consistent with the quasi-seasonal trend in brightness of Uranus identified in the long-term lightcurve \cite{Lockwood2006}. Professional and amateur brightness measurements of Uranus are also consistent with a small seasonal change in the B-V color index.\\

$\bullet$ \textbf{Uranus spectroscopy.}  Several amateurs have obtained spectra of Uranus of diverse quality. Figure~\ref{fig:uranus_SPECTRUM} shows a spectrum of Uranus obtained by Frank Melillo in the wavelength range between about 450 and 950 nm.  This observer has carried out measurements of this type for several years between 1999 and 2011 \cite{Schmude2012} with spectral resolutions between 10 nm (1999) and 3 nm (2011). Several absorption features are clearly observable and temporal variation can be studied from detailed comparison of several spectra.\\

$\bullet$ \textbf{Uranus system occultations.}  On 8 September, 2008, Uranus' moon Titania occulted the star HIP 106829.  Observations of this event -- both by professionals and by amateurs using telescopes as small as 5 cm in aperture -- were used to constrain the size, shape, ephemeris, and atmosphere of Titania \cite{Widemann2009}.  Given the paucity of information about the Uranian moons, such events are of great importance.  Though these opportunities are rare, amateurs are encouraged to participate in occultations observations whenever possible, not only of ice giant satellites but of the planets themselves.

\subsubsection{Neptune}
\label{Neptune}

$\bullet$ \textbf{Neptune visual and imaging studies.}  Neptune imaging is extraordinarily difficult, and requires exquisite seeing and true dedication on the part of the amateur.   That said, it is within the reach of a 35-cm telescope. For instance, D. Peach recorded realistic details in a RGB image on September 25, 2012 (see Fig.~\ref{fig:uranus_neptune_IMAGES}) during the same observing series when he recorded no distinct features on Uranus.  C. Pellier likewise obtained a Neptune image with a resolved disk on August 11, 2012.  Given the paucity of time for Neptune observations on professional telescopes, observations by amateurs are needed in order to trigger special observing opportunities for this planet.\\

$\bullet$ \textbf{Photoelectric Photometry.}  For Neptune, a bigger area of PRO-AM collaboration has been photoelectric photometry.  The coordinator of the ALPO Remote Planets Section has received 683 brightness measurements of Neptune as of October 2012 \cite{Schmude2012}. The distribution across filters is: B = 30\%, V = 58\%, R = 6\%, and I = 6\%.  The lower panel of Fig.~\ref{fig:uranus_neptune_PHOTOMETRY} compares the normalized magnitude of Neptune since 1991 with the long-term observations from Lowell Observatory.  Between 1991 and 2000, that planet brightened by 0.1 magnitude (or ~0.01 magnitude/year).  Since then, it has maintained a roughly constant brightness.

\subsubsection{Tips for observing Uranus and Neptune}

Several resources are available for amateurs who are interested in observing the ice giants \cite{Arditti2009,Schmude2010}. Interested observers should consult those resources and we provide a few tips here.  Given the very small apparent disks, unusually good seeing is often critical. In order to achieve the best images, the instrumentation should be perfectly aligned. The larger the aperture, the better the chances are of seeing atmospheric detail. A productive avenue is the growing use of professional-sized instrument (1 meter or larger) by amateurs. Excellent images of Uranus were taken in 2012 by mixed PRO-AM teams at the Pic-du-Midi Observatory, as well as the C2PU 1--meter telescope at the Calern Observatory in the southeast of France. Images in near-infrared filters are more likely to show albedo features than those in visible light filters.  The ice giants' methane absorption bands are relatively broad, allowing the visibility of belts even with so-called R+IR filters ($>$ 600 nm). Up to now the best results have been obtained with IR-pass filters from 685 nm, where the contrast is greater. In all cases, the orientation of the image should be known accurately and precisely. If intriguing details appear, tests must be applied to the observation to discard potential artifacts or processing effects (verify the orientation, assess whether the behaviour is as expected due to rotation, etc).  Simultaneous independent observations are especially important in order to confirm ambiguous observations. ALPO has designated the 15th of each month to be a special time when people should try to image Uranus and Neptune.  Negative data is also important.  For example, in the case of a suspected occultation, a non-event (an appulse) is valuable for timing and ephemerides purposes, and thus should be reported. Brightness measurements should be corrected for both atmospheric extinction and color transformation. Professional astronomers are often interested in the relative intensities across the disk of Uranus and Neptune, therefore contrast should not be stretched unless the stretch is noted quantitatively. Any changes in limb darkening or albedo features are important.  Very long exposures are required to make out belts on images (up to 30--45 mins), therefore the use of de-rotation techniques such as WinJupos would allow amateurs to correct the drifting of potential spots with time. Such de-rotation techniques have also been introduced with great success in analyses of Keck observations of Uranus \cite{Fry12}.

\section{Comets}
\label{phil}

Comets, with their roughly round comae and their long tails, have been observed since ancient times. Thousands of them have been discovered. A comet mostly consists of a small, kilometer-sized nucleus, built up of ices (mostly water ice) and dust particles. As a nucleus, on its elongated elliptic orbit, approaches the inner Solar System, its surface is heated enough to make it active, with the sublimation of some ices that triggers the ejection of dust. Gaseous molecules and dust particles, through respectively fluorescence and solar light scattering, form a bright coma that hides the nucleus. Molecules, dissociated and ionized by solar radiation, are dispersed by the solar wind to form long and narrow plasma tails. Fluffy dust particles, which progressively fragment, are driven back by the solar radiation pressure, to form broader dust tails. Cometary orbits are perturbed by gravitational and non-gravitational forces, the latter ones resulting from a combination of the nucleus rotation and activity.

Amateur astronomers have always played an important role in the observation of comets. For many years, amateurs discovered most of the new comets, and they continue to contribute actively to discovery and imaging of comets. Moreover, they also still provide most of photometric and astrometric data on comets. With the improvement of their instruments and the development of CCDs and digital cameras, they can provide accurate measurements for the different databases available to the community.

\subsection{Search for new comets and comets imaging}

In the history of astronomy, amateur comet hunters have played an important role for discovering new comets. From J.-L. Pons, initially caretaker at Marseille observatory, who discovered 37 comets from 1801 to 1827, to D. H. Levy, who has contributed to the discovery of about 23 comets, including famous comet Shoemaker-Levy 9, the fragments of which impacted Jupiter in 1994, many amateur astronomers managed to associate their name with different comets. A comet is indeed usually named after its (up to three) independent discoverers, who can be the observers or simply the telescopes/faciities used by a team of astronomers. 

\subsubsection{Discovery of comets}

Comets were discovered for a long time via visual observations, but  nearly all recent discoveries are made through automated CCD searches. The competition with professional instruments became stronger in the 1990s with the development of different automatic surveys mainly designed for discovering NEOs (see also Sec. \ref{aste}), such as LINEAR, NEAT, LONEOS, Spacewatch, Catalina Sky Survey or Pan-STARRS (this last one in the beginning of 2010s). The Edgar Wilson Award, which celebrates amateur cometary discoverers, was given in 2012 to five amateurs for their discoveries of comets. The names of three of these discoverers were indeed given to comets C/2010 X1 Elenin, C/2011 W3 Lovejoy (a spectacular sungrazer) and C/2012 C2 Bruenjes. The comet spotted by A. Novichonok and V. Gerke, in images from the International Scientific Optical Network, has been named C/2012 S1 ISON.

A close examination of the overall statistics (Fig.~\ref{f:stat}) reveals, nevertheless, that, despite a significant increase of the discovery rate in the mid-1990s, the absolute number of discoveries by amateur astronomers or by small telescopes (up to 50~cm in diameter) is, more or less, stable. For telescopes up to 50~cm, i.e., with instruments available in the amateur astronomers community, the number of discoveries has been steady at typically about 15 per year since mid-2000s, despite the beginning of Pan-STARRS observations. 

Such a constant discovery rate for amateur astronomers and small telescopes can be explained by different factors. First, they have adapted their strategy to search for new objects in the regions poorly covered by these telescopes. Second, a main advantage of small telescopes is their ability to scan quickly regions of sky close to the Sun, typically with elongation below about 100$^{\circ}$. 

Figure~\ref{f:elong} presents the discovery magnitudes as a function of elongations both for Pan-STARRS comets and small telescopes. It shows that small telescopes used both by amateur and professional astronomers manage to discover new comets with lower elongations and smaller magnitudes than those discovered by Pan-STARRS. 

An alternative amateur comet-hunting opportunity involves SOHO. This space observatory, in orbit since December 1995, is unique for cometary observations. First, because of its halo orbit around Sun-Earth L1 point, it may provide observations close to the Sun; secondly, a significant part of the comets discovered with SOHO instruments, mostly LASCO coronograph and SWAN, are due to amateur astronomers who process the observational data available on internet\footnote{http://comethunter.lamost.org/SOHO/rank.htm}. SOHO is, in fact, the top discoverer of comets to date with a 
total of 2437 discoveries as of January 8, 2012, most of them corresponding to sungazing comets, including the fragments of one particular comet, known as the Kreutz group. These comets are mostly detected in the close vicinity of the Sun, with the LASCO coronograph. The SWAN instrument, in comparison, covers the whole sky in the Lyman~$\alpha$ line (allowing observations of huge hydrogen halos around comae), and has a sensitivity limited to comets of magnitude 10 or less. Although this magnitude is much smaller than the range of magnitudes corresponding to the discoveries performed with small telescopes, it has nevertheless, thanks to amateurs from Australia and California, allowed the discovery of three new comets. Amateur astronomers interested in the best strategy for such a ``hunting'' can find useful information in M. McKenna's website\footnote{http://www.nightskyhunter.com/index.html}. Finally, amateurs may also contribute to the discovery of comets hidden within the asteroids population (see Sec. \ref{main}).

\subsubsection{Imaging of comae and tails }

Amateur astronomers have been able to study structures in cometary comae, dust tails and plasma tails in the past by drawing accurate sketches of their visual observations, and today with flat-field corrected CCD images. For example, systematic amateur observations of comet 1P/Halley (at the great refractor of Paris-Meudon Observatory) from October to December 1985 visually revealed some faint coma and tail structures \cite{berge1988,levasseur1988}.

Structures in plasma tails, such as disconnection events, are tracers of the solar wind and interplanetary magnetic field properties. In dust tails, subtle striations (as retrieved in March 2013 on comet C/2011 L4 (PANSTARRS), can be used to provide information on the size of the dust particles\footnote{http://apod.nasa.gov/apod/ap130330.html}. Finally, in dust comae, jet-like or spiral structures are clues to the presence of active regions on the nucleus and to nuclear rotation. In such domains, cooperation between amateurs and professionals is always fruitful. 

\subsection{Astrometry}

Astrometry of comets is similar to that of asteroids (see Sec. \ref{pipe}). The best results are obtained with long focal lengths (minimum 2~meters) and with exposure times as short as possible while maintaining a good signal-to-noise ratio. As compared to asteroidal astrometry, cometary astrometry differs in a few aspects: (i) the photocenter of the coma is not always located at the nucleus center position. The brightest region is the one (usually sunslit) where the largest amount of gas and dust is released, leading to significant errors for the nucleus position. (ii) Cometary orbits can be influenced by non-gravitational forces when activity is important, leading to the necessity of more observational data during the period of activity. (iii) Some cometary orbits are highly eccentric (e~$\simeq1$) and a high accuracy is needed to distinguish parabolic orbits from highly elliptic ones in the period following their discovery. 

The MPC collects the astrometric observations. The magnitude of the object is indicated for each observation, but is not very accurate because  it depends on the instrumentation and the aperture used for the measurement. Nevertheless, these magnitudes are most often used to analyze the activity of comets.

Astrometric data are mostly obtained without any filter for faint comets, and with a R filter when the flux is high enough (providing a better accuracy). Different astronomical software packages that include astrometric functions can be used (e.g., Astrometrica, Prism or Au-dela), before the data are sent to the MPC\footnote{See, e.g., http://www.britastro.org/projectalcock/CCD\%20Astrometry\%20and\%20Photometry.htm for more details.}.

The data collected by the MPC are used by some other institutes, such as the Institut de M\'ecanique C\'eleste et Calcul d'Eph\'em\'erides (IMCCE\footnote{http://www.imcce.fr/langues/en/}). The IMCCE computes, e.g., the difference between the observed and the calculated position (O-C) for the data collected by the MPC\footnote{See, e.g., http://www.imcce.fr/fr/ephemerides/donnees/comets/FICH/OMCF0835.php}. Such O-C calculations permit an estimation of the quality of the data for each observer and of the problems associated with a bright coma, for active objects. For such comets there is no special method to perform astrometric observations (compared to asteroids). If the difference with the theoretical trajectory is too large the MPC rejects automatically the observational data.

\subsection{Photometry and activity monitoring}

Comets' behavior is controlled by the solar flux received from the Sun and their physical properties, which differ greatly from one comet to another. During their period of activity, which can last typically a couple of years, this behavior is often unpredictable. Both the overall activity and unusual events -- such as outbursts or splittings -- need to be monitored frequently (typically several times per month). 

Because of their large amounts of observational time, the contribution of amateur astronomers for photometry and monitoring of unusual events is of vital importance for cometary science. Such works perfectly complements professional observations performed with state-of-the-art scientific instruments during very short periods of time. As an example, amateur astronomers extensively observed comet 9P/Tempel~1 in 2005, prior to and after the Deep Impact mission, pointing out some outbursts and providing hundreds of CCD images. The amateur-astronomer observations are important both for modeling the nucleus' physical properties and helping professional astronomers to prepare their observations (either to know in advance the activity level of a target for standard observational proposals or to request observing time in emergency in case of an unusual event).

The different parameters whose determinations are accessible to amateur astronomers are:

\begin{enumerate} 

\item The visual magnitude (for the photometric center region, close to the nucleus and/or total magnitude);

\item The appearance: size, tail direction, appearance of the central condensation, coma diameter, the degree of condensation (DC, on a scale of 0 to 9, where 0 is completely diffuse and 9 is completely stellar in appearance);

\item  $Af\rho$, as defined by \cite{ahearn1984}. It represents the product of the albedo $A$, the filling factor $f$ (i.e., the ratio of the total cross section of dust grains within the field of view divided by the area of the field of view), and the linear radius $\rho$ of the field of view at the comet. It is homogeneous to a distance and is usually expressed in cm. It has the main advantage of being, more or less, independent of the field of view, thus providing a simple comparison of the cometary activity monitored by different observers.

\end{enumerate}

\noindent
In addition to these parameters, amateur astronomers can contribute to the field by triggering an alert for a given unusual event (see, e.g., \cite{buzzi2007} for comet 17P/Holmes outburst on 24 October 2007).

\subsubsection{Instrumentation}

Ideally, the choice of an instrument depends on the interest of the observer for the type of cometary science. For photometry the observers will prefer shorter focal lengths (compared to astrometry) with longer exposure time (without saturating the photocenter) to detect the coma and its extension, and follow its temporal evolution. Both for astrometry and photometry, the use of ``large'' sensors is usually  preferable. The choice of the best focal length for aperture photometry depends on the apparent size of the comet, because the overall coma must be observable through the instrument. The brightness and apparent size of the comet at the time of observation are the two important parameters that have to be taken into account. However, a simple camera lens will be better for comets with very large angular sizes (e.g., C 2006 P1 McNaught or 17P/Holmes). Such cameras have a large field of view that includes both the comet and several bright stars, mandatory for a correct photometric reduction. It is also possible to use the green channel of digital cameras to match a V filter and the human eye response.

Regular monitoring of fainter comets is mainly limited by the available instrument. A telescope with a diameter of 20~cm can be used for monitoring almost all observable comets up to a magnitude of about 20 (this magnitude being reached by such a telescope in about 8$\times$120~s of exposure time).

Although not totally essential, the use of filters is clearly recommended because the determination of dust properties, e.g., $Af\rho$, requires filters, and tentatively narrow-band filters, to avoid contamination by the gaseous emissions which are specially significant in the blue and green domains. Unfiltered observations can be useful if the goal is simply to record a lightcurve for amplitude and period determination or to detect faint objects. Unfiltered observations can in some cases be combined with measurements in a standard system, i.e., a photometric system with a set of well-defined passbands (or filters), with a known sensitivity to incident radiation. The most commonly used filters are BVRI filters. They can be chosen to approach as well as possible the classic Johnson-Cousins filters. Another important criterion for the choice is the spectral response of the sensor. This spectral response must be optimized, as far as possible, to the passband of interest.

The photometric reduction needed to convert instrumental magnitudes into magnitudes expressed in a standard photometric system is an essential step in order to make a scientific use of the data\footnote{See, e.g., http://www.observatorij.org/CCDPhot/iwca5.html or http://www.icq.eps.harvard.edu/CCDmags.html}. Reference stars are chosen in standard photometric catalogs such as Loneos\footnote{ftp://ftp.lowell.edu/pub/bas/starcats/loneos.phot}. Atmospheric extinction must be taken into account in the calculations by observing two separate fields at different airmasses. The fields must not be too low above the horizon (airmass $\leq$ 2). The conversion to a photometric system (2 or 3 filters) is performed by calculating photometric coefficients (this is a complex calculation that can be done by most softwares designed for processing astronomical data). Reference stars must be selected with color indices as different as possible. For a given location and instrument, the color terms are almost constant; only the extinction coefficients and zero points have to be determined for each night\footnote{See, e.g., http://www.aavso.org/sites/default/files/Transforms-Sarty.pdf for more details.}.

\subsubsection{Databases}

The ICQ (\textit{International Comet Quaterly}\footnote{http://www.icq.eps.harvard.edu/icq.html}) collects mainly visual observations, which are vitally important to link information about comets observed in the past and in the present. These observations include the magnitude and the appearance of comets. All the details about the instrumentation, the conditions and the method of observation, and reference stars (visual magnitude) are required to submit the observation. The format of observations sent to the ICQ is carefully codified\footnote{See this note on ICQ: http://www.icq.eps.harvard.edu/ICQFormat.html}. Since 2002, a new extended format has been introduced for the CCD observations. It provides a means for sending observation reports of fainter comets.

Other databases such as the CARA (\textit{Cometary ARchive for Afrho}\footnote{http://cara.uai.it/}) are devoted to the collection of $Af\rho$ measurements. Unusual events are often announced in different mailing lists that include both amateur and professional astronomers\footnote{See, e.g., \textit {Comets Mailing List} at http://tech.groups.yahoo.com/group/comets-ml/}, in order to trigger an alert for follow-up observations.
 
\subsection{The future}

Amateur astronomers will probably remain active in the field of cometary observations, providing useful data for professional astronomers. Amateurs' contribution to the monitoring of photometric parameters, to acquire astrometric data, and to trigger an alert for unusual events is complementary to professional astronomical observations with large telescopes that are focused on specific scientific issues. With the constant improvement of amateur astronomers' instruments, some new fields might be opened to this community, among them:

\begin{enumerate} 

\item The monitoring of cometary activity at large heliocentric distance. Many comets or Centaurs are now known to exhibit cometary activity at large heliocentric distance (i.e., above 5~AU). This cometary activity differs from normal activity because it is not driven by water sublimation, and different scientific issues about the physical nature of cometary nuclei can be addressed by monitoring such activity. Such objects are faint (V$\simeq$15-20) and need telescopes larger than 50~cm. However, even measurements of $Af\rho$ parameter with large uncertainties would be very useful since it is difficult for professional astronomers to obtain observing time with a large telescope for monitoring this type of phenomenon;

\item Photometric measurements by using narrow-band filters centered either on bright emission bands (mainly C$_2$ and CN) or on the dust continuum. Such observations allow measuring absolute production rates and radial profiles of the main radicals present in the coma;

\item Polarimetric imaging of dust comae by using, together with a filter (e.g., broad-band red Bessel filter), a rotating polaroid (fast axis oriented along four directions at 45$^{\circ}$ from one another). With telescopes larger than 50~cm, such observations on bright comets (for short duration series of measurements) reveal changes in the dust properties that are independent of the dust concentration \cite{Hadamcik2010};

\item Long-slit spectroscopy with low-resolution (or even medium-resolution) spectrometers. Such spectra can be used to measure absolute production rates of the main radical, as well as their radial profile along the slit.

\end{enumerate} 


\section{Observation of Kuiper Belt Objects and Centaurs}
\label{Kuiper}


\subsection{Direct observations}
\label{dirobs}

\subsubsection{What is observable ?}

All but a handful of KBOs are fainter than magnitude 20, making them unreachable except with the largest telescopes (bigger than 1~m in aperture) and/or the best sky conditions. Since Centaurs are closer to the Sun than KBOs, a greater fraction of them are brighter than magnitude 20. But they represent a rather sparse population, as they are on unstable orbits, in transit between the Outer Solar System and the active comet region. 

Amateur astronomers have asked about the possibility of finding more {\it big} KBOs, or rather {\it bright} ones. M. Brown, quoted on the Minor Planet Mailing List\footnote{http://tech.groups.yahoo.com/group/mpml/message/27762} gives the following response: ``The short answer is that we are complete in the North to about 20th mag and in the south to about 19 (probably final analysis still finishing up). If there are any more bright ones left to be found, the only place left to hide them is the galactic plane or within about 15$^\circ$ of the celestial poles.'' This statement applies to the KBOs, but probably also to the close-in Centaurs. The current expectation is that there should be one more very big object as large as Pluto or Eris, with an absolute magnitude\footnote{The absolute magnitude is the magnitude that an object would have if it were at 1~AU from the Sun, 1~AU from the observer, and seen with a phase angle of 0$^\circ$. This is a good proxy for the size of an object, $H$ decreasing as the object gets bigger.} $H \sim -1$. As people observing asteroids know, a bright object does not need to be big. It may simply be close. Similarly, a big object may be faint, so long as it is far enough from the Sun. Different populations have different magnitude variations along an orbit. Apparent magnitude varies with time much more for objects with large eccentricity than for those with low eccentricity. 

As can be inferred from Table~\ref{tab:delta_m}, Centaurs of moderate sizes ($7 < H < 10$ or sizes from 60 to 250~km) are found with apparent magnitudes within reach of amateur telescopes. This population is relatively undersampled compared to the KBOs and asteroids. A few new objects are regularly discovered, mostly by Pan-STARRS, but not tracked\footnote{A list of these objects is available at http://www.minorplanetcenter.net/db{$_-$}{\rm search}}. The global properties of Centaurs are $a_{\rm max} = 30$~AU and minimum perihelion distance at 7~AU. Tracking those objects would be a very valuable contribution. Setting up a survey to detect all Centaurs brighter than magnitude 20 or 21 would be even better, but it is likely that Pan-STARRS would scoop such a survey, except in the Southern hemisphere. 

One important question currently investigated in KBO and Centaur science is the shape of the size distribution. There are hints at a change in shape around $H = 9$. This is accessible to amateur telescopes for Centaurs between 5 and 15~AU. whether for searching for big KBOs or smaller, closer Centaurs, amateur astronomer E. Ansbro suggests surveys at large ecliptic latitudes to magnitude 21\footnote{http://tech.groups.yahoo.com/group/mpml/message/27846}, in line with the previous suggestions.

\subsubsection{Observing}

The basic method for observing KBOs and Centaurs is similar to that of asteroids: acquire two or more images at some interval of time and compare them. Stars will be fixed, while the object of interest should move from one frame to the other. The first characteristic for an observer to consider is the rate of motion on the sky, how this will limit the techniques to be used, and how to advantage of it.

The first problem is to avoid being confused with asteroids. When observing within 20$^\circ$ of opposition, objects at different distances from the Sun have different apparent motions.  Between 25$^\circ$ and 55$^\circ$ from opposition, asteroids have similar apparent on-sky motions to those of more distant objects, making the distinction between different dynamical classes impossible and increasing the risk of confusion. Between 60$^\circ$ (2 month) and 90$^\circ$ (3 month) from opposition, distant objects separate again from asteroids, being close to their stationary point. One can again observe them, although with less optimal conditions.

The KBOs and Centaurs are faint objects, hence a dark sky is essential. All these objects tend to be neutral to very red in color. In particular, they are redder than the Moon, and brightest where most of the CCD detectors are most efficient. So it is beneficial to use an R (or r, or r') broadband filter. Cutting the blue to green part of the spectrum removes more of the Moon and sky background flux than that of the object. Also, it is important that the filter has a sharp cut-off at long wavelengths around 750~nm. Having a transmission of even only a few percent up to 900~nm may reduce the limiting magnitude by a few tenths of a magnitude up to 0.5~mag. Another effect of too wide a filter (V+R) or no filter at all is differential refraction. This behaves like increased seeing, reducing the limiting magnitude, even though it also increases the flux from the object. The need for an R filter is more stringent for: a site with sub-arcsecond seeing; small pixel size (images are well over-sampled); and targets that are fainter in the seeing area than the sky background in the same area. However, in all cases, it is beneficial to cut wavelengths longer than about 750~nm.

The exposure time of each image must be limited to avoid trailing the object which would result in a decrease of SNR. This time limit sets the maximum depth achievable in a single image. To get around this limitation, one can collect several images of the same field and then add the images, a technique known as {\it pencil-beam} (see Sec. \ref{astdet}). If one simply aligns the images with respect to the stars and adds them, then the stars will be brighter but the signal of the object will be spread over a trail. If one instead shifts the registered images at the displacement rate of the object and adds them, the stars will be trailed, but the signal from the object will add up on the same pixels and thus increase the SNR (see Fig.~\ref{fig:1-8.1}). One can even improve the detection limit by {\it suppressing} the stars. To do this, in the last step, instead of adding the images (or taking the mean), for each pixel of the image one can take the median of all images for that shifted pixel. If the sequence is long enough, then the contribution of the stars will almost disappear while the contribution of the object remains the same.

\subsubsection{Contributing but not observing}

Amateur astronomers, or even the public at large, can participate in KBO science by performing analysis tasks that cannot be fully automated and that professional astronomers do not have the {resources} to achieve. A nice example of such public involvement was the IceHunters\footnote{http://www.icehunters.org/} project which helped the New Horizon KBO search team to look for potential targets for a flyby by the New Horizon probe after its close approach to Pluto in July 2015. {The Ice Hunters project was replaced in 2012 by its successor Ice Investigators, which was supposed to search the data from the spring and summer 2013 observations.} These projects, like PlanetHunters\footnote{http://www.planethunters.org/} or Moon Mappers, Vesta Mappers or Mercury Mappers\footnote{http://cosmoquest.org/} use the image analysis skills of the human eye and brain to go through a huge amount of data and detect specific patterns of interest. Not all surveys can use this approach to involve the general public. Large-area surveys for KBOs or Centaurs need a fast analysis and detection pathway to be able to acquire follow-up observations in a timely fashion. Work achieved by an open community cannot guarantee an almost {\it real time} analysis; and these surveys must rely on their own forces to achieve their goals. {In its initial form, the New Horizon KBO search was well adapted to open-community involvement as it is interested only in objects that will be within reach of the probe and part of a well populated component of the Kuiper Belt. Looking at a restricted area without propagation is well known meaning that even a 6-month to 1-year delay in detecting the KBOs was not a problem. Now, with the approach of the Pluto encounter and the New Horizon maneuver, any delay in data processing becomes problematic. Recent developments in data reduction pipeline have proved efficient in detecting KBOs and Centaurs. However, the need for an accurate calibration of the detection efficiency for large surveys requires a lot of visual inspection of the pipeline proposed candidates. This could be achieved by a dedicated open community if they were to commit to performing the task in a given time lapse.}

\subsubsection{What about photometry?}

Because the objects are big, they tend to be round and the lightcurve is generally rather flat. So the requirement on the photometric precision is tighter than for most asteroids; it should be 0.05~mag or better, or signal-to-noise ratio (SNR) of 20 or more. This can be achieved only for objects that are of 2 orders of magnitude brighter than the limiting magnitude of the telescope (all other things being equal).

Except for this stronger limitation, photometry of KBOs and Centaurs is similar to that of asteroids. Given the strong SNR constraint, amateur contribution to photometric studies is even more limited.

\subsection{Stellar occultations}
\label{occultno}

There is a considerable lack of information about distant objects in our Solar System. The properties of KBOs are more or less unknown apart from a few prominent examples. One of the reasons is their large distance from the Sun and their size in general (less than the size of the dwarf planet Pluto). Their physical properties can only be addressed by very sophisticated indirect techniques, such as infrared observations from space by the Herschel Space Telescope. Ground-based spectroscopy and photometry are other tools which may help. Space probes to these far distant worlds are not available for the next decades with the exception of the ``New Horizons'' mission to Pluto and beyond. Observations of stellar occultations can provide some more insight into our outer Solar System, from defining the shapes of the bodies to the detection of possible atmospheres.

The observation of occultations by dwarf planets and KBOs is a little bit more challenging than observing occultations by main-belt asteroids. The small angular diameters of the bodies combined with the large distances (more than 40 AU) need ultra-precision astrometry for predicting possible occultation events. Astrometry sets the limitation for successful observation campaigns. From the distance of Pluto, the Earth has an apparent diameter of a little more than 400 milliarcseconds (mas). Even with optimal astrometry the final error can be 1,000 km or more projected onto the Earth. The KBOs are in general very faint; the Pluto system is an exception. The need for high precision in combination with faintness of the object limits the astrometric work more or less to professional stations, and is beyond the scope of typical amateur work.

On the other hand, the recording of stars with magnitudes between 11 and 19 (V, R, or I Band) with exposure times less than 5 seconds is today possible with small- to medium-sized instruments. The cost of such kind of instrumentation is within the reach of many amateur stations or public observatories. Because KBOs may have thin atmospheres (Pluto and Neptune's satellite Triton are examples), a slightly different approach for photometry is useful (discussed in the next Section). If this can be provided, the detection of atmospheres down to a surface pressure of less than 10$^{-2}$ Pa is feasible. Therefore, occultation astronomy is an ideal topic, where professional and amateur astronomers can work together. Many aspects have already been discussed in \ref{occul} and \ref{dirobs}. However, the faintness of the stars and the related astrometric problems are somewhat different for KBOs than for asteroids of the main belt.

\subsubsection{The pipeline}

The full-recursive pipeline, from astrometry to the publication of data and back to astrometry is shown in Fig.~\ref{fig:1-8.2}. Astrometry as the starting point is an absolute critical task. Because of the faintness of stars and objects (typical less than 16th magnitude) it is mostly a task for professional astronomy. Observatories with smaller telescopes and less experience can very well do astrometry in the range down to about 30 mas precision, but not for such faint objects. If astrometry of the KBO and the star is provided, the occultation track on Earth is calculated similar to the case for asteroids, as described in chapter \ref{occul}. Once the catalog of the Gaia mission is available for astrometry, the precision of predictions will vastly improve.

The final occultation tracks must be distributed to all possible observatories in the appropriate part of the world. Depending on the brightness of the star, even small observatories with instruments of 20 cm diameter should be informed. The concept for this is the same as for main-belt asteroids, described already in \ref{organ}. For many campaigns, mobile groups are sent out either by car or even by airplane to distant sites where a prediction has been issued (details of one of the largest worldwide campaigns can be found in \cite{Widemann2009}). Portability of large instruments is a problem, the limitation for air-transport may be about 35 cm diameter if no special devices are used (fold-up Dobsonian telescopes etc). Webpages and mailing lists are the main distribution media to interconnect the network of observers; the information structures as described in 5.3.1 are typically used. Web 2.0 activities may add some extra information sources in the future. Even after good ``last minute astrometry'', one-sigma areas of occultation probability can be thousands of kilometers for KBOs. This increases the effort of traveling and transportation. Observing stations can be informed precisely, and mobile stations can be sent out either by car or by plane to even distant parts of the world, if the time gap between prediction and the event is not too small.

After the observations, lightcurves must be extracted from the images. If the detection of atmospheres is the goal, the stellar light has to be followed precisely during emersion and immersion. Often the full images must be sent to trained people who do photometry on these images. From these lightcurves just as described in \ref{occul}, the diameter and shape of the KBO is determined. From the post-event results, an update of the astrometry can be evaluated and again fed into the pipeline (the dotted red line in Fig.~\ref{fig:1-8.2}) to improve the orbit of the KBO for further occultations \cite{Assafin10}. When lightcurves are available, one can look for traces of an atmosphere of the sampled KBO, or for atmospheric changes in known atmospheres, as has been done successfully for Triton \cite{Elliot98} and Pluto \cite{Sicardy03}.

\subsubsection{Technology of detection}

The techniques used for occultation by Main Belt asteroids can in principle be used for KBOs. However, the objects are fainter. This can be compensated by larger exposure times. Typical velocities of KBOs with respect to the Earth are usually in the range of 20 to 25 km/sec, but with some events near quadrature as slow as a few km/sec. Because most of the objects have unknown shapes, a determination with an accuracy of only +/- 100 km can provide valuable information of their albedo and density as well. Exposure times of up to 5 seconds can be used. This allows the use of standard astronomical cameras of the amateur market, built for other purposes, such as deep space object recording (see Sec. \ref{ccd}). Timing has to be as good as for main belt objects, but even with much lower time-resolution in the second range, valuable research can be done.

It has been found that, under good atmospheric conditions (sky background less than 21 mag per arcsec$^2$, scintillation less than 1.5 arcsec) with an instrument of about 0.4 m diameter a focal ratio of 1:4 and the use of a camera with a sensitive CCD chip (for example Sony Exview Chips, commercially available in many cameras), an exposure time of 1 second is long enough to clearly record a star with 17 mag (R Band), if no filter is used. The interval time between two images has to be small or near zero, 0.2 seconds may be acceptable. Otherwise it reduces the effective quantum efficiency of the camera. Even more, for evaluation of the structure of an atmosphere, it is necessary to record the full light curve with time. Short spikes of light can occur in the lightcurve, which may be lost during dead time of the camera. If higher acquisition speed is necessary, EMCCD cameras with their low read-out noise are very valuable (see Sec. \ref{ccd}), but with a high price tag ($>$ 6000 Euro).

For atmospheric detection, the resolution of the analog-digital converters of the cameras should be more than 8 bits and a high linearity is required. To compensate for changes in the Earth's atmospheric transparency during the occultation, reference stars near the object should be recorded in the field of the CCD chip on each frame to be used as an internal intensity references.

If a detailed analysis of the atmosphere is the goal, it is necessary to determine the intensity of the KBO and the occulted star independently. Because of possible rotational albedo variation of the KBO, comparison photometry has to be done before and/or after the event, as soon as it is possible to separate the star from the occulting body. This can take hours or more, if the relative movement of the object versus the star is small and the focal length of the used telescope is small too. If this cannot be done, extra parameters have to be defined in a fitting algorithm for atmospheric determination.

\subsubsection{Post event analysis and atmospheric determination}

Images and video recordings are analyzed using various software packages, such as IRIS, MIDAS, IRAF, or IDL programs. The final curve of light intensity versus UTC time has to be normalized with respect to the occulted star. Full-light intensity of the star is set to unity, and totally occulted stellar light is set to zero. By using reference stars in the images, a change of light transmission in the Earth's atmosphere (aerosols due to clouds etc.) during the occultation can be removed.

To detect a possible atmosphere from the lightcurve, the star's disappearance and reappearance times are not enough. The precise light intensities relative to full stellar intensity for each data point is also needed. Extracting atmospheric details from occultation lightcurves has been done since about 1953, when Baum and Code recorded an occultation by Jupiter \cite{Baum53}. Since then, atmospheric details by occultation astronomy have been determined for Mars, Jupiter, Saturn, and Titan \cite{Sicardy06a}, Uranus, Neptune and Triton, and Pluto \cite{Sicardy03}, often in PRO-AM co-operations. For Charon \cite{Sicardy06b} and Titania \cite{Widemann2009}, where no atmosphere could be found, the minimum detection level was less than 10$^{-2}$ Pa.

Density and temperature profiles, as well as details such as wind speeds in planetary atmospheres, can be evaluated by mathematical methods such as inversion techniques or ray tracing \cite{Sicardy06a,Vapillon73}. The light rays are bent by refraction through the atmosphere, which distributes the light on a larger area in the plane of the observer the deeper the light gets into the atmosphere \cite{Vapillon73}. If the atmosphere is dense enough, the stellar light may not disappear at all. In case of a central occultation, i.e., an occultation where the center of the occulting body passes exactly in front of the star, a so called "central flash" can be observed, a short increase of light at midtime of the occultation \cite{Sicardy06a}. If many observers are placed close to each other near the central occultation track, a two-dimensional intensity profile of the central flash or caustics can be recorded, as it has been done already for Titan \cite{Sicardy06a}, or modeled for Pluto \cite{Lellouch09} and other objects. Data from a central flash are especially valuable to determine the oblateness of the atmosphere caused by strong winds. From the peak height of the central flash intensity, estimating the absorptions due to aerosols or dust is possible \cite{Sicardy06a}. For demonstration, Fig.~\ref{fig:2-8.2} shows examples for an occultation by an object without (left) and with (right) an atmosphere.

A special approach to distant objects are the observations of serendipitous occultations. In this case, one or more stars are observed continuously with high photometric and time resolution to detect occultations just by chance \cite{Roques08}. Ground-based projects dedicated to this approach exist, like the MIOSOTYS instrument attached to a 1.93-m telescope at the Observatoire de Haute-Provence (OHP) and to the 1.5-m telescope in Calar Alto. After extensive observations (3 years, 3,400 hours observation), observers recorded 6 candidate events. The possibility exists that this technique could be also spread out into the amateur community.

\subsubsection{The Pluto System}

In 1985 and 1988, the confirmation of an atmosphere around Pluto was made by occultation astronomy. Fourteen years later, in 2002, two international campaigns with professional and amateur astronomers found the expansion of Pluto's atmosphere compared with 1988 \cite{Sicardy03} by a factor of more than 2. Since then, many more campaigns have been organized, which confirmed the expansion and showed at the same time that no further considerable increase of the surface pressure on Pluto took place. Corrections with respect to the JPL ephemeris could be calculated from the observed occultations, which improved the precision of predictions considerably. Pluto has been moving across the Milky Way, giving the chance to observe in principle several occultations per year with stellar magnitudes brighter than 16. In 2005 an occultation by Charon was observed from more than one station. The diameter of Charon was determined \cite{Sicardy06b}. Occultations by Pluto and Charon of the same star have been observed allowing to define Charon's orbital parameters \cite{Sicardy11b}.

\subsubsection{Quaoar, Eris, Makemake and 2003AZ84}

After many misses, where campaigns did not succeed either because of astrometry problems and/or poor weather conditions, diameters for Eris, Makemake, 2003 AZ84 and Quaoar were finally determined in 2010 from observations in southern America. For Eris, a radius of 1163~$\pm$~6 km (spherical solution) has been determined by a campaign for the occultation on the 6th of November 2010 \cite{Sicardy11a}. It is to date the farthest object ever probed by an occultation, at $\sim$95.7 AU from Earth. For Makemake, an occultation campaign on the 23rd of April, 2010 gave an elliptical solution for its shape with axes of 1430~$\pm$~9 km and 1502~$\pm$~45km (each 1 sigma limit) \cite{Ortiz12}. For 2003AZ84, only one positive and one negative occultation chord have been determined from a campaign on the 8th of January, 2011. It gives a lower limit for its diameter of 573~$\pm$~21km \cite{Braga-Ribas11}. The stellar magnitude in the R-band was around 18mag. (50000) Quaoar was observed occulting a magnitude 16 star (R-band) on 4th of May, 2011, using 16 stations distributed in Argentine, Brazil, Chile and Uruguay. The longest chord had an equivalent length of 1170~km \cite{Braga-Ribas11}.

\section{Exoplanets: research and characterization}
\label{Exo}

Extrasolar planets are an important field of planetary science since they provide a comparative view of our Solar System with respect to other planetary systems. Even if the level of detail reached cannot be compared to planets of the Solar System, studies of exoplanets permit the exploration of planetary diversity in terms of planet mass, radius, density, orbital period, eccentricity, obliquity, host star physical parameters, and planetary atmosphere properties and composition. The discovery and characterization of extrasolar planets is also providing elements to understand planet occurrence and to constrain planetary formation, migration and evolution models. Among the techniques used by professionals to discover new exoplanets, we focus here on two of them that explore two different regions of the galaxy: the transit and microlensing methods. These two techniques are performed using a very wide range of instruments, including small-aperture photometric telescopes and amateur telescopes.

\subsection{Transiting exoplanets}
\label{hunt}

A primary transit occurs when an extrasolar planet passes in front of its host star as seen from the observer. It occurs once per orbital period of the planet with a typical duration of several hours, and the decrease of luminosity of the star is typically of $\sim$1\% for a Jupiter mass planet transiting a Sun-like star. Two space missions, \textit{CoRoT} and \textit{Kepler}, and several ground-based observatories, like HATNet, MEARTH, OGLE, SuperWASP and others, have been dedicated to finding transiting planets. The majority of transiting exoplanets discovered so far are giant planets orbiting at short orbital period (a few days). Thanks to space-based photometry from CoRoT and Kepler, we are discovering more and more transiting exoplanets with a lower mass and/or a longer-orbital period. These low-mass or long-orbital period planets seem to be more common in multiple systems \cite{Lissauer2011}. These planets in multiple systems exhibit variation in their transit time due to gravitational perturbation from the other planets in the system. These transit time variations (TTV) have a typical amplitude of a few minutes \cite{Ballard11}. In some particular configurations, i.e., when the planets are close to the orbital resonance, the TTV amplitude can reach the level of a few hours \cite{Fabrycky12}.

Transits of giant planets with depth at the level of about 1\% ($\sim$10 mmag) are within the reach of amateur photometry. We discuss here three different cases where their contributions can be significant.

\subsubsection{Maintaining ephemeris of known transiting exoplanets}

Complementary study and observations of transiting exoplanets (TEPs), such as Rossiter-McLaughlin observations or transit spectrophotometry, require precise ephemerides on transit epochs. For TEPs that have not been observed for a long time, the uncertainty on the transit epoch can be large, depending on the quality and timescale of the photometry used for the planet discovery. In some cases, the uncertainty on the transit epoch is even larger than the transit duration. This strongly limits complementary observations with professional telescopes. To avoid that, transits should be observed quite frequently, at least once per year, in order to refine the planet's ephemeris. With the increasing number of known TEPs, only amateur astronomers may be able to perform such follow-up observations of all the TEP within reach of their instruments.

The ephemeris of each known TEPs is provided and kept up-to-date on the Czech Republic's Exoplanet Transit Database\footnote{ETD: \url{http://var2.astro.cz/ETD/index.php}} and on the US' Amateur eXoplanet Archive\footnote{AXA: \url{http://brucegary.net/AXA/x.htm}}. The former database also provides a finding chart for the host star of the exoplanet and tools to fit the data.

\subsubsection{Searching for transit time variations}

TTV can be used to discover new planets in known planetary systems \cite{Holman05}. According to \textit{Kepler} statistics, most planetary systems contain Neptune-- or Earth--sized planets for which transits are out of reach of amateur telescopes. Some studies have found TTV on a few short-period giant TEPs (e.g., \cite{Macie11}) but they were later unconfirmed \cite{Barros13}. Thanks to \textit{Kepler} data, we now know that there is a lack of TTV for short-period giant TEPs, in contrast with long-period ones \cite{Steffen12}. Looking for TTV on short-period giant TEP with an amateur telescope will not be fruitful. Nevertheless, a few long-period giant TEP in multiple system present significant TTV, thanks to \textit{Kepler} long timescale data. Those systems are Kepler-30 \cite{Fabrycky12}, Kepler-46 \cite{Nesvorny12}, and KOI-1474 \cite{Dawson12}. They present TTV with an amplitude up to one day. After the end of the \textit{Kepler} mission, expected for 2016, it might be interesting to follow these systems up with a network of amateur astronomers. This could permit the systems to be better characterized: only upper-limits on planets' masses have been constrained so far. However, a serious difficulty is combining non-uniform datasets for TTV studies. The correct approach is to have a homogenous dataset, observed in the same band, by the same telescopes over a long period of time. Such projects would require close coordinations between amateurs and professionals to be fruitful.

\subsubsection{Hunting for new transiting planets, photometric follow-up of non-transiting planets}

Among all the extrasolar planets discovered using the radial velocity technique, only a small fraction is known today to be transiting their host star. Most of the planets with long periods have not been searched yet for transits. This requires photometric follow-up observations around the expected transit epoch. For giant planets, this photometric follow-up might be done using amateur telescopes. 

This PRO-AM collaboration has already led to publications \cite{Barbieri07,Moutou11,Anderson11,Hebrard12}. A superb example has been the detection of the primary transit of a planet on a very excentric orbit, HD80606b seen by four different teams simultaneously. Ironically the best photometric data set was obtained by a 30-cm telescope in the suburbs of London \cite{Fossey09}, rather than with the 120-cm telescope at OHP \cite{Moutou09} and a 0.6-m telescope \cite{Garcia-Melendo09b}. A network of amateur astronomers spread in longitude can follow up planets discovered by radial velocity with a high transit probability \cite{Szabo10}. Such a network will unambiguously discover new transits among the radial velocity-discovered planets.

Ephemerides of radial velocity planets are listed on the TransitSearch website\footnote{TransitSearch: \url{http://www.transitsearch.org/}} and international campaigns are reported on the AXA web page\footnote{AXA: \url{http://brucegary.net/AXA/TransitSearch/TransitSearchLC.htm}}.

\subsubsection{Observation requirements}

There is no limitation in the instrumental setup to observe a transit,since some amateurs have already caught a transit light-curve using a DSLR camera mounted on a telephoto lens. Moreover, most of transiting giant planets discovered to date have been detected using 10-cm class telescopes. To achieve enough photometric precision for PRO-AM collaborations, telescopes with aperture greater than 20--25 cm are required (see Fig. \ref{HATP8}). The use of a monochrome CCD camera without anti-blooming system with pixel size smaller than 1'' (depending on the average seeing of the observatory) would be a nominal choice for these projects.

Transits of planets have a typical duration of several hours. For a good characterization of the transit shape and epoch, observations must also include at least one hour of out-of-transit data, obtained just before and after the transit. These out-of-transit observations are needed to rigorously normalize the out-of-transit flux. Observing a full transit thus requires almost a whole night. Transits also need good time sampling. For this reason, we recommend using exposure times of about one minute, up to two or three minutes, as constant as possible during the whole night.

To perform high-accuracy photometric observations at the level of a few milli-magnitudes, relative photometry is needed. A field of view of several tens of arc-minutes will secure several quiet reference stars. To limit differential atmospheric refraction between the target and the references stars, wide-band red filters must be used. Most amateur cameras are not so efficient in the near infrared, we thus recommend the standardized sloan r' filters for such observations. To improve photometric accuracy, observations should be performed slightly out of focus. Stars' PSF (Point Spread Function) must be spread onto about 10 to 15 pixels to average the CCD read-out noise. Some precautions should be taken to avoid blending stars. Photometric precision can be improved with a very good guiding of the stars on the same pixels during the whole night. This limits errors made during the flat-field correction due to inter-pixel sensitivity differences.

Data reduction should take into account the flat-field, dark and bias corrections. We recommend the use of the Muniwin software\footnote{http://c-munipack.sourceforge.net/} to reduce the raw data and to perform the aperture photometry. We also recommend the book ``Exoplanet Observing for Amateur: Second Edition'' by B. Gary\footnote{Available in free download at: http://brucegary.net/book\_EOA/x.htm}.

\subsection{Microlensing}
\label{micro}

Gravitational microlensing is based on Einstein's theory of general relativity: a massive object (the lens) will bend the light of a bright background object (the source). This can generate multiple distorted, magnified, and brightened images of the background source. When the lens is a star, these images are unresolved and the brightness of the background star is amplified. The source's apparent brightness varies as the alignment changes due to relative proper motion of the source with respect to the lens. This lightcurve is monitored to detect and study the event. Thus, a microlensing event is a transient phenomenon with a typical time scale $t_E = 20 \sqrt {(M /M_\odot)}$ days, where $M$ and $M_\odot$ are the masses of the lens and the source, respectively. If the lens is not a single star (binary star or star with a planet), the companion will distort the gravitational lens creating regions of enhanced magnification (caustics), which introduce anomalies in the lightcurve, lasting for about a day for a Jupiter-mass planet and less than two hours for an Earth mass planet. Microlensing is a rare phenomenon (a probability of $\sim 10^{-6}$ at a given time for a star of the galactic Bulge to be magnified).  Therefore a two-step approach has been adopted since the 1990s. First, wide-field imagers are monitoring a very large number of stars in order to detect real-time ongoing microlensing events and send out public alerts (OGLE and MOA collaborations). The second step is to have a network of telescopes (mainly PLANET, $\mu$FUN, RoboNET, Mindstep) doing a follow up of a selected sample of the events with the highest sensitivity to exoplanets. From a network of 4 telescopes in 2002, there are now up to 50 telescopes available on alert, ranging from robotic 2-m telescopes to 30-cm amateur telescopes in a backyard. In some cases, more than 20 telescopes have been collecting scientifically useful data on a given microlensing event \cite{Batista09}. This includes cold super Earths \cite{Beaulieu06,Muraki11,Bennett08,Kubas12}, cold Neptunes \cite{Sumi10}, Saturns \cite{Bachelet12,Miyake12}, Saturns in the Bulge \cite{Janczak10}, and multiple planet systems \cite{Gaudi08}. Brown dwarfs orbiting M dwarfs \cite{Bachelet12} and 4 massive Jupiters orbiting M dwarfs \cite{Street13} that are not predicted by the core accretion theory \cite{Alibert05} have also been detected. On the other hand, gravitational instability can form large planets around M dwarfs \cite{Boss06}, but typically  farther out. Planets formed by such mechanism would have to migrate significantly. Although the number of microlensing planets is relatively modest compared with that discovered by the radial velocity method and by Kepler, this technique probes a part of the parameter space (host separation vs. planet mass) which is not accessible currently to other methods.

Of the 19 planets detected by microlensing and published today, amateur telescopes had a significant scientific contribution to a number of them. In 2005, New Zealand amateurs G. Christie (Auckland Observatory) and J. McCormick (Farm Cove) reacted to the public microlensing alert  on a high magnification event with potential sensitivity to extrasolar planets. They observed the planetary anomaly and contributed significantly to the discovery of a massive Jupiter orbiting an M dwarf \cite{Udalski05}. Note that they monitored continuously the fraction of the lightcurve when the magnitude was in the range $I=15-16$. These observations were done by amateurs even though the alerts aimed primarly professional astronomers. After this, amateurs joined the community of microlensers answering the alerts and acquiring scientific useful data. 

The system OGLE 2008-BLG-109 \cite{Gaudi08} is a very complex microlensing event with five subsequent short-lived anomalies. It has a scale 1/2 of our Solar System, composed of a $0.5 M_\odot$ star with two gaseous planets in orbit. Amateurs contributed to key data in different parts of the lightcurves, which allowed the detection of a super Earth of $\sim$10 $M_\oplus$ orbiting a $\sim$0.8 $M_\odot$ star \cite{Muraki11}. The microlensing source being a Bulge giant, it was a relatively easy target and B. Monard (Bronberg Observatory) started observations 5 hours after the anomaly had been detected, followed by 5 professional observatories. 

There is one case where most of the data showing the presence of a Saturn-mass planet orbiting a star in the Bulge of the galaxy \cite{Janczak10} have been collected mainly by one amateur, Monard at the Bromberg Observatory. Based on these amateur data, a target of opportunity was triggered at the VLT to obtain complementary adaptative observations with NACO.

 \subsubsection{Amateur contribution to microlensing : how does it work in practice}
 \label{lenswork}

The  $\mu FUN$ collaboration has been advocating strongly for amateur observations in microlensing: doing the coordination of the network and issuing the microlensing alerts. The observing strategy has been summarized by S. Gaudi as being ``Wait, wait, wait, ..., panic!''. Professional telescopes (OGLE, MOA, CTIO, PLANET) are monitoring a large number of microlensing events to detect which ones will become high magnification events. These very rare events, where the flux of the source is amplified by a factor of 100, have two advantages. First, the source star (usually faint I=18--22) could become very bright thanks to the lensing effect. Secondly, the stronger the amplification, the more sensitive to extrasolar planets the event is. As a consequence, typically once a week, an alert for potential high magnification alert is issued on a number of targets, with a request to ``observe continuously until further notice'' with magnitude estimates and potential behavior for the coming hours. Extra care is taken into making sure not to ask telescopes to follow targets that are too faint for them. Generally, alerts are sent for magnification over 100, with a magnitude brighter than I=16, the ideal scenario being when the magnitude is brighter than I=14. The telescopes observe continuously until the alert is called off by the coordinator, based on data collected with the professional telescopes and the real time modeling. Usually, data after the alert are required to be able to calibrate the lightcurve. 

The amateur telescopes answering the alert are typically in the range 30--50 cm, equipped with CCD cameras with a well sampled PSF (FWHM of 2--3 pixels minimum), and guiding systems allowing them to take exposures up to a few minutes. It is also important to have a GPS clock in order to record exposure times precise to the second. An R or I filter is needed since the targets in the galactic bulge are red stars and obscured by extinction. Some of the smallest telescopes observe in white light, which turns out  to complicate significantly the data analysis process.

Standard procedures (bias and flatfielding) have to be performed to calibrate the images. Once a series of images have been acquired, the amateur immediately informs the coordinator about the timing of the observations. Usually, a day or two later, the bias subtracted and flat fielded images are sent to the coordinator, with information about conditions of the observing run. Some amateurs have their own photometric packages and are providing a first set of reduced photometry. However, microlensing fields being very crowded, it requires sophisticated pipelines for the final version of the analysis. All the amateurs who contributed critical data are included in the publication, in some cases among the first authors. The microlensing community is very grateful to the contribution of the amateurs and always considers it to be fair to add them as co-authors. As a consequence, there is now a large group of amateurs following the microlensing alerts and trying to contribute. The strongest nodes are currently in New Zealand and South Africa. 

\subsubsection{Amateur contribution to microlensing : 2013 and beyond}
\label{lensfut}

In 2012, up to 50 telescopes were answering microlensing alerts in order to provide complete coverage of high magnification events with high sensitivity to extrasolar planets. Twenty-two planets have been discovered but, contrary to earlier years, the major contribution has been the wide field imagers on professional telescopes. Nevertheless, in this new era where a world wide network of 1.3--1.8 m telescopes with cameras of 0.5--5 degrees$^2$ exists, there is still a niche for monitoring by amateur astronomers. First, a wide coverage in longitude with a fleet of telescopes might still be useful to cover critical sections of the exoplanet lightcurves. Secondly, it is also possible to detect time differences in a handful of very high magnification events between different observing sites \cite{Gould09,Gould13}. Such measurements provide the means to measure the mass of the microlensing lens very precisely. We could envision that in the era 2013--2018, the monitoring of microlensing events with amateur telescopes can still bring interesting results. With microlensing, amateurs are real partners to professional astronomers in discovering new extrasolar planets.

\section{Conclusions}
\label{cls}

In this paper, we have presented a review of the different PRO-AM collaborations that already exist and from which many papers have been published in the field of planetary science. We have also discussed the instruments, detectors, softwares and methodologies typically used by amateur astronomers to collect the scientific data in the different domains of interest. The equipment and the personnel needed to take and reduce the observations are perfectly within the possibilities of most amateur astronomers and colleges. 

The topics addressed in this review could also motivate some experiments under the guidance of teachers for science education, typically at the high school or college levels. Many schools already own good astronomical equipment and it would be easy for science teachers to propose monitoring programs of the Moon and the giant planets in the framework of networks coordinated by professionals or experienced amateurs. The measurement of asteroid lightcurves, comet research, or the characterization of transiting exoplanets are also perfectly within the reach of school programs but would require the guidance of professionals or amateurs experienced in the field. Unfortunately, facilities providing access to professional-level telescopes and supplying free resources for science education already exist\footnote{http://www.faulkes-telescope.com/} but remain somewhat difficult to access due to their limited number.

\begin{acknowledgement}
O.M. acknowledges support from CNES. RH, ASL and SPH were supported by the Spanish MICIIN project AYA2009-10701 and AYA2012-36666 with FEDER support, Grupos Gobierno Vasco IT-464-07 and UPV/EHU UFI11/55. A.S. acknowledges the support by the European Research Council/European Community under the FP7 through Starting Grant agreement number 239953. D.B. and S.B are supported by the Partenariat Hubert Curien/French--Moroccon volubis program PHC 24675QJ. G.F. was supported by the Austrian Science Fund FWF, project P24325-N16. G.O. was supported by awards from NASA to JPL/Caltech.

\end{acknowledgement}

\clearpage

\begin{landscape}
\begin{center}
\tablecaption{Summary of science goals, required instruments, detectors and filters for the different targets.}\label{tabsum}
\tablehead{%
\hline
\bf Target & \bf Science & \bf Instrument & \bf Detector & \bf Filter & \bf Timing requirements & \bf Comments & \bf Section \\
&&&&&&& \\
\hline}
\tabletail{\hline}
\begin{supertabular}{|p{1.5cm}|p{3cm}|p{2.5cm}|p{2.5cm}|p{2cm}|p{2.5cm}|p{2.5cm}|p{1.5cm}|}
%
&&&&&&& \\
Venus & Image dayside UV markings, measure superrotation & Open-tube telescope (Newton, Cassegrain, Dall-Kirkham) or apochromatic refractor (\diameter $\ge 15$ cm) & CCD black and white video camera with small pixels (5--8 $\mu$m) and high frame rates (15 fps up to 60--100 fps) & Near UV narrow band (350--360 nm, FWHM $<100$ nm) or Wratten 47 and IR-blocking & 0.1 min time stamping accuracy &Treatment with Registax or Autostakkert & ~~3.1.3 \\
&&&&&&& \\
& NIR dayside imaging (lower atmospheric level) & & & NIR long pass with transmission cut-on at $\sim 800$ nm & & & ~~3.1.3 \\
&&&&&&& \\
& IR thermal emission from surface \& CO$_2$ thermal emission & & & IR, centered around 1000 nm (Sch\"ott RG 1000) & & & ~~3.1.3 \\
&&&&&&& \\
& Visible dayside lowcontrast details & & & RGB & & & ~~3.1.3 \\
&&&&&&& \\
\hline
&&&&&&& \\
Mars & Weather and clouds, dust storms, unusual high clouds on the limb, long-term evolution of polar caps & & & R or I for surface features, B for clouds and fog, R for dust clouds and storms & 0.1 min time stamping accuracy & 10 month observing period around opposition & ~~3.2\\
&&&&&&& \\
\hline
&&&&&&& \\
Meteor showers & Observation and (continuous) monitoring & Naked eye or digital camera  & High-sensitivity video camera & & 0.1 min time stamping accuracy (naked eye) / 0.02 sec time interval accuracy (camera) & Knowledge of the sky required & ~~4.1.3 \\
&&&&&&& \\
& &  & & & & Use of detection software (UFOCapture, MeteorScan, ASGARD) & ~~4.1.3 \\
&&&&&&& \\
Giant planet impacts & Continuous monitoring \& flash light curves & \diameter $>12$ cm & CCD black and white video camera with small pixels (5--8 $\mu$m) and high frame rates (15 fps up to 60--100 fps) & RGB & 0.02 sec time stamping accuracy & Essential for short flash detection. Use of detection software & ~~4.3.1 \\
&&&&&&& \\
& Observation of dark debris & Any telescope &  & Narrow band methane at 890 nm & & & ~~4.3.1 \\
&&&&&&& \\
\hline
&&&&&&& \\
Moon impacts & Flash (impact) detection & Short focal telescope (0.7 -- 1 m), Newton \diameter $15--25$ cm, $F/D=4$, or Schmidt-Cassegrain \diameter 20 cm with $F/6.3 $focal reducer or C14 with Hyperstar optical system ($F/1.9$) &  & None & 0.02 sec time stamping accuracy & Observe the non-illuminated part of the lunar disk between the last and first quarter. Short events (generally less than 1s). Detection software : LunarScan or UFOCapture & ~~4.4.3 \\
&&&&&&& \\
Near-Earth asteroids (NEAs) & Discovery \& astrometry & \diameter $>30$ cm & CCD $60'\times 60'$ field or more & BVRI and broad band V+R & 1 sec time stamping accuracy & $20-240$ sec exposure & ~~5.1.4 \\
&&&&&&& \\
Asteroids & Photometry / lightcurves & \diameter $>20$ cm & CCD (no anti-blooming) & Broadband R or V  & 1 sec time stamping accuracy & Better than 0.05 to 0.1 mag precision (1 min exposure for a V = 12 target) & ~~5.2.2 \\
&&&&&&& \\
& Stellar occultations (size and shape determination) & \diameter $>20$ cm, $F/3.3$ (Schmidt-Cassegrain + focal reducer) & Analog video camera (10 fps) & & 0.01 sec time stamping accuracy & 0.04 sec integration for V $\sim 12$ & ~~5.3.2 \\
&&&&&&& \\
& & Compact camera lens & Digital camera (CCD) & & & 0.32 sec integration for V $\sim 11$ &~~5.3.2 \\
&&&&&&& \\
\hline
&&&&&&& \\
Giant planets (in general) 	& Atmospheric dynamics \& details on Jupiter's satellites\ 	& Schmidt-Cassegrain, \diameter $>15$ cm (30 cm for satellite details)	 & CCD black and white video camera with small pixels (5--8 $\mu$m) and high frame rates (15 fps up to 60--100 fps) 	& RGB (Wratten 21 or 80A) 	& 0.1 min time stamping accuracy & Use of Barlow lens, perfect collimation and thermalization required, treatment with Registax or Autostakkert 	& ~~6.1\\
						& Trace higher atmospheric levels 				& \diameter $>25$ cm (Jupiter) or 30 cm (Saturn) 							& 											& Narrow band UV or Near IR methane at 890 nm & & & ~~6.1 \\
&&&&&&& \\
Jupiter & Photometry (albedo \& color changes)					 		& \diameter $>15$ cm 	& 				& UV (Johnson U) or Near IR methane at 890 nm 		& 											& & ~~6.4.3 \\
&&&&&&& \\
& 						& \diameter $>35$ cm 						& 					& Narrow band methane at 725 nm and 750 nm continuum 				& 											& & ~~6.4.3 \\
&&&&&&& \\
Saturn 					& Monitor storm activity and Great White Spot; Observation of spokes in the rings and of the north polar hexagon 		& \diameter $>15$ cm 	&	 														& 					& 						& & ~~6.5 \\
&&&&&&& \\
Uranus, Neptune & Imaging of atmospheric bands and discrete features 			& \diameter $> 30-50$ cm or more & CCD (``lucky'' imaging not possible with typical amateur telescopes)		& IR cut-off at 680 nm, R+IR ($>600$ nm), IR-pass from 685 nm & & 35--45 min acquisition time 		& 6.1 \& 6.6 \\
&&&&&&& \\
& Stellar occultations in Uranus system & & & & & & ~~6.6.1\\
&&&&&&& \\
& Photometry & & & & & & 6.6.1 \& 6.6.2\\
&&&&&&& \\
Uranus, Neptune, Titan & Spectroscopy & & Spectrometer \& CCD & & & & 6.2 \& 6.6.1 \\
&&&&&&& \\
\hline
&&&&&&& \\
Comets 		& Search for comets hidden in the asteroid population 			& \diameter $>40$ cm 				& CCD & R filter 			& 1 sec time stamping accuracy & SNR $>$ 10, with no trailing effect on Individual images. Typically, series of 30 images of 30 to 120s each		& ~~5.4.2			 																		\\
&&&&&&& \\
			& Search 				& \diameter $>30$ cm ($F/D\ge 3$) 		& CCD (large format) 				& No filter 					& 1 sec time stamping accuracy & Visual: binoculars to search in the twilight 	& ~~7.1 																						\\
			& Astrometry 			& \diameter $>20$ cm ($F>1$ m) 		& CCD (no anti blooming) 			& No filter of R (depending on the signal-to-noise) 						& 1 sec time stamping accuracy & Short exposure time (2 min max.), R filter best & ~~7.2 													\\
&&&&&&& \\
&&&&&&& \\			& Photometry 											& Telescope: to be adjusted according to the brightness of the comet 	& CCD (no anti blooming) 	& IRVB 		 & 1 min or less, depending on the exposure time & Visual estimation: eyes, binoculars. Telescope adjusted to the brightness of the comet, possibly Swan filter 	& ~~7.3 \\
&&&&&&& \\
Kuiper-Belt objects (KBOs) & Astrometry \& photometry & \diameter $>40$ cm & CCD & R broadband, sharp cut-off above 750 nm & 1 sec time stamping accuracy & Exposure time limited to avoid trailing & ~~8.1.2 \\
&&&&&&& \\
& Stellar occultations & \diameter $>40$ cm & Sensitive CCD (Sony Exview chip) or EMCCD & & 0.1 sec or better time stamping accuracy & Exposure time up to 5 sec & ~~8.2.2 \\
&&&&&&& \\
\hline
&&&&&&& \\
Exoplanets & Transit observations & \diameter $>20$ to 30 cm (minimum configuration) & Monochrome CCD camera, no anti-blooming, pixel size $<1''$ & Sloan r' & 1 sec time stamping accuracy & Relative photometry needed to achieve milli-magnitude accuracy. Exposure 30 sec to 2--3 min. Defocusing required & ~~9.1.4 \\
&&&&&&& \\
& Microlensing & \diameter 30 to 50 cm & CCD camera, well-sampled PSF & R or I &  1 sec time stamping accuracy & & ~~ \ref{lenswork} \\
&&&&&&& \\
\end{supertabular}
\end{center}
\end{landscape}

\clearpage

\begin{table}[h]
\caption{List of stars that can be used to calibrate reflectance spectra.
\label{tab:stars}
}
\centering
\begin{tabular}{|p{1.5cm}|p{1cm}|p{1cm}|p{1cm}|p{1cm}|p{1cm}|p{1cm}|}
\hline
HD		&	RA		&	DEC		&	Sp		&	V		& 	B--V		& 	V--R		\\
\hline
1835		&	00 22.9	&	-12 13	&	G2V		&	6.402	&	0.660	&	0.537	\\
4915		&	00 51.2	& 	-05 02	&	G0V		&	6.982	&	0.666	&	0.543	\\
8262		&	01 22.3	& 	+18 41	&	G3V		&	6.973	&	0.630	&	0.513	\\
10307	&	01 41.8	& 	+42 37	&	G1.5V	&	4.965	&	0.623	&	0.499	\\
20630	&	03 19.3	& 	+03 22	&	G5V		&	4.83		&	0.68		&			\\
28099	&	04 26.7	& 	+16 45	&	G2V		&	8.09		&	0.657	&			\\
34411	&	05 19.1	& 	+40 06	&	G2V		&	4.705	&	0.622	&	0.499	\\
44594	&	06 20.1	& 	-48 44	&	G3V		&	6.61		&	0.66		&			\\
76151	&	08 54.3	& 	-05 26	&	G2V		&	6.01		&	0.68		&			\\
78418	&	09 08.8	& 	+26 38	&	G5IV		&	5.98		&	0.65		&			\\
86728	&	10 01.0	& 	+31 55	&	G3V		&	5.40		&	0.65		&			\\
89010	&	10 16.5	& 	+23 30	&	G1.5V	&	5.968	&	0.668	&	0.529	\\
95128	&	10 59.5	& 	+40 26	&	G0V		&	5.037	&	0.622	&	0.505	\\
126053	&	14 23.3	& 	+01 14	&	G1V		&	6.266	&	0.644	&	0.527	\\
133002	&	14 50.3	& 	+82 31	&	F9V		&	5.643	&	0.682	&	0.555	\\
141004	&	15 46.4	& 	+07 21	&	G0V		&	4.419	&	0.611	&	0.494	\\
144585	&	16 07.1	& 	-14 04	&	G5V		&	6.32		&	0.66		&			\\
146233	&	16 15.6	& 	-08 22	&	G2V		&	5.499	&	0.650	&	0.524	\\
159222	&	17 32.0	& 	+34 16	&	G5V		&	6.537	&	0.646	&	0.510	\\
177082	&	19 02.6	& 	+14 34	&	G2V		&	6.895	&	0.641	&	0.518	\\
181655	&	19 19.6	& 	+37 20	&	G8V		&	6.31		&	0.68		&			\\
186408	&	19 41.8	& 	+50 32	&	G1.5V	&	5.986	&	0.659	&	0.521	\\
186427	&	19 41.9	& 	+50 31	&	G2.5V	&	6.244	&	0.671	&	0.531	\\
187237	&	19 48.0	& 	+27 52	&	GIII		&	6.896	&	0.654	&	0.512	\\
191854	&	20 10.2	& 	+43 56	&	G5V		&	7.45		&	0.56		&			\\
193664	&	20 17.5	& 	+66 51	&	G3V		&	5.932	&	0.601	&	0.497	\\
197076 	&	20 40.8	& 	+19 56	&	G5V		&	6.444	&	0.628	&	0.505	\\
217014	&	22 57.5	& 	+20 46	&	G2.IV	&	5.459	&	0.676	&	0.517	\\
222143	&	23 38.0	& 	+46 12	&	G5		&	6.591	&	0.652	&	0.522	\\
\hline
\end{tabular}
\end{table}

\clearpage

\begin{table}[h]
\caption{Difference between absolute magnitude $H$ and apparent magnitude $M$ at opposition for various heliocentric distances $r$.
\label{tab:delta_m}
}
\centering
{\begin{tabular}{|@{\ \ }c@{\ \ }|@{\ \ }c@{\ \ }|@{\ \ }c@{\ \ }|@{\ \ }c@{\ \ }|@{\ \ }c@{\ \ }|@{\ \ }c@{\ \ }|@{\ \ }c@{\ \ }|@{\ \ }c@{\ \ }|@{\ \ }c@{\ \ }|@{\ \ }c@{\ \ }|@{\ \ }c@{\ \ }|@{\ \ }c@{\ \ }|@{\ \ }c@{\ \ }|@{\ \ }c@{\ \ }|}
\hline
\ & \ & \ & \ & \ & \ & \ & \ & \ & \ & \ & \ & \ & \ \\ [-1ex]
$R$ (AU) & 5 & 10 & 15 & 20 & 25 & 30 & 35 & 40 & 45 & 50 & 60 & 80 & 100 \\ [1ex]
\hline
\ & \ & \ & \ & \ & \ & \ & \ & \ & \ & \ & \ & \ & \ \\ [-1ex]
$M-H$ & 6.5 & 9.8 & 11.6 & 12.9 & 13.9 & 14.7 & 15.4 & 16.0 & 16.5 & 16.9 &
17.7 & 19.0 & 20.0 \\ [1ex]
\hline
\end{tabular}
}
\end{table}

\clearpage

\clearpage

\clearpage

\clearpage

\clearpage

\clearpage

\clearpage

\clearpage

\clearpage

\clearpage

\clearpage

\clearpage

\clearpage

\clearpage

\clearpage

\clearpage

\clearpage

\clearpage

\clearpage

\clearpage

\begin{figure}
\begin{center}
\includegraphics[scale=.27]{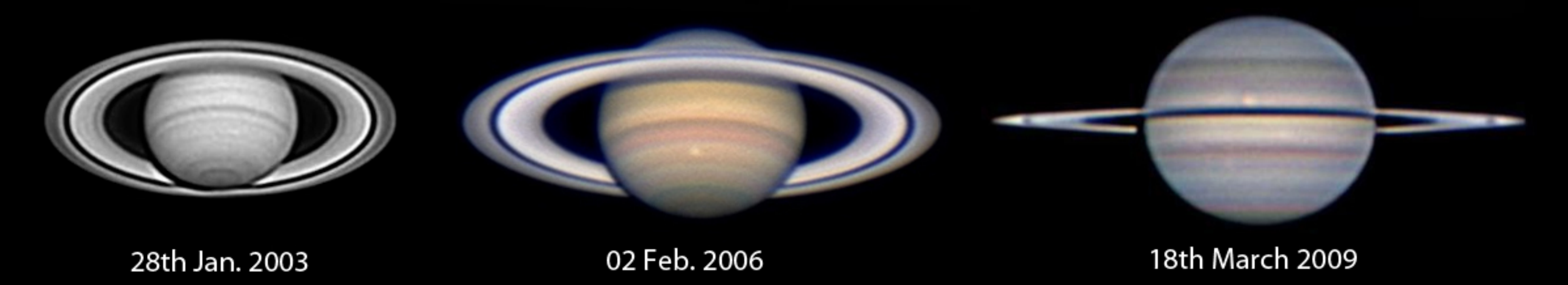}
\caption{Examples of images used to survey the global convective activity of Saturn. {\bf From left to right:} credits D. Peach (small storm at the ``storm-alley''), J. R. S\'anchez (storm imaged by Cassini and nicknamed dragon storm), and M. Lecompte (unusual equatorial convective feature). North is up.}
\end{center}
\label{Saturn1}       
\end{figure}

\clearpage

\begin{figure}
\begin{center}
\includegraphics[scale=.190]{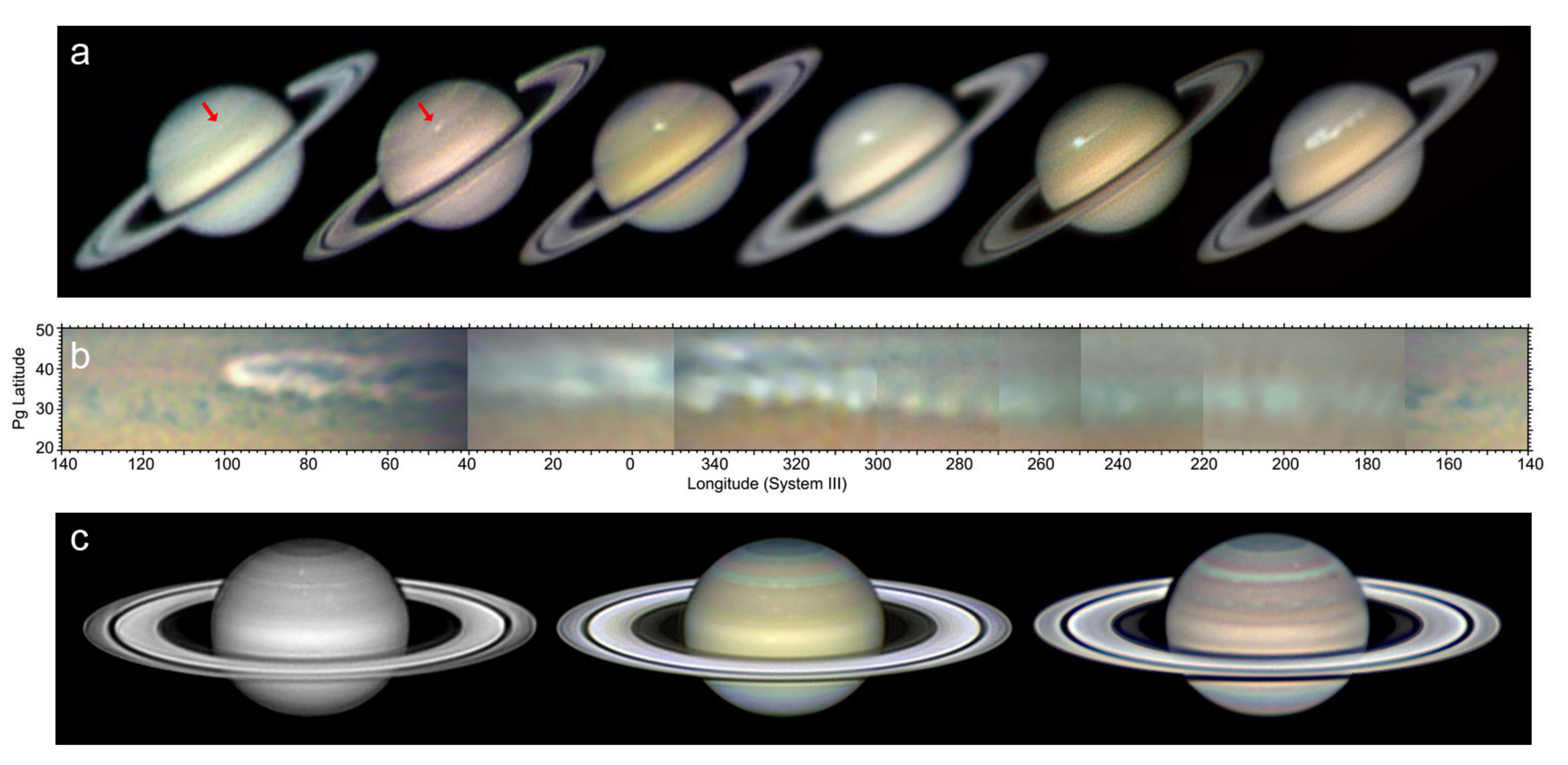}
\caption{{\bf a)} Early development of the Great White Spot, showing its growth and zonal expansion over December 2010. {\bf From left to right:} images by (day in December indicated) T. Ikemura (5th), S. Ghomizadeh (8th), T. Kumamori (9th), A. Wesley (10th), C. Go (13rd), T. Akutsu (26th). {\bf b)} Mature stage of the storm on February 19, 2011 as imaged by J. Hottershall (Australia), E. Morales (Puerto Rico) and G. Walker (USA). Complete longitudinal cover of the planet is only attainable from combined observations obtained from distant points on the Earth. {\bf c)} Observations of Saturn over 2012 after the storm had weakened and disappeared. {\bf From left to right:} images by A. Wesley (April 12, 2012), D. Peach (April 21, 2012) and D. Parker (May 7, 2012).}
\end{center}
\label{fig:Saturn_Figure3}       
\end{figure}

\clearpage

\begin{figure}
\centering
\includegraphics[scale=.480]{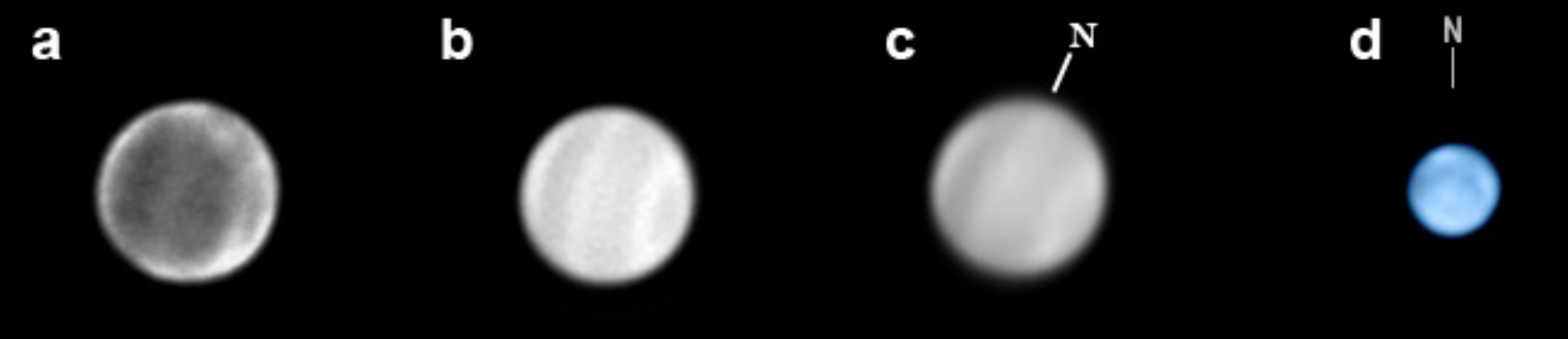}
\caption{Uranus and Neptune observed by amateur astronomers. {\bf a)} Uranus observed on October 30, 2011 {by a PRO-AM team} at the {1.05-m} telescope of Pic du Midi observatory, France (credit F. Colas and J. L. Dauvergne). Cloud banding is apparent as well as a convective bright feature on the upward limb. {\bf b)} Uranus observed on August 8, 2012 at the same telescope (credit F. Colas, J. L. Dauvergne, M. Delcroix, T. Legault and C. Viladrich). {\bf c)} Uranus observed on October 6, 2012 on one of the 1-meter C2PU telescopes at Calern, France (credit J.P. Prost and D. Vernet). Banding is apparent in all images. Images were acquired with R+IR filters that transmit light at longer wavelengths than 685 nm. {\bf d)} Neptune observed on September 25, 2010 with a 14'' telescope at visual wavelengths demonstrating the capability to detect features in the planet (credit D. Peach).}
\label{fig:uranus_neptune_IMAGES}       
\end{figure}

\clearpage

\begin{figure}
\centering
\includegraphics[scale=.35]{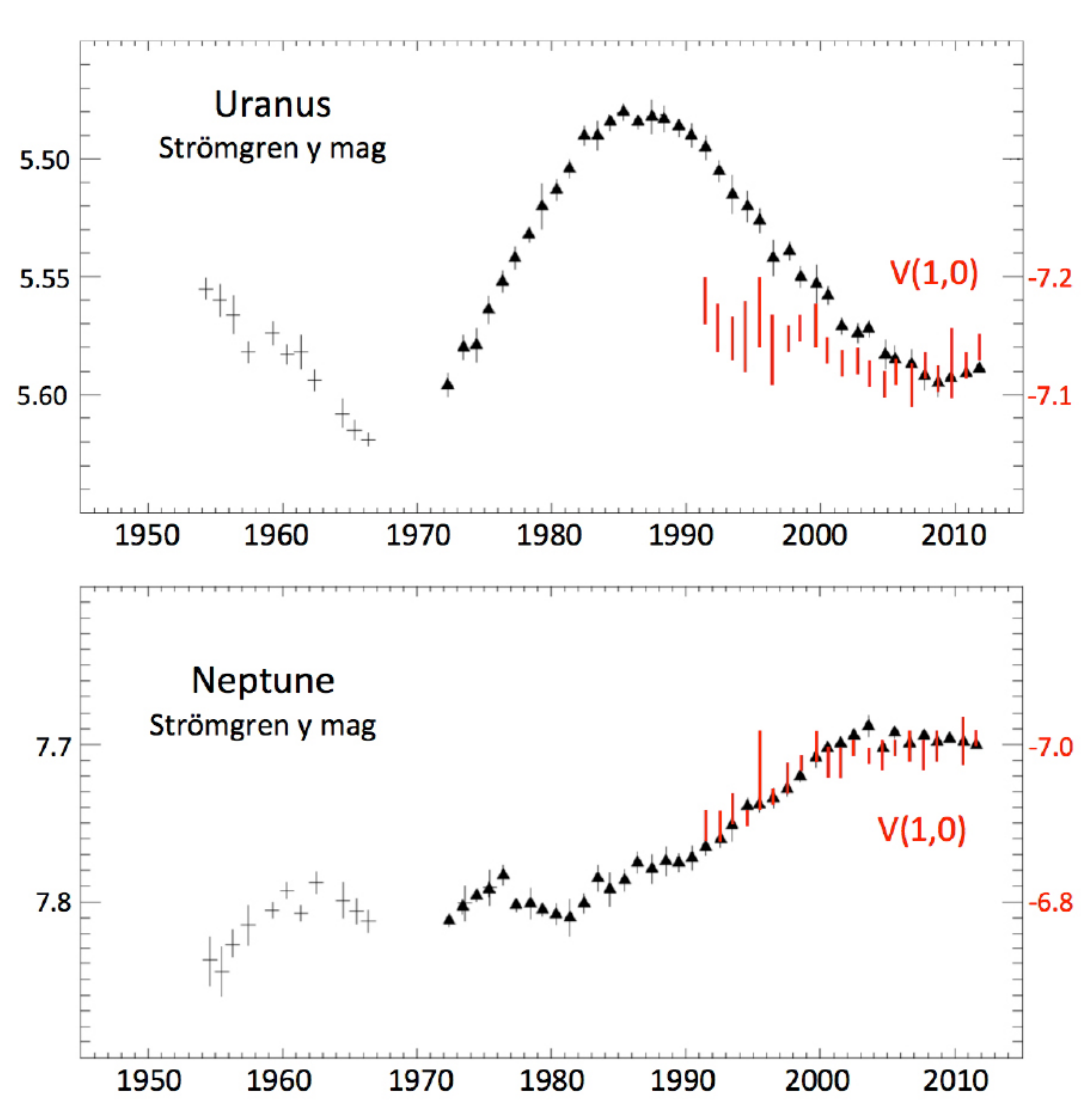}
\caption{Uranus and Neptune photometry.  Upper panel: The black symbols are from Lowell Observatory, and represent observations at (or converted to) the Stromgren y bandpass (updates provided by personal communication from W. Lockwood). The red symbols correspond to amateur data and represent the annual average of the normalized magnitude of Uranus transformed to the Johnson V system.  The normalized magnitude V(1,0) is the brightness the planet would have if it were 1 AU from both the Earth and Sun at a solar phase angle of zero degree.  The height of the bar is the uncertainty range of each measurement.  The lower panel presents similar data sets for Neptune.}
\label{fig:uranus_neptune_PHOTOMETRY}       
\end{figure}

\clearpage

\begin{figure}
\centering
\includegraphics[scale=1.2]{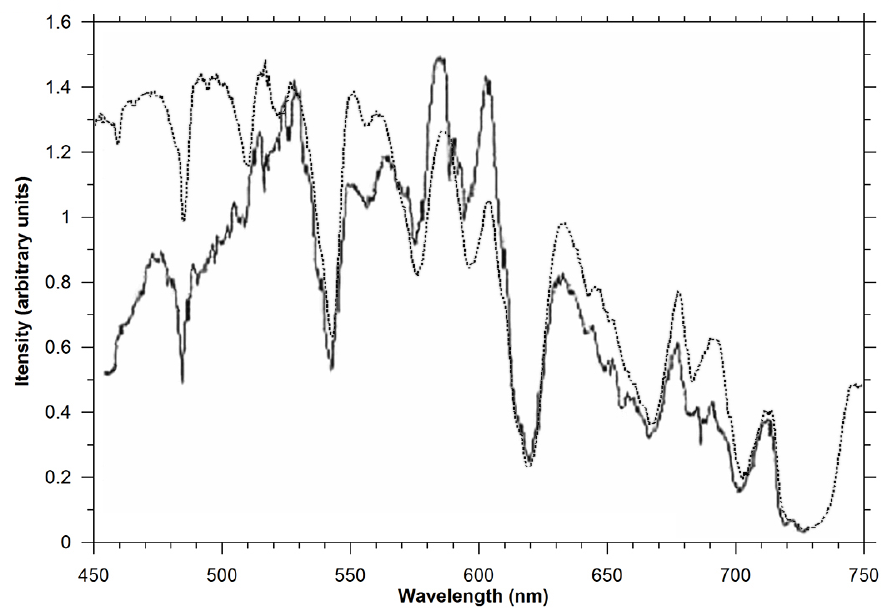}
\caption{Amateur (dotted line) and professional (solid line) spectra of Uranus. The amateur spectrum was recorded at visible-wavelengths on October 24, 2011 with a 0.25-m aperture telescope (credit  F. Melillo). The professional spectrum was acquired in 1995 by the 1.52-m ESO telescope \cite{Karkos98}. Differences between both spectra are apparent and due to calibration issues of the amateur data. The main bands are observable and identifiable. }
\label{fig:uranus_SPECTRUM}       
\end{figure}

\clearpage

\begin{figure}
\centering
\includegraphics[scale=0.45,angle=0]{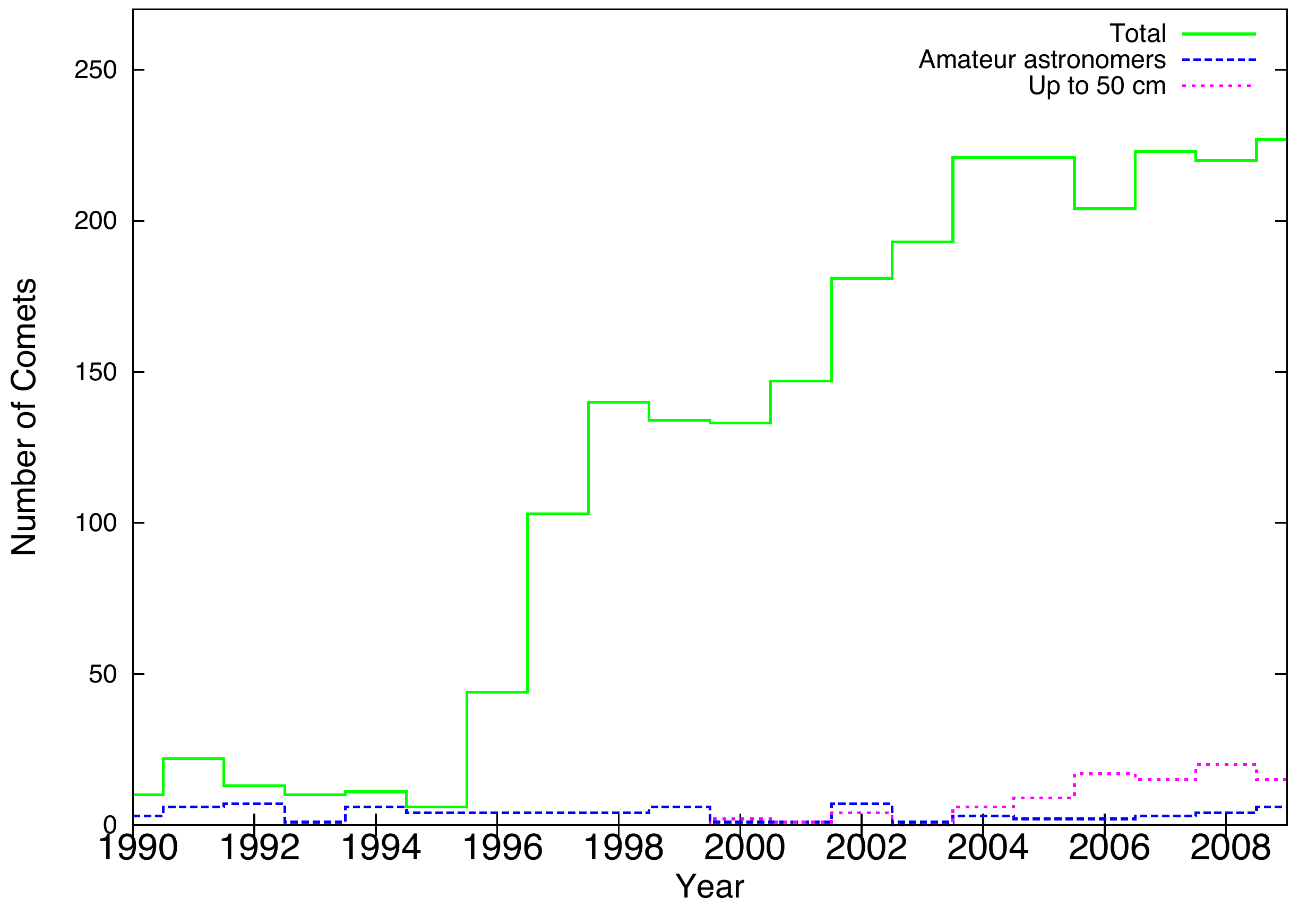}
\caption{Number of comets discovered per year for the period 1990--2009. This plot is based on the data available at the COCD website (Catalogue of Comet Discoveries, see: www.comethunter.de/). The discoveries classified as ``Amateur astronomers'' correspond to the data provided in the COCD. The data with telescope size up to 50~cm does not distinguish professional or amateur astronomers results. SOHO discoveries of comets started in 1996, LINEAR discoveries in 1998, Catalina sky survey discoveries in 2003 and
the one of Pan-STARRS in 2010. {In the 2000s, SOHO discoveries represent about 80\% of the total number of discoveries.}}
\label{f:stat}
\end{figure}

\clearpage

\begin{figure}
\centering
\includegraphics[scale=0.45,angle=0]{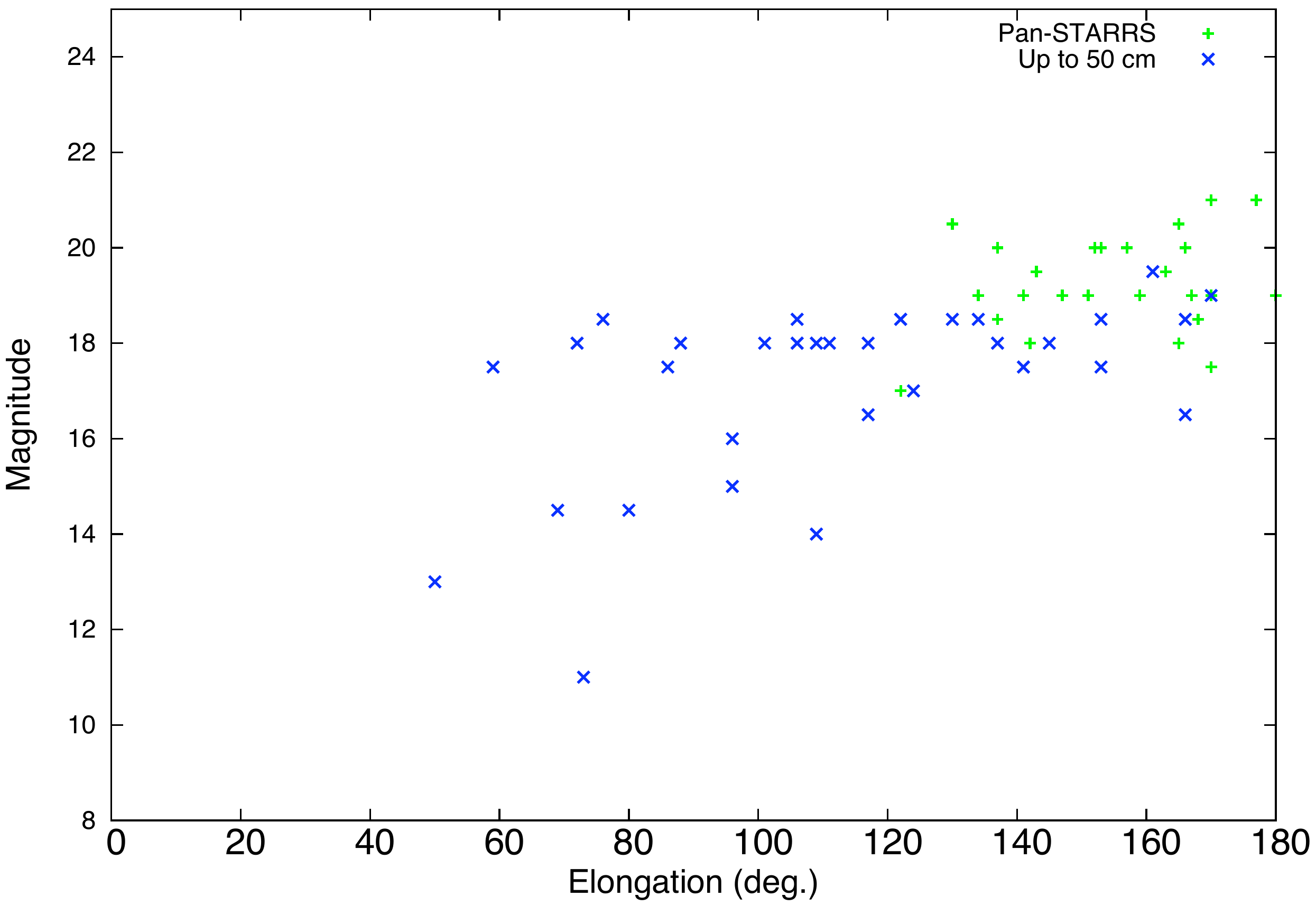}
\caption{{Apparent magnitude of comets at the time of discovery} as a function of solar elongation. This plot is based on data avilable at the COCD website (see above) and takes into account all the discoveries in the period 2010--2012.}
\label{f:elong}
\end{figure}

\clearpage

\begin{figure}
\begin{center}
\includegraphics[scale=.70]{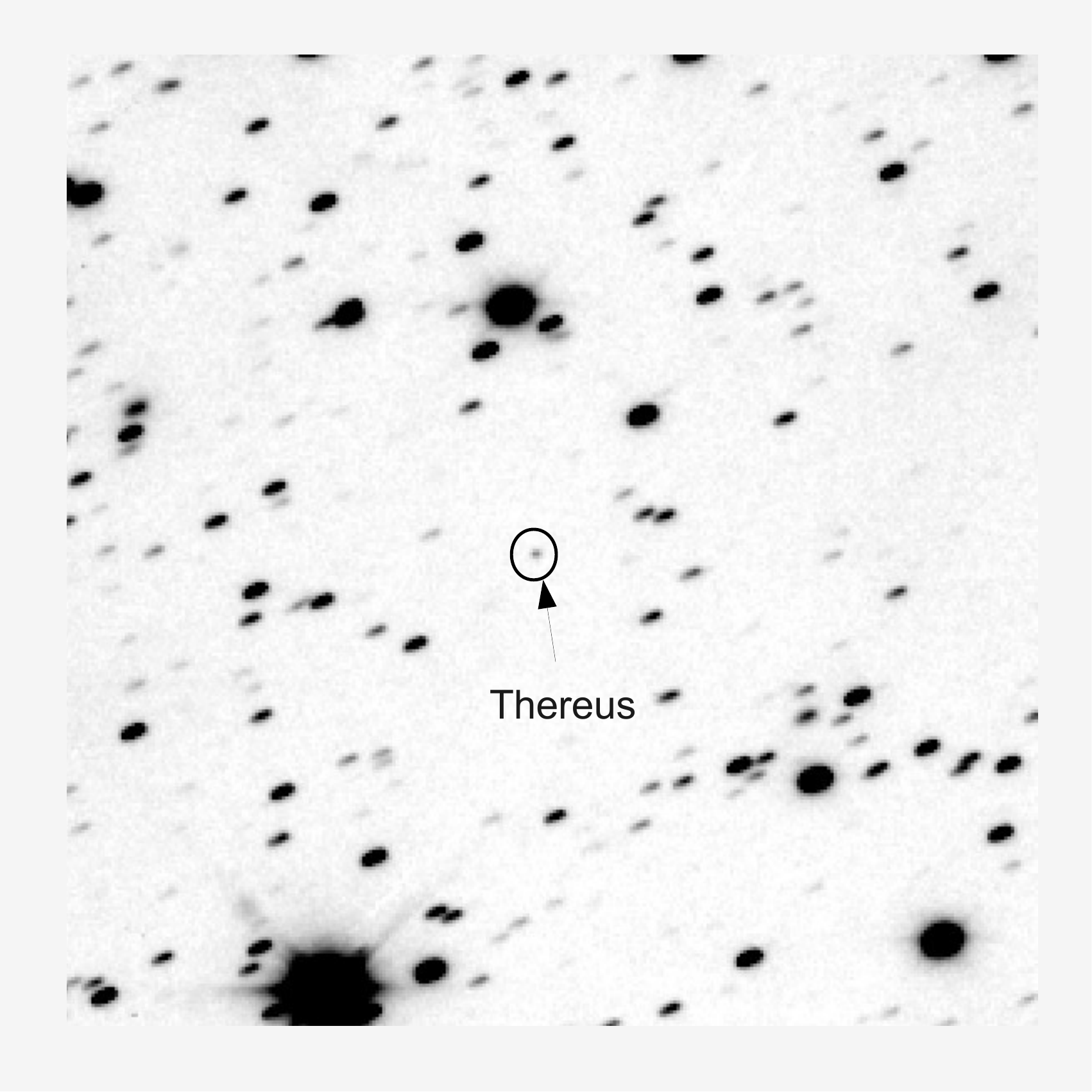}
\caption{Image of Thereus, a magnitude-20.4 Centaur, taken from Oukaimeden Observatory, Morocco, on the evening of 2013-01-31, with a 50-cm aperture, F/3 telescope, obstruction 35\%, STL 11000 camera, pixel size of 1.24''/pixel, and a measured seeing of about 3'' (credit C. Rinner). The image is the shifted sum of 25 2-min exposures. The images were shifted at the object's motion of 5.82''/h, direction 292$^\circ$ from North, positive to East. Thereus is the point like source in the middle of the frame, while the stars appear trailed.}
\label{fig:1-8.1}
\end{center}
\end{figure}

\clearpage

\begin{figure}
\begin{center}
\sidecaption
\includegraphics[scale=.18]{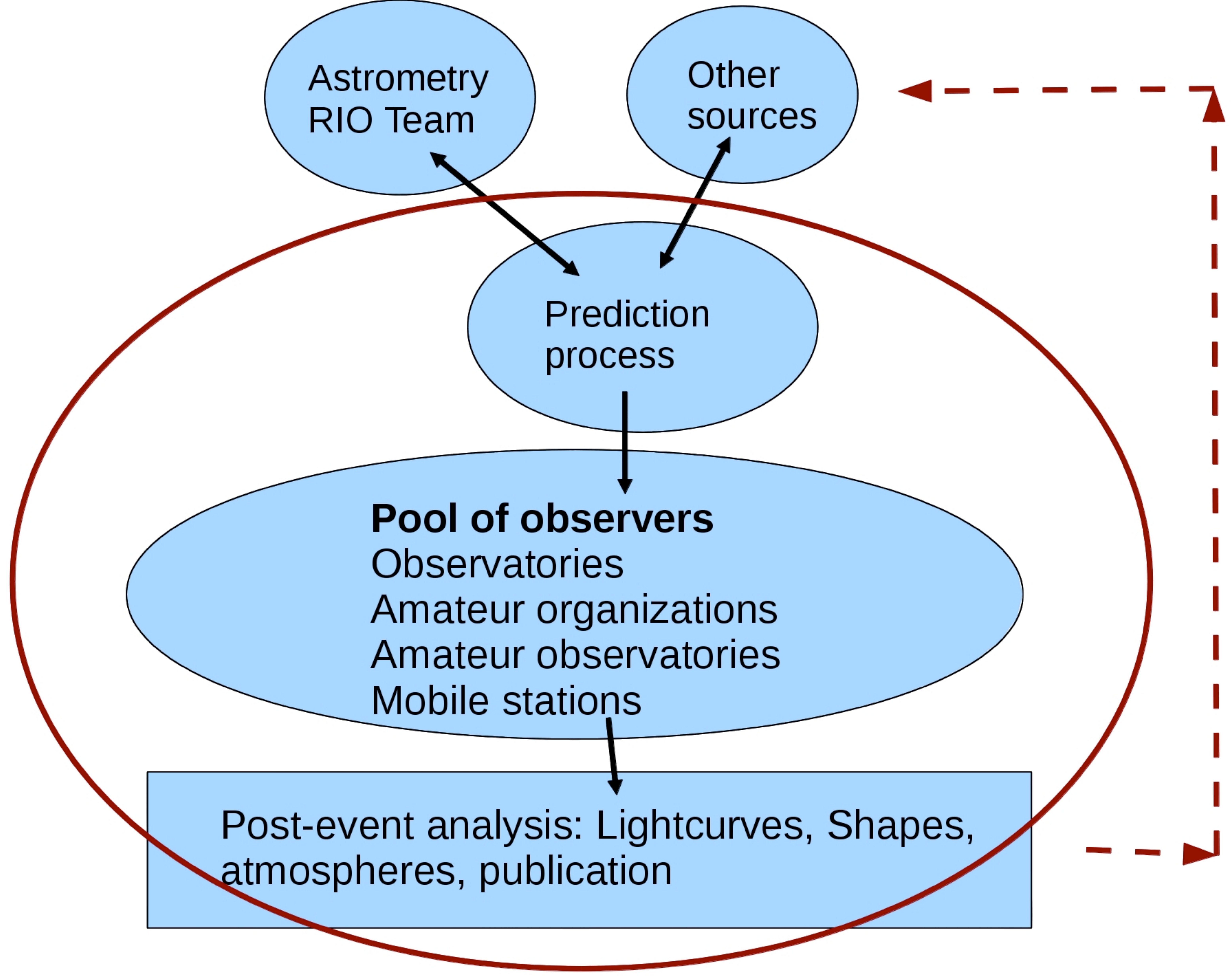}
\caption{The pipeline of the full project. {The red circle assembles the activity domains where amateur astronomers take part. Mostly they contribute to observation campaigns, but to a lesser extent they also participate in the prediction process and the lightcurve analysis and shape determination. The red arrow describes the feedback between results derived from one occultation for astrometry improvement for a new upcoming event. The observed occultations provide extremely precise positions of the occulting body at the time of occultation, which refines the orbit determination of the TNO.}}
\label{fig:1-8.2}      
\end{center}
\end{figure}

\clearpage

\begin{figure}
\begin{center}
\includegraphics[scale=.17]{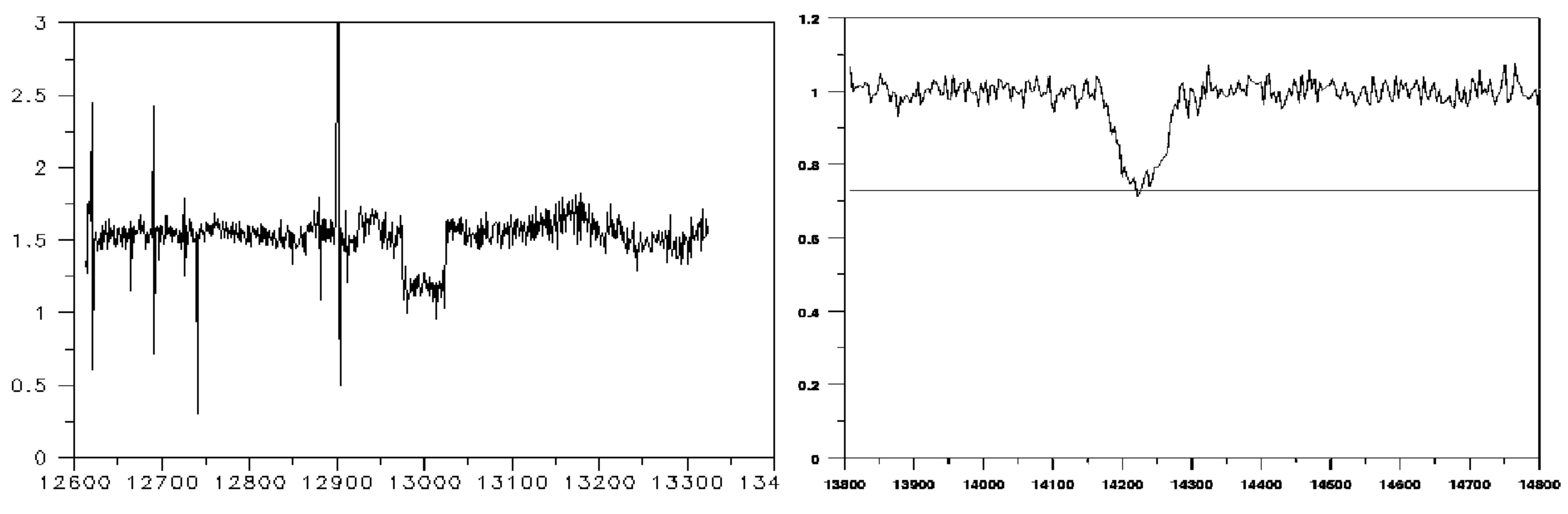}
\caption{{\bf Left:} Occultation of a 15 mag star by Charon as observed with the 2~m telescope at Complejo Astronomico El Leoncito (CASLEO), Argentina. {\bf Right:} Occultation of a 16 mag star by Triton observed with a 50-cm telescope at the Internationale Amateursternwarte (IAS) in Hakos, Namibia.}
\label{fig:2-8.2}      
\end{center}
\end{figure}

\clearpage

\begin{figure}[h]
\begin{center}
\includegraphics[scale=.6]{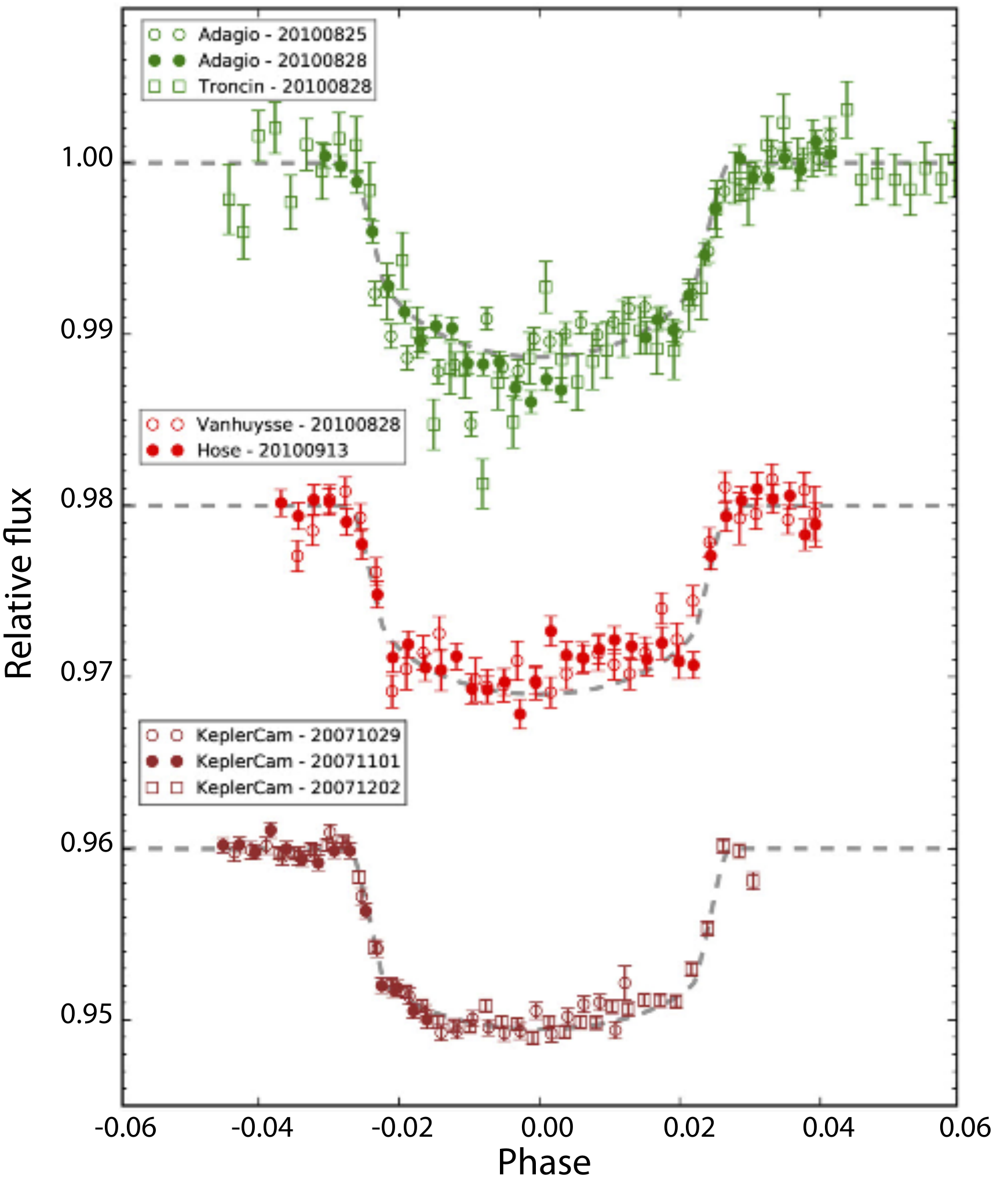}
\caption{Comparison of transit lightcurves of HAT-P-8~b obtained from amateur and professional observatories \cite{Moutou11}. From top to bottom: 82-cm amateur telescope, SBIG STL-6303E camera and V-band filter (credit Adagio association); OHP 120-cm professional telescope, V-band filter (credit J.-P. Troncin); 35-cm amateur telescope, SBIG STL-1001e camera and R-band filter (credit M. Vanhuysse); 32-cm amateur telescope, QSI 516wsg camera, R-band filter (credit K. Hose); KeplerCam: FLWO 1.2-m professional telescope, z-band filter (data binned to 10 minutes).}
\label{HATP8}
\end{center}
\end{figure}

\end{document}